\documentclass[a4paper,11pt]{article}
\usepackage{amsmath}
\usepackage{braket}
\usepackage{amssymb}
\usepackage{slashed}
\usepackage[normalem]{ulem}
\usepackage{color,xcolor}
\pdfoutput=1 

\usepackage{jheppub} 
\usepackage{soul}
\usepackage[T1]{fontenc} 

\title{On the breakdown of eikonal approximation and survival of Reggeization in presence of dimension-5 Higgs-gluon coupling}

\author[a,1]{Michael Fucilla,\note{Corresponding author.}}
\author[a]{Maxim A. Nefedov,}
\author[b,c]{Alessandro Papa}

\affiliation[a]{Université Paris-Saclay, CNRS, IJCLab, 91405 Orsay, France}
\affiliation[b]{Dipartimento di Fisica, Universit\`a della Calabria, I-87036 Arcavacata di Rende, Cosenza, Italy}
\affiliation[c]{Istituto Nazionale di Fisica Nucleare, Gruppo collegato di Cosenza, I-87036 Arcavacata di Rende, Cosenza, Italy}

\emailAdd{michael.fucilla@ijclab.in2p3.fr}
\emailAdd{maxim.nefedov@ijclab.in2p3.fr}
\emailAdd{alessandro.papa@fis.unical.it}

\abstract{
We consider the one-loop effective vertex for the interaction of a gluon with a Reggeized gluon and a Higgs boson in the infinite-top-mass limit, which is described by a dimension-5 non-renormalizable operator. This vertex enters the calculation of differential cross sections for the forward inclusive production of a Higgs boson in high-energy proton-proton collisions, possibly in association with a backward jet or identified hadron, in a framework where next-to-leading logarithms of the energy are resummed to all orders.  The effective vertex is extracted from the high-energy behavior of two-to-two amplitudes for the Higgs production in parton-parton collisions and relies on the validity of the Regge form for these amplitudes. We find that the usual eikonal approximation (Gribov prescription) for the Regge limit and the known region-expansion technique in this limit lead to an incomplete result for the amplitude. The discrepancy is traced back to the non-renormalizable nature of the involved operator. However, the Regge limit of the exact QCD amplitude agrees with the Regge-pole exchange form at one loop, nontrivially supporting the Reggeization hypothesis. }


\arxivnumber{...}

\begin{document} 
\maketitle
\flushbottom

\clearpage

\section{Introduction}
\label{sec:intro}

The discovery potential of New Physics at current colliders, such as the Large Hadron Collider (LHC), and at next-generation machines, such as the Electron-Ion Collider (EIC) or the Future Circular Collider (FCC), crucially depends on our ability to achieve a detailed knowledge of the structure of protons and nuclei and to produce precise predictions by the perturbative treatment of the theory of strong interactions, Quantum Chromodynamics (QCD). In some kinematic regimes, the standard approach based on {\em collinear factorization} between {\em coefficient functions}, describing the hard partonic scattering and calculated at fixed order in perturbation theory, and {\em nonperturbative functions}, describing the partonic distribution (PDFs) in the colliding objects and the parton fragmentation (FFs) into hadronic states, is insufficient to reach the needed level of precision and must be enhanced by all-order {\em resummations}. One of such kinematic regimes is the so-called {\em semi-hard} one, characterized by a neat scale hierarchy, $s\gg\{Q_i^2\}\gg\Lambda_{\rm QCD}^2$, where $s$ is the squared center-of-mass energy, $\{Q_i\}$ is a set of hard scales typical of the process and $\Lambda_{\rm QCD}$ is the QCD mass scale. Here, the necessary resummation is that of energy logarithms, which enter the perturbative calculation with a power increasing along with the perturbative order and thus compensate the smallness of the coupling $\alpha_s$, ensured, in its turn, by the hardness of the process. 

The framework for the resummation of energy logarithms was established long ago by Balitsky-Fadin-Kuraev-Lipatov (BFKL)~\cite{Fadin:1975cb,Kuraev:1976ge,Kuraev:1977fs,Balitsky:1978ic}; it allows the systematic resummation of all terms proportional to $[\alpha_s\ln(s)]^n$ (leading-logarithmic approximation or LLA), and of those of the form $\alpha_s[\alpha_s\ln(s)]^n$ (next-to-leading logarithmic approximation or NLLA). At the basis of the BFKL framework is the property of gluon Reggeization in QCD~\cite{Grisaru:1973vw,Grisaru:1973wbb,Lipatov:1976zz}, which means that there is a Reggeon with gluon quantum numbers, negative signature and trajectory $j(t) = 1 + \omega(t)$ passing through 1 at $t = 0$, which gives the leading contribution  to amplitudes with gluon quantum numbers in the $t$-channel, in each order of perturbation theory.
This remarkable property appeared first in direct calculations at fixed order~\cite{Lipatov:1976zz,Kuraev:1976ge}. Then, it was proved, both in the LLA~\cite{Balitsky:1979ap} and in the NLLA (see~\cite{Fadin:2015ym} and references therein) using bootstrap relations~\cite{Fadin:2006pr} following from the requirement of compatibility of the pole Regge form with the $s$-channel unitarity. 

In the BFKL approach, (differential) cross sections can be written as the convolution, in the 
space of transverse momenta, of two process-dependent impact factors, describing the transition of each colliding particle to a definite state in its fragmentation region, and a universal, process-independent Green's function, which encodes the resummation of energy logarithms. The BFKL Green's function is determined by an integral equation, whose kernel is known up to next-to-leading order (NLO), both for forward scattering ({\it i.e.} for $t = 0$ and color singlet in the $t$-channel)~\cite{Fadin:1998py,Ciafaloni:1998gs} and for any fixed, not growing with $s$, momentum transfer $t$ and any possible two-gluon colored exchange in the $t$-channel~\cite{Fadin:1998jv,Fadin:2000kx,Fadin:2000hu,Fadin:2004zq,Fadin:2005zj}.
Recent years have witnessed some promising steps towards the calculation of ingredients of the next-to-NLO kernel, in $\mathcal{N}=4$ SYM~\cite{Byrne:2022wzk}, in pure-gauge QCD~\cite{DelDuca:2021vjq} and in full QCD~\cite{Caola:2021izf,Falcioni:2021dgr,Fadin:2023roz}.
The perturbative calculation of impact factors with NLO accuracy is a challenging task and, indeed, only a few of them are known with this accuracy: (\emph{i}) quark and gluon impact factors~\cite{Fadin:1999de,Fadin:1999df,Ciafaloni:1998kx,Ciafaloni:1998hu,Ciafaloni:2000sq}, which are at the basis of the calculation of the (\emph{ii}) forward-jet~\cite{Bartels:2001ge,Bartels:2002yj,Caporale:2011cc,Ivanov:2012ms,Colferai:2015zfa} and (\emph{iii}) forward light-hadron~\cite{Ivanov:2012iv} impact factors, (\emph{iv}) the impact factor for the light vector-meson electroproduction, (\emph{v}) the ($\gamma^* \to \gamma^*$) impact factor~\cite{Bartels:2000gt,Bartels:2001mv,Bartels:2002uz,Bartels:2003zi,Bartels:2004bi,Fadin:2001ap,Balitsky:2012bs}, and (\emph{vi}) the one for the forward-Higgs production from an incoming proton in the infinite-top-mass limit~\cite{Hentschinski:2020tbi,Celiberto:2022fgx,Nefedov:2019mrg}. Predictions for several observables have been built with full NLLA, by combining NLO impact factors with the NLLA Green's function, or with only a partial inclusion of NLLA effects, by convoluting the NLLA Green's with one or both the impact factors taken at the leading order (LO), up to the NLO corrections
dictated by renormalization group invariance (see e.g. ~\cite{Celiberto:2020wpk,Celiberto:2022keu} and references therein).

The availability of the NLO Higgs impact factor opens new interesting channels whereby testing high-energy effects on the Higgs production mechanism. These channels comprise the inclusive forward production of a Higgs in concomitance with a backward jet or an identified hadron (a {\it forward-backward} process) or the single inclusive production of a Higgs (a {\it single forward} process).
In the first case the theoretical description is based on the convolution of the Higgs impact factor with the BFKL Green's function and the impact factor for the production of the backward object; in the second case the Higgs impact factor needs to be convoluted with an {\it unintegrated gluon distribution}~(see for instance \cite{Bolognino:2018rhb}). So far, theoretical predictions for the forward-backward Higgs plus jet production have been carried out with partial inclusion on NLLA corrections~\cite{Celiberto:2020tmb,Celiberto:2022zdg,Celiberto:2023rtu}, complemented in the most recent investigations by a matching procedure with fixed-order calculations~\cite{Celiberto:2023uuk,Celiberto:2023eba}. The comparison with predictions based on fixed-order Monte Carlo shows deviations and further motivates the extension of the analysis with the full inclusion of NLLA effects, {\it i.e.} by using NLO impact factors for both the forward Higgs and the backward identified object, and by relaxing the infinite-top-mass approximation. \\

In this paper, we concentrate on some subtle technical issues which appeared in the calculation of the NLO Higgs impact factor in the infinite-top-mass limit~\cite{Celiberto:2022fgx}, which are closely related to the use of the effective 5-dimensional Lagrangian for the direct coupling of gluons to Higgs. Indeed, it is known that a local Higgs-gluon coupling can give rise to peculiar perturbative behaviors~\cite{Catani:1990xk, Catani:1990eg, Hautmann:2002tu}. The NLO Higgs impact factor takes contribution from the so-called {\it real} corrections, related with the production of a parton in the association with the Higgs in the forward region, and the {\it virtual} ones, which take into account loop corrections. The main ingredient of the latter is the effective gluon-Reggeon-Higgs vertex at one-loop order, denoted by $\Gamma_{gH}^{(1)}$ (up to color indices) in what follows. In the BFKL framework, the procedure to determine a particle-Reggeon-particle vertex is by calculating the high-energy limit of an amplitude with gluon quantum number exchange in the $t$-channel where this vertex enters, and to extract it by comparing the calculated amplitude with its predicted form according to the hypothesis of gluon Reggeization. As discussed above, this hypothesis has been proven in QCD, but is not guaranteed to work in presence of an extra vertex coming from a non-renormalizable effective Lagrangian. Indeed, in Ref.~\cite{Celiberto:2022fgx}, the one-loop gluon-Reggeon-Higgs vertex was extracted {\it assuming}
that the Regge form holds for the amplitude gluon + quark to Higgs + quark, $\mathcal{A}_{gq \rightarrow Hq}$. More precisely, it was assumed that the {\it interference} between the Born and the one-loop correction terms of that amplitude is as dictated by the Regge form. 
The considerations in Ref.~\cite{Celiberto:2022fgx} were made at the level of the interference between the Born and the one-loop amplitude because it was observed that the one-loop amplitude contains unexpected leading-$s$ contributions, which come from the transverse part of some $t$-channel gluon propagators in covariant Feynman gauge. These contributions are absent in pure QCD calculations, where $t$-channel gluons are dominated by a longitudinal component, which ultimately leads to what are known as {\it eikonal approximation} in high-energy scattering or  {\it Gribov} (sometimes called ``\textit{nonsense}'') polarizations for the Reggeized gluons. The unexpected, {\it ``non-Gribov''} contributions which appear in presence of a dimension-5 operator could endanger the Regge form of the one-loop amplitude, which therefore must be checked explicitly. This is the main aim of the present paper.

Another subtle point, which arose in the calculation of $\Gamma_{gH}^{(1)}$ in~\cite{Celiberto:2022fgx}, is related to the contributions to the scattering amplitude coming from exchange of two gluons in the $t$-channel. In QCD, a powerful technique has been developed~\cite{Fadin:1999qc,Fadin:2001dc} to calculate the high-energy limit of the box (and crossed) diagrams which appear in this part of the calculation, which we refer to as {\it method of rapidity regions}. It exploits one of the natural characteristics of high-energy scattering, the rapidity factorization. In~\cite{Celiberto:2022fgx} it was found that the naive implementation of this technique fails. Indeed, in that work the two-gluon exchange part of the gluon + quark to Higgs + quark amplitude was first calculated by brute force and then the high-energy limit was taken to extract $\Gamma_{gH}^{(1)}$. In the present paper we identify the culprit in the failure of the method of regions for the dimension-5 Higgs effective vertex and demonstrate that this latter also spoils the rapidity partitioning of the amplitude. The unexpected {\it factorization-breaking} terms arise in the ``target'' rapidity region (region B) of the $gq\to Hq$ and $gg\to Hg$ amplitudes, when the method of regions with Gribov's prescription is used for the computation. However the {\it non-Gribov} terms cancel this factorisation-violating contributions, restoring the expected rapidity partitioning and, ultimately, the Reggeization. 

Another goal of the paper is to verify that the effective one-loop vertex $\Gamma_{gH}^{(1)}$ is unambiguous, by checking that its extraction from the gluon + gluon to Higgs + gluon scattering amplitude, $\mathcal{A}_{gg \rightarrow Hg}$, leads to the same result as the extraction from $\mathcal{A}_{gq \rightarrow Hq}$.
This also allows to verify the agreement between the calculation performed in Ref.~\cite{Celiberto:2022fgx} and that carried out within the Lipatov's effective action framework (Lipatov's EFT) in Ref.~\cite{Nefedov:2019mrg}. Indeed, the calculation of the effective vertex $\Gamma_{gH}^{(1)}$ was verified, both in Ref.~\cite{Celiberto:2022fgx} and in Ref.~\cite{Nefedov:2019mrg}, by using the large-$s$ limit of one-loop scattering amplitudes taken from the literature. The point is that in Ref.~\cite{Celiberto:2022fgx} the gluon-quark channel was used ({\it i.e.} the $\mathcal{A}_{gq \rightarrow Hq}$ amplitude), while in Ref.~\cite{Nefedov:2019mrg} the gluon-gluon one was considered ({\it i.e.} the $\mathcal{A}_{gg \rightarrow Hg}$ amplitude).

The results presented in this paper, though obtained to fix some issues related to the presence of a specific non-renormalizable interaction, can turn to be useful in the application of high-energy perturbative techniques in contexts different from the one considered here, since they uncover the subtle interplay between renormalisability, eikonal approximation and gluon Reggeization. \\

The paper contains six sections. In section~\ref{sec:ThepoleReggeform}, we introduce the notation and review the standard Regge-pole form of the amplitude. In section~\ref{sec:GribovViolation}, we explain the appearance of non-Gribov contributions and discuss the one-loop Regge form of the gluon-quark into Higgs-quark scattering amplitude in their presence. In section~\ref{sec:RapidityRegions}, we investigate the lack of rapidity partitioning for this amplitude and factorisation-breaking terms. In section~\ref{sec:EFT}, we investigate the gluon-gluon into Higgs-gluon amplitude, demonstrating the non-ambiguity of the gluon-Reggeon-Higgs vertex and the agreement between the result obtained in the standard BFKL approach and the one obtained in the Lipatov's EFT framework. Section~\ref{sec:conclusions} contains our conclusions and outlook.

\section{The Regge-pole form}
\label{sec:ThepoleReggeform}

The pole Regge form of amplitudes in QCD, valid to all orders in perturbation theory both in the LLA and in the NLLA, implies that, in the high-energy limit $s \gg |t|$, scattering amplitudes $A+B \to A'+B'$ with exchange of gluon quantum numbers in the $t$-channel\footnote{Negative signature and color octet.} assume the factorized form\footnote{For the definition of transverse Minkowskian four-vector see Eq.~(\ref{GenericSudDec}).}
\begin{equation}
        \Gamma_{A'A} \; \frac{s}{t} \left[ \left( \frac{s}{-t} \right)^{\omega (t)} + \left( \frac{-s}{-t} \right)^{\omega (t)} \right] \Gamma_{B'B} \; ,
\label{Pole:Eq:ReggeForm} 
\end{equation}
where $1 + \omega (t)$ is the Regge trajectory, whose the expression for $\omega(t)$ at one-loop and in a space-time with dimension $D=4-2\epsilon$ reads~\cite{Lipatov:1976zz}
\begin{equation}
    \omega^{(1)}(t) = \frac{g^2 t}{(2 \pi)^{D-1}} \frac{N}{2} \int \frac{d^{D-2}k_{\perp}}{k_{\perp}^2 (q-k)_{\perp}^{2}} = - \frac{g^2 C_A \Gamma(1+\epsilon)(\vec{q}^{\; 2})^{-\epsilon}}{(4 \pi)^{2-\epsilon}} \frac{\Gamma^2(-\epsilon)}{\Gamma(-2\epsilon)} \; , \;\;\; t=q^2\simeq -\vec{q}^{\;2}\;,
    \label{ReggeTraj}
\end{equation}
and $\Gamma_{A'A}$ and $\Gamma_{B'B}$ are the coupling between external particles and the $t$-channel Reggeized gluon\footnote{Since in the following we consider just the Reggeized gluon, we will refer to it as Reggeon for simplicity.}. The $s$-behaviour of the amplitude in Eq.~(\ref{Pole:Eq:ReggeForm}) comes together with the Regge trajectory, while the couplings $\Gamma_{AA'}$ and $\Gamma_{BB'}$ can depend only on the Mandelstam variable~$t$. The Born quark-Reggeon-quark and gluon-Reggeon-gluon effective vertices can be both written as 
\begin{equation}
    \Gamma_{AA'}^{c (0)} = g \delta_{\lambda_{A'} \lambda_{A}} \braket{A'|T^c|A} \; ,
\end{equation}
where $\lambda_{A'}$ ($\lambda_{A}$) is the helicity of the outgoing (incoming) parton, while $\braket{A'|T^c|A}$ stands for the matrix element of the color group generator in the corresponding representation. \\ 

At one-loop level, particle-Reggeon-particle vertices can be extracted by considering the high-energy limit of a reference amplitude with gluon quantum numbers exchanged in the $t$-channel and comparing the fixed-order perturbative calculation with the pole Regge form~(\ref{Pole:Eq:ReggeForm}). For instance, to get the one-loop correction to the quark-Reggeon-quark vertex, one can calculate any elastic quark-quark scattering with gluon quantum numbers exchange in the $t$-channel at one-loop in the high-energy limit and compare it with its one-loop-expanded Regge form,
$$
{\cal A}_{q q \rightarrow q q}^{(8,-)} = \Gamma_{q q}^{c} \frac{s}{t}\left[ \left( \frac{s}{-t} \right)^{\omega(t)} + \left(
\frac{-s}{-t} \right)^{\omega(t)} \right] \Gamma_{qq}^{c} \approx 
\Gamma_{q q}^{c(0)} \frac{2s}{t} \Gamma_{q q}^{c(0)}
$$
\begin{equation}
+ \Gamma_{qq}^{c(0)} \frac{s}{t}\omega^{(1)}(t)
\left[ \ln\left( \frac{s}{-t} \right) + \ln\left(
\frac{-s}{-t} \right) \right] \Gamma_{q q}^{c(0)} + \Gamma_{qq}^{c(0)} \frac{2s}{t}\Gamma_{q q}^{c(1)} +
\Gamma_{qq}^{c(1)} \frac{2s}{t} \Gamma_{qq}^{c(0)}  \; .
\label{Pole:Eq:ReggeFormEx1}
\end{equation}
Using the known expression for the one-loop Regge trajectory in Eq.~(\ref{ReggeTraj}) and for the Born quark-Reggeon-quark vertex, one can easily find $\Gamma_{qq}^{c(1)}$.
Since this vertex will be useful in the following, we give here its expression (see, for instance,~\cite{Fadin:2001dc}):
\begin{equation}
    \Gamma_{qq}^{c(1)} = \Gamma_{qq}^{c(0)}  \delta_{Q} \; ,
    \label{Pole:Eq:QuarkCorr}
\end{equation}
with
\begin{gather}
\delta_Q=\omega^{(1)}(t) \frac{1}{2}\left[-\frac{1}{\epsilon}+\psi(1+\epsilon)+\psi(1)-2 \psi(1-\epsilon)+\frac{2-\epsilon}{2(1-2 \epsilon)(3-2 \epsilon)}\right. \nonumber \\ \left.-\frac{1}{2 C_A^2}\left(1 - \frac{2}{\epsilon(1 - 2 \epsilon)}\right)-\frac{n_f}{C_A} \frac{(1 - \epsilon)}{(1 - 2 \epsilon)(3 - 2 \epsilon)}\right]\;.
\label{Pole:Eq:QuarkCorrDelta}
\end{gather}
Similarly, one can find the one-loop correction to the gluon-Reggeon-gluon vertex, $\Gamma_{g g}^{c(1)}$, starting from the elastic gluon-gluon scattering amplitude, $\mathcal{A}_{gg \rightarrow gg}^{(8,-)}$~\cite{Fadin:2001dc}. The validity of gluon Reggeization then implies that the high-energy limit of the elastic quark-gluon amplitude within the NLLA must be reproduced by using the known expressions for $\omega^{(1)}(t), \Gamma_{q q}^{c(1)}$ and $\Gamma_{g g}^{c(1)}$. This is a manifestation of the {\it universal} behavior of particle-Reggeon-particle vertices and is induced by the property of gluon Reggeization, which is valid in QCD.

\section{The violation of Gribov's trick for a dimension 5-operator}
\label{sec:GribovViolation}
\subsection{Quark-quark elastic amplitude at one-loop}
\begin{figure}
      \centering
      \includegraphics[scale=0.50]{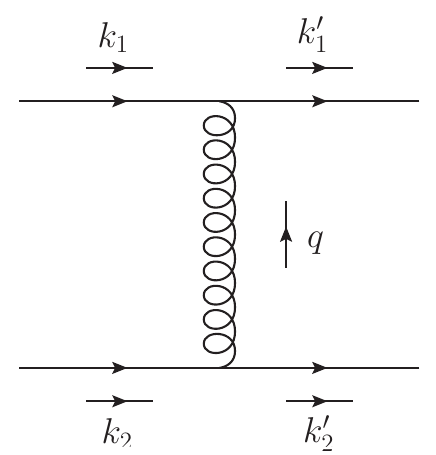}
      \caption{Dominant contributions to the quark-quark into quark-quark scattering amplitude in the high-energy limit.}
      \label{Fig:Gribov:QuarkQuark}
\end{figure}
In order to extract the high-energy behaviour of scattering amplitudes in QCD, a series of simplifications can be applied. One of the most useful is known as {\it eikonal approximation} or  {\it Gribov's trick} and, in order to illustrate it, let us first consider the Born quark-quark into quark-quark scattering amplitude (see Fig.~\ref{Fig:Gribov:QuarkQuark}). The amplitude is 
\begin{equation}
    \mathcal{A}_{q_1 q_{1'} \rightarrow q_2 q_{2'}}^{(8,-)} = (-i)  \bar{u} (k_1+q) (-i) g t_{ji}^a \gamma^{\mu} u (k_1) \left( \frac{-i g_{\mu \nu} \delta^{ab}}{q^2} \right) \bar{u} (k_2-q) (-i) g t_{kn}^b \gamma^{\nu} u (k_2) \; .
    \label{Int:Eq:GeneralQuarkAmp}
\end{equation}
We work in light-cone coordinate, using as basis the momenta $k_1$ and $k_2$ of the incoming particles\footnote{The incoming particles in the whole paper are always on the light-cone.}. Therefore, a generic $D$-vector $p$ is decomposed as
\begin{equation}
   p = \beta_p k_1 + \alpha_p k_2 + p_{\perp} \; ,
   \label{GenericSudDec}
\end{equation}
where $p_{\perp}$ is a $D$-vector orthogonal to the plane identified by $k_1$ and $k_2$. The Gribov's trick consists in replacing the numerator of all $t$-channel gluon propagators as 
\begin{equation}
    g^{\mu \nu} =  g_{\perp \perp}^{\mu \nu} + 2 \frac{k_2^{\mu} k_1^{\nu} + k_2^{\mu} k_1^{\nu} }{s} \longrightarrow \frac{2 k_2^{\mu} k_1^{\nu} }{s},
    \label{Gribov:Eq:GribovsTrick}
\end{equation}
in order to extract the high-energy behaviour of the amplitude in the leading power approximation in the Regge limit, {\it i.e.} up to corrections power-suppressed in $(-t)/s\ll 1$. In this way, one gets exactly the LO expansion of the Regge amplitude in Eq.~(\ref{Pole:Eq:ReggeForm}), {\it i.e.}
\begin{equation}
    \Gamma_{q_1 q_{1'}}^{c(0)} \frac{2s}{t} \Gamma_{q_2 q_{2'}}^{c(0)} \;,
\end{equation}
with 
\begin{equation}
    \Gamma_{q_1,q_{1'}}^{c (0)} = g t_{ji}^c \bar{u} (k_1+q) \frac{\slashed{k}_2}{s} u(k_1) \; \simeq \; g t_{ji}^c \delta_{\lambda_{q_{1'}} \lambda_{q_1}}
\end{equation}
and
\begin{equation}
    \Gamma_{q_2,q_2'}^{c (0)} = g t_{kn}^c \bar{u} (k_2-q) \frac{\slashed{k}_1}{s} u(k_2) \; \simeq \; g t_{kn}^c \delta_{\lambda_{q_{2'}} \lambda_{q_2}} \; .
\end{equation}
The possibility of using the Gribov's trick is just a manifestation of the \textit{eikonal} nature of the interaction between the upper (lower) external particles and the $t$-channel gluon in the Regge limit, which is exploited in all existing approaches to the physics of high-energy scattering in QCD. \\
\begin{figure}
  \begin{center}
  \includegraphics[scale=0.5]{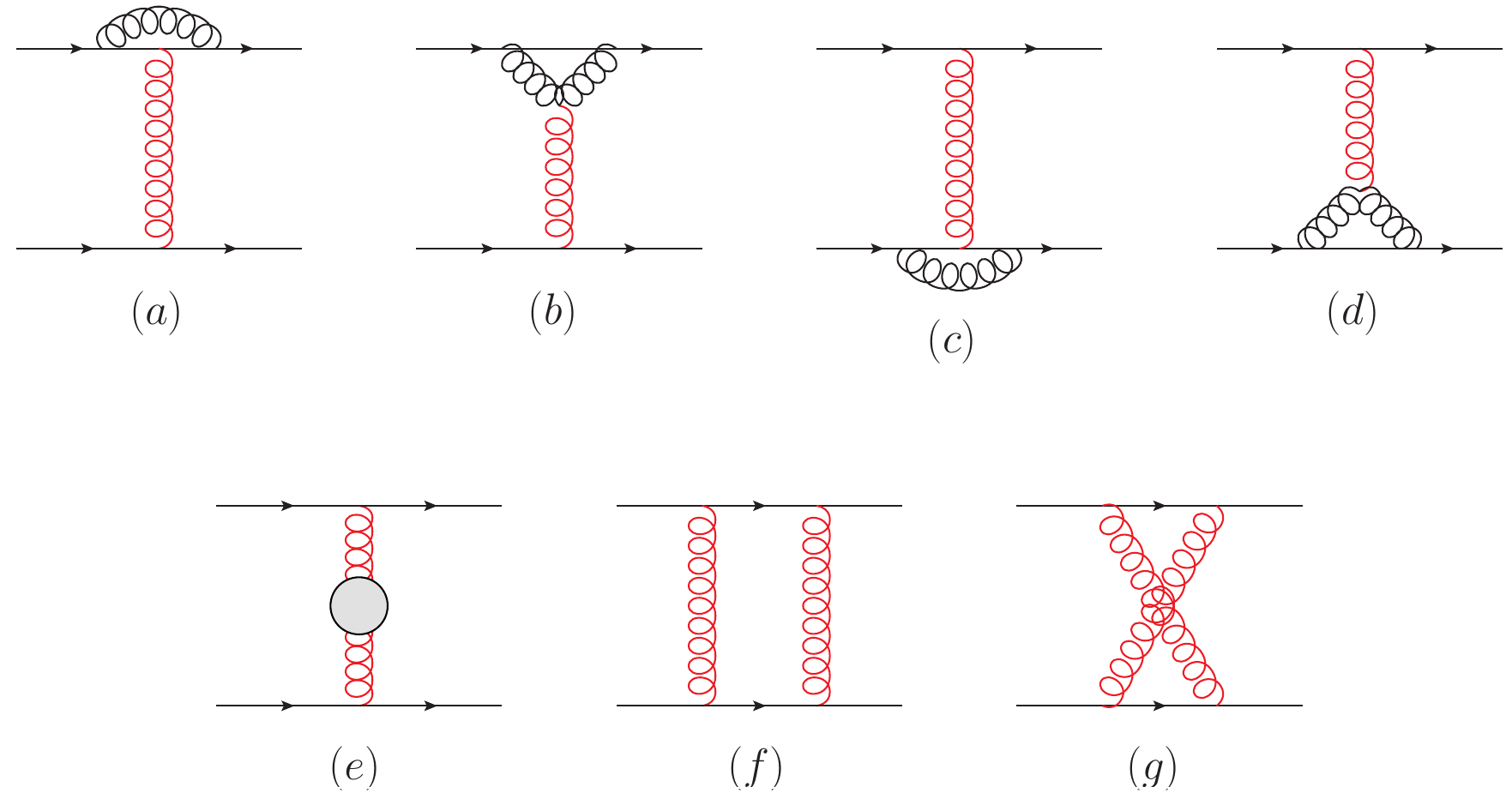}
  \end{center}
  \caption{Dominant contributions to the quark-quark into quark-quark scattering amplitude in the high-energy limit at one-loop level. The blob in the diagram $(e)$ incorporates quark, gluon and ghost loop contributions. All propagators in red are the ones that can be approximated à la Gribov.}
  \label{Fig:Gribov:GribovTrickQuark}
\end{figure}

The Gribov's trick becomes an essential tool at the NLO. Considering again the previous amplitude but at one-loop accuracy, we have nine contributing diagrams\footnote{Other diagrams are suppressed in the high-energy approximation.}
(see Fig.~\ref{Fig:Gribov:GribovTrickQuark}). Following the usual convention, we will refer to diagrams $(a),(b),(c),(d)$ and $(e)$ of Fig.~\ref{Fig:Gribov:GribovTrickQuark} as \textit{single-gluon} $t$-channel diagrams and to diagrams $(f)$ and $(g)$ of Fig.~\ref{Fig:Gribov:GribovTrickQuark} as \textit{two-gluon} $t$-channel diagrams. The blob in the diagram $(e)$ incorporates quark, gluon and ghost loop contributions.  One easily finds that, for all the gluon propagators connecting vertices with momenta predominantly along $k_1$ and $k_2$, respectively (the propagators in red in Fig.~\ref{Fig:Gribov:GribovTrickQuark}), the Gribov's trick can be applied and enables to find the correct high-energy behaviour of the amplitude. Using this the Gribov's trick and performing a projection onto the negative signature and octet color state in the $t$-channel, one finds exactly the form in Eqs.~(\ref{Pole:Eq:ReggeFormEx1}), (\ref{Pole:Eq:QuarkCorr}), (\ref{Pole:Eq:QuarkCorrDelta}). 

\subsection{$gq\to Hq$ amplitude at one-loop}
\begin{figure}
      \centering
      \includegraphics[scale=0.50]{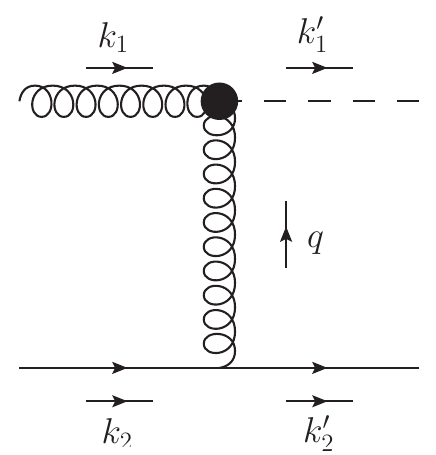}
      \caption{Dominant contribution to the gluon-quark into Higgs-quark scattering amplitude in the high-energy limit.}
      \label{Fig:Gribov:GluonHiggs}
\end{figure}

\begin{figure}
  \begin{center}
  \includegraphics[scale=0.5]{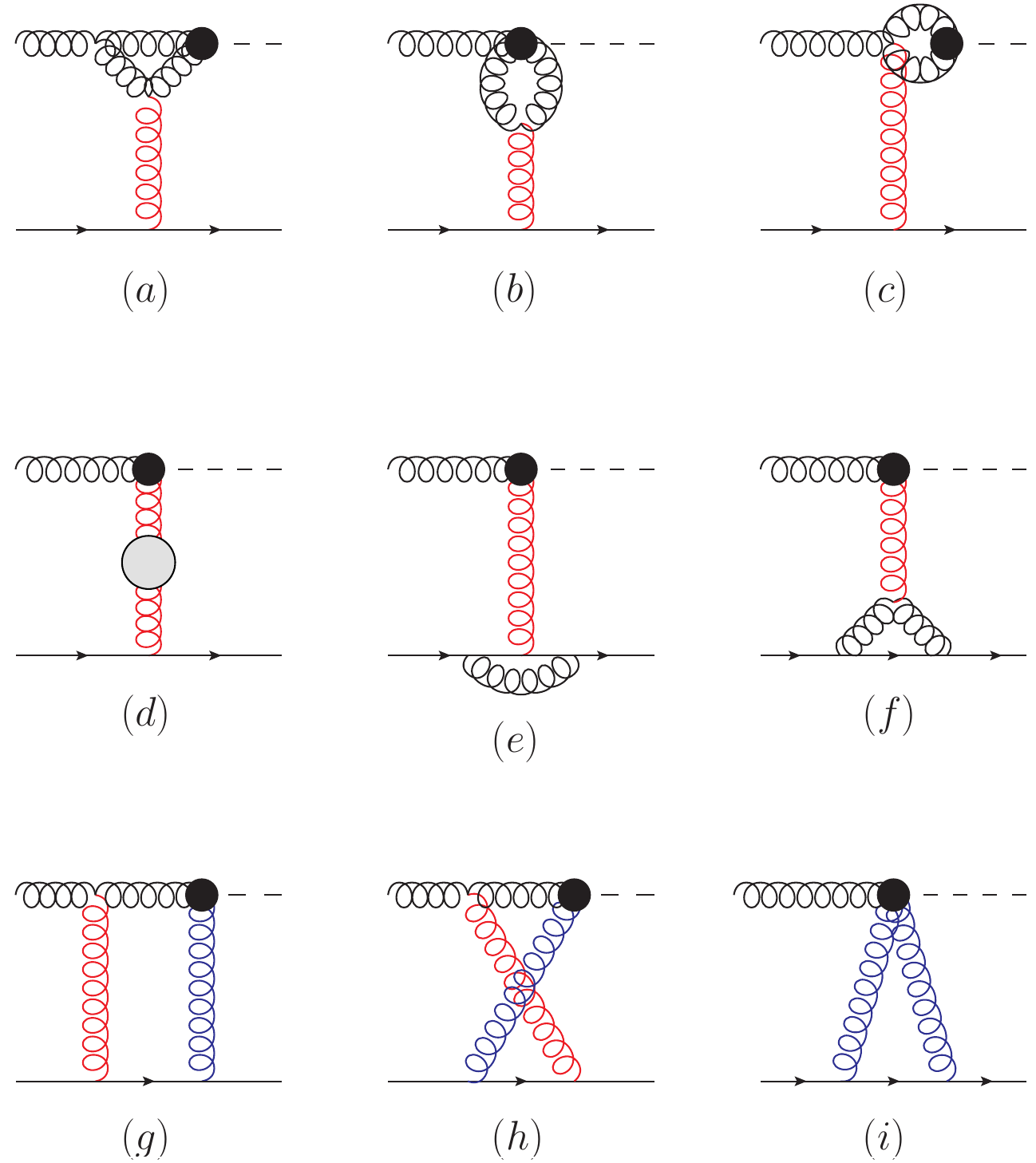}
  \end{center}
  \caption{Dominant contribution to the gluon-quark into Higgs-quark scattering amplitude in the high-energy limit at the next-to-leading order. The blob in diagram $(d)$ contains the summed contribution of quark, gluon and ghost loop. All propagators in red are the one that can be approximated {\it \`a la} Gribov, while the propagators in blue must be approximated as in Eq.~(\ref{Gribov:Eq:GribovsTrickMod}).}
  \label{Fig:Gribov:GribovTrickHiggs}
\end{figure}

We would like to apply the procedure outlined in the previous section to the case when QCD is extended by the inclusion of an effective dimension-5 interaction, such as the gluon-Higgs effective interaction in the infinite top-mass limit. The gluon-Higgs effective Lagrangian and the relative Feynman rules are given in Appendix~\ref{AppendixA}. In particular, our aim is to extract, with one-loop accuracy, the gluon-Reggeon-Higgs effective vertex, $\Gamma_{gH}$, which is one of the ingredients of the NLO Higgs impact factor~\cite{Celiberto:2022fgx}. For this purpose, we can consider as reference amplitude the scattering of a gluon off a quark to produce a Higgs plus a quark. The amplitude, $\mathcal{A}_{gq \rightarrow Hq}$, must be calculated with one-loop accuracy in the high-energy limit and compared with the Regge form~(\ref{Pole:Eq:ReggeFormEx1}), {\it assuming} its validity in the extended theory formed by QCD plus the gluon-Higgs effective interaction in the infinite-top-mass limit.
The leading order amplitude, see Fig.~\ref{Fig:Gribov:GluonHiggs}, reads
\begin{gather*}
    \mathcal{A}_{gq \rightarrow Hq}^{(8,-)} = (-i) \varepsilon_{\perp, \rho} (k_1) i g_H \delta^{ac} H^{\rho \mu} (-k_1,-q) \left( \frac{-i g_{\mu \nu} \delta^{bc}}{q^2} \right) \bar{u} (k_2-q) (-i)g t_{ji}^b \gamma^{\nu} u(k_2) \nonumber \\
    =-\frac{g g_H \varepsilon_{\perp, \rho} \delta^{ac} t_{ji}^c}{q^2} \left( g^{\mu \rho} k_1 \cdot q - k_1^{\mu} q^{\rho} \right) \bar{u} (k_2-q) \gamma_{\mu} u(k_2) \; ,
\end{gather*}
where we have adopted the following gauge choice for the external gluon:
\begin{equation}
     \varepsilon (k_1) \cdot k_2 = 0 \implies \varepsilon_{\rho} (k_1) = \varepsilon_{\perp, \rho} (k_1) \; .
     \label{GaugeChoice}
\end{equation}
Now, the first term of the first round bracket produces a contribution,
\begin{equation*}
    -\frac{g g_H t_{ji}^a (k_1 \cdot q)}{q^2} \bar{u} (k_2-q) \slashed{\varepsilon}_{\perp} u(k_2) \; ,
\end{equation*}
which is suppressed in the high-energy approximation, while the second term in the same round bracket produces the actual energy-leading term,
\begin{equation}
   \frac{g_H \delta^{ac} \varepsilon_{\perp} \cdot q_{\perp}}{2} \left( \frac{2s}{t} \right) g t_{ji}^a \bar{u} (k_2-q) \frac{\slashed{k_1}}{s} u(k_2). \label{eq:amp_gH-qq_Regge-limit}
\end{equation}
Hence, we find that the Born amplitude can be put in the LO Regge form
\begin{equation}
   \mathcal{A}_{gq \rightarrow Hq}^{(8,-)}  = \Gamma_{gH}^{ac(0)} \left( \frac{2s}{t} \right) \Gamma_{qq}^{c(0)} \; ,
\end{equation}
with the leading gluon-Reggeon-Higgs vertex given by
\begin{equation}
    \Gamma_{gH}^{ac(0)} = \frac{g_H \delta^{ac} \varepsilon_{\perp} \cdot q_{\perp}}{2} \; .
\end{equation}
One can easily check that this is the result that we would have immediately obtained by applying the Gribov's trick. 
Let us move on now to the one-loop level, where the diagrams contributing to the amplitude are the ones shown in Fig.~\ref{Fig:Gribov:GribovTrickHiggs}. In Ref.~\cite{Celiberto:2022fgx}, it was found that the Gribov's prescription (again all propagators in red) can be applied to \textit{single-gluon} $t$-channel diagrams (from $(a)$ to $(f)$), while it has to be suitably modified for the \textit{two-gluon} $t$-channel diagrams (from $(g)$ to $(i)$). In this latter case, to obtain the correct high-energy behaviour of the amplitude, for any $t$-channel propagator connected directly to the Higgs vertex (in blue in Fig.~\ref{Fig:Gribov:GribovTrickHiggs}), the prescription to be used is
\begin{equation}
    g^{\mu \nu} =  g_{\perp \perp}^{\mu \nu} + 2 \frac{k_2^{\mu} k_1^{\nu} + k_2^{\mu} k_1^{\nu} }{s} \longrightarrow  g_{\perp \perp}^{\mu \nu} + \frac{2 k_2^{\mu} k_1^{\nu} }{s} \; ,
    \label{Gribov:Eq:GribovsTrickMod}
\end{equation}
which contains an additional transverse term, $g_{\perp\perp}^{\mu\nu}$, with respect to the standard Gribov's trick in Eq. (\ref{Gribov:Eq:GribovsTrick}).

By itself, this seems to be a simple modification of the computational technique, but it actually has a much more profound impact. The  additional terms thus arising in separate diagrams, which we will refer to from now on as \textit{non-Gribov terms}, apparently, lead to new Lorentz structures in the amplitude, which do not allow to obtain the factorized pole Regge form introduced in section~\ref{sec:ThepoleReggeform}. We will show an explicit computation in the next section, but we anticipate one of the results here for illustrative purposes. The diagram $(i)$, which in the pure Gribov approximation would be nullified by the gauge choice for the external gluon in Eq.~(\ref{GaugeChoice}), gives instead a contribution 
\begin{equation}
    g^3 g_H \frac{C_A}{2} \delta^{ac} t_{ji}^c \left( \frac{2s}{t} \right) \bar{u} (k_2-q) \slashed{\varepsilon}_{\perp} (k_1) u(k_2) i \int \frac{d^D k}{(2 \pi)^D} \frac{1}{k^2 (k-q)^2} \;.
    \label{Gribov:Eq:FirstNonGrib}
\end{equation}
We see that in this term neither $\Gamma_{gH}^{ac(0)}$ nor $ \Gamma_{qq}^{a(0)}$ can be factorized out due to the coupling between helicities of the gluon and quark in the Lorentz structure $\bar{u} (k_2-q) \slashed{\varepsilon}_{\perp} (k_1) u(k_2)$, which is absent in Eq.~(\ref{eq:amp_gH-qq_Regge-limit}). This means that, in the high-energy approximation within NLLA, the form of the amplitudes should be
\begin{gather}
{\cal A}_{g q \rightarrow H q}^{(8,-)} \approx
\Gamma_{g H}^{ac(0)} \frac{2s}{t} \Gamma_{q, q}^{c(0)} + \Gamma_{g H}^{ac(0)} \frac{s}{t}\omega^{(1)}(t)
\left[ \ln\left( \frac{s}{-t} \right) + \ln\left(
\frac{-s}{-t} \right) \right] \Gamma_{q q}^{c(0)} \nonumber \\ 
+ \Gamma_{gH}^{ac(0)} \frac{2s}{t}\Gamma_{q q}^{c(1)} +
\Gamma_{gH}^{ac(1)} \frac{2s}{t} \Gamma_{qq}^{c(0)} + \text{non-Gribov contributions} \;,
\end{gather}
thus apparently invalidating the Regge form, as the mere consequence of the presence of the dimension-5 operator in the Lagrangian. However the analysis in section~\ref{sec:anom-hel-cans} shows that the new {\it Lorentz} structures arising in the {\it non-Gribov terms} can be decomposed as a sum of the helicity structure of the Born amplitude plus the {\it anomalous} helicity structure. It turns out that the {\it anomalous} helicity structure cancels among different diagrams and therefore the Regge-pole form (\ref{Pole:Eq:ReggeForm}) of the amplitude is preserved at one loop.  \\

\subsection{The appearance of non-Gribov terms}

\subsubsection{Non-Gribov terms from the triangular diagram}

\begin{figure}
  \begin{center}
  \includegraphics[scale=0.5]{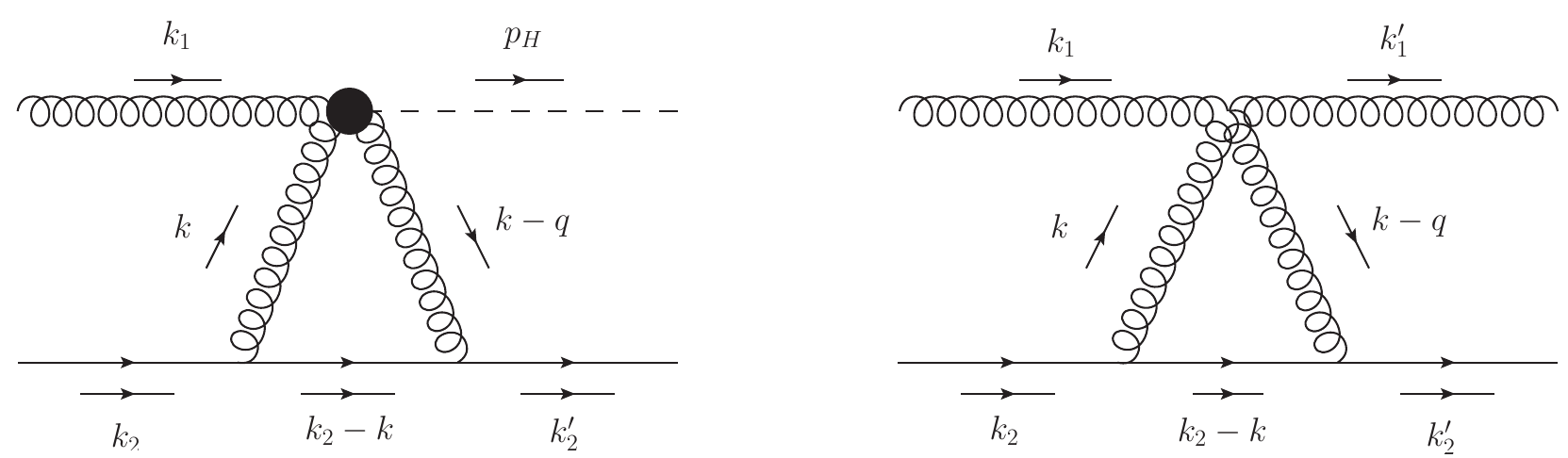}
  \end{center}
  \caption{Triangular diagrams in the $\mathcal{A}_{gq \rightarrow Hq}$ (left) and in the $\mathcal{A}_{gq \rightarrow gq}$ (right) amplitudes.}
  \label{Fig:Gribov:NonGribovTri}
\end{figure}

We start our discussion on non-Gribov terms, reproducing the result of the diagram $(i)$ in Fig.~\ref{Fig:Gribov:GribovTrickHiggs} (same as the left diagram in Fig.~\ref{Fig:Gribov:NonGribovTri}). We write down the amplitude approximating the $t$-channel gluons with the prescription in~(\ref{Gribov:Eq:GribovsTrickMod}), {\it i.e.}
\begin{gather*}
    \mathcal{A}_{gq \rightarrow Hq}^{\text{n.G.} (i)} = (-i) \varepsilon_{\perp, \mu} (k_1) \int \frac{d^D k}{(2 \pi)^D} g g_H f^{abc} V_{\; \; \delta \alpha}^{\mu} (-k_1, -k, k-q) \frac{(-i) \delta^{db} \left( \frac{2 k_1^{\nu} k_2^{\delta}}{s} + g^{\delta \nu}_{\perp \perp} \right)}{k^2} \\
    \times \frac{(-i) \delta^{ec} \left( \frac{2 k_1^{\beta} k_2^{\alpha}}{s} + g^{\alpha \beta}_{\perp \perp} \right)}{(k-q)^2} \bar{u} (k_2') (-i) g t_{jm}^e \gamma_{\beta} \frac{i \delta_{mn} (\slashed{k}_2 - \slashed{k})}{(k_2-k)^2} (-i) g t_{ni}^d \gamma_{\nu} u(k_2) \; .
\end{gather*}
The product of the two Lorentz structures coming from $t$-channel propagators is
\begin{equation*}
   \frac{4}{s^2} k_1^{\beta} k_2^{\alpha} k_1^{\nu} k_2^{\delta} + \frac{2 k_1^{\nu} k_2^{\delta}}{s} g^{\alpha \beta}_{\perp \perp} + \frac{2 k_1^{\beta} k_2^{\alpha}}{s} g^{\delta \nu}_{\perp \perp} + g^{\alpha \beta}_{\perp \perp} g^{\delta \nu}_{\perp \perp} \; .
\end{equation*}
The first term is the usual Gribov term and it is nullified by our gauge choice for external gluon polarization, while the last one is suppressed in the high-energy approximation. We want to investigate the second and the third terms which, in a pure QCD calculation, would lead to suppressed contributions. After some algebra, we find 
\begin{gather*}
    \mathcal{A}_{gq \rightarrow Hq}^{\text{n.G.} \; (i)} = - g_H g^3 \frac{C_A \delta^{ac} t_{ji}^c}{2} \bar{u}(k_2') \left[ \slashed{\varepsilon}_{\perp} (k_1) \gamma^{\mu} \slashed{k}_1 - \slashed{k}_1 \gamma^{\mu} \slashed{\varepsilon}_{\perp} (k_1) \right] u (k_2) \\ \times i \int \frac{d^D k}{(2 \pi)^D}  \frac{k_{\mu}}{k^2 (k-k_2)^2 (k-k_2')^2}\;,
\end{gather*}
up to energy-suppressed contributions.
Taking into account that the two external quark momenta are on-shell ($k_2^2 = k_2'^{\; 2} = 0$), we decompose the integral as 
\begin{equation}
    \int \frac{d^D k}{(2 \pi)^D} \frac{k^{\mu}}{k^2 (k-k_2)^2 (k-k_2')^2} = \frac{q^\mu - 2 k_2^\mu}{q^2} \int \frac{d^D k}{(2 \pi)^D} \frac{1}{k^2 (k-q)^2}\;.
    \label{Gribov:Eq:TensDecompo}
\end{equation}
In principle, both the contributions in $q^\mu$ and in $k_2^\mu$ are non-Gribov terms. The first one gives a suppressed contribution when taking the interference with the Born and hence we ignore it for the current discussion. 
It is easy to see that the second contribution in Eq.~(\ref{Gribov:Eq:TensDecompo}) gives
\begin{equation}
    - \frac{g_H \delta^{ac}}{2} \left( \frac{2s}{t} \right) g t_{ji}^c \bar{u}(k_2') \slashed{\varepsilon}_{\perp} (k_1) u(k_2) g^2 C_A \int \frac{d^D k}{i (2 \pi)^D} \frac{1}{k^2 (k-q)^2} \; ,
\end{equation}
which is exactly the term we anticipated in Eq.~(\ref{Gribov:Eq:FirstNonGrib}). This term is not in the Regge form (\ref{Pole:Eq:ReggeForm}), nevertheless, if we compute the interference between this correction and the Born $gq \rightarrow H q$ amplitude, we find\footnote{The sum over quark and gluon polarization is understood.}
\begin{equation}
    \delta_{gq \rightarrow H q}^{\text{n.G.} \; (i)} \equiv \frac{ \mathcal{A}_{gq \rightarrow Hq}^{(0)*} \mathcal{A}_{gq \rightarrow Hq}^{\text{n.G.} \; (i)}}{|\mathcal{A}_{gq \rightarrow Hq}^{(0)}|^2} = g^2 C_A B_0(q^2) \; .
    \label{Eq:Gribov:NonGrib(i)}
\end{equation}
The definition and the value of the integral $B_0$, together with all the other scalar integrals that appear in the calculation, is given in the Appendix~\ref{UsefulIntegrals}. This is exactly the one-loop correction to the Higgs impact factor coming from this diagram found in~\cite{Celiberto:2022fgx}. \\

The reason why this type of contributions appears becomes even clearer when we compare the diagram under examination with a similar contribution to the amplitude $\mathcal{A}_{gq \rightarrow gq}$ (diagram $(b)$ in Fig.~\ref{Fig:Gribov:NonGribovTri}). In this latter case, the Gribov term contains the following expression:
\begin{equation}
    k_2^{\delta} k_2^{\alpha} \epsilon^{\mu} (k_1) \epsilon^{\gamma} (k_1') Y_{\delta \mu \gamma \alpha}^{abcf} \; ,
\end{equation}
where $Y_{\delta \mu \gamma \alpha}^{abcf}$ is the four-gluon vertex, and it is thus nullified by the on-shell condition of the incoming quark line, $k_2^2=0$, and the gauge choice, $\epsilon (k_1) \cdot k_2 = 0$. In this case, however, also the two non-Gribov terms suffer the same fate; indeed, they contain the structure like the following
\begin{equation}
    k_2^{\delta} g_{\perp \perp}^{\alpha \beta} \epsilon^{\mu} (k_1) \epsilon^{\gamma} (k_1') Y_{\delta \mu \gamma \alpha}^{abcf} \; ,
\end{equation}
and there are no non-zero contractions available for the four-vector $k_2$ (with the gauge choices $\epsilon (k_1) \cdot k_2 = 0$ and $\epsilon (k_1') \cdot k_2 = 0$). 

In the scattering of a gluon off a quark to produce a Higgs plus a gluon, $\mathcal{A}_{gq \rightarrow Hq}$ (see diagram $(i)$ in Fig.~\ref{Fig:Gribov:GribovTrickHiggs} or diagram $(a)$ in Fig.~\ref{Fig:Gribov:NonGribovTri}) the relevant contraction is
\begin{equation}
\varepsilon_{\perp, \mu} (k_1) V_{\; \; \delta \alpha}^{\mu} (-k_1, -k, k-q) \left[ \frac{2 k_1^{\nu} k_2^{\delta}}{s} g^{\alpha \beta}_{\perp \perp} + \frac{2 k_1^{\beta} k_2^{\alpha}}{s} g^{\delta \nu}_{\perp \perp} \right] \bar{u} (k_2') \gamma_{\beta} (\slashed{k}_2-\slashed{k}) \gamma_{\nu} u (k_2) \; .
\end{equation}
Let us focus only on the first of the two terms (the second follows a similar path), we have 
\begin{equation}
\varepsilon_{\perp, \mu} (k_1) \left[ (k-k_1)^{\alpha} g^{\mu \delta} + (q - 2 k)^{\mu} g^{\delta \alpha} + (k-q+k_1)^{\delta} g^{\alpha \mu} \right] \frac{2  k_{2, \delta} }{s}  \bar{u} (k_2') \gamma_{\perp, \alpha} (\slashed{k}_2-\slashed{k}) \slashed{k}_1 u (k_2) \; .
\end{equation}
Since only the last term survive, we get\footnote{We neglect $q\cdot k_2 \sim s^0$ with respect to $k_1 \cdot k_2 \sim s$.}
\begin{equation}
   \frac{2 k_2 \cdot (k+k_1)}{s} \bar{u} (k_2') \slashed{\varepsilon}_{\perp} (k_1) (\slashed{k}_2-\slashed{k}) \slashed{k}_1 u (k_2) \; .
\end{equation}
This term, as might be expected, contains one less $\slashed{k}_1$ in the Dirac structure than a Gribov term and, instead, has a completely transverse object ($\epsilon_{\perp, \mu}$). On the other side, the $gggH$-vertex, being linear in the momenta, had generated a contraction of the order of $s$ making the non-Gribov term dominant in energy. We want to stress this last point: it was not enough for the contraction to be non-zero, it was also necessary for the vertex to generate additional power of $s$. This becomes even clearer if we consider the non-Gribov terms of a box-like diagram.

\subsubsection{Non-Gribov terms from the box-type diagrams}
\begin{figure}
  \begin{center}
  \includegraphics[scale=0.5]{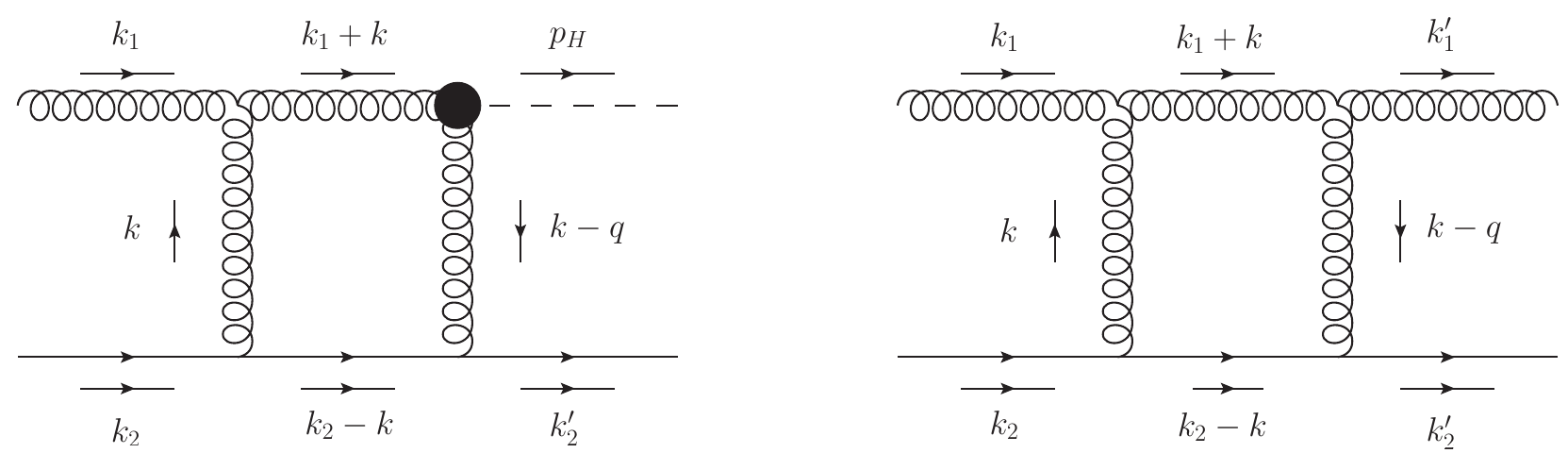}
  \end{center}
  \caption{Box diagrams in the $\mathcal{A}_{gq \rightarrow Hq}$ (left) and in the $\mathcal{A}_{gq \rightarrow gq}$ (right) amplitudes.}
  \label{Fig:Gribov:NonGribovBox}
\end{figure}

As noted in Ref.~\cite{Celiberto:2022fgx}, also the box and the crossed diagrams (see $(g)$ and $(h)$ in Fig.~\ref{Fig:Gribov:GribovTrickHiggs}) contain non-Gribov contributions. As for the triangular diagram, we want to perform a comparison between the diagram $(g)$ (see also the left diagram in Fig.~\ref{Fig:Gribov:NonGribovBox}) and a similar diagram in the $\mathcal{A}_{gq \rightarrow gq}$ amplitude (right diagram in Fig.~\ref{Fig:Gribov:NonGribovBox}). As explained before, the $t$-channel gluon propagator which is not connected to the $ggH$-vertex can be approximated by the standard Gribov approximation, Eq.~(\ref{Gribov:Eq:GribovsTrick}), while the one directly connected to the vertex should be approximated as in Eq.~(\ref{Gribov:Eq:GribovsTrickMod}). The product of these two structures yields a single non-Gribov term,  
\begin{gather*}
    \mathcal{A}_{gq \rightarrow Hq}^{\text{n.G.} \; (g)} = g^3 g_H \varepsilon_{\perp, \mu} (k_1) f^{abc} (t^b t^c)_{ji} \frac{2}{s} \int \frac{d^D k}{(2 \pi)^D} \frac{1}{k^2 (k-q)^2 (k+k_1)^2 (k-k_2)^2} \\
  \times  A^{\mu \rho \nu} (-k, k+k_1) k_{2, \rho} \; g_{\sigma \xi}^{\perp \perp} H_{\nu}^{\; \; \sigma} (-k-k_1, k-q) \bar{u} (k_2') \gamma^{\xi} (\slashed{k}_2-\slashed{k}) \slashed{k}_1 u (k_2) \; ,
  \label{Eq:Gribov:BoxDiagram}
\end{gather*}
where $A^{\mu \rho \nu} (-k, k+k_1)$ is the Lorentz structure in the three-gluon vertex. Since we are mainly interested in comparing the behaviour in the power of $s$ of diagrams in Fig.~\ref{Fig:Gribov:NonGribovBox}, we focus on specific terms which are dominant in the high-energy limit. To this aim, in the part of the amplitude containing the Dirac structure, we consider the term 
\begin{equation}
    \bar{u} (k_2') \gamma^{\xi} \slashed{k}_2 \slashed{k}_1 u (k_2) = s \bar{u} (k_2') \gamma^{\xi} u (k_2)
\end{equation}
which is the leading one in powers of $s$. Then, we perform the contraction
\begin{gather}
   \varepsilon_{\perp, \mu} (k_1)  k_{2, \rho} A^{\mu \rho \nu} (-k, k+k_1) = 2 (k_{\perp} \cdot \varepsilon_{\perp} (k_1)) k_2^{\nu} - (2 k_1 + k) \cdot k_2 \; \varepsilon_{\perp}^{\nu} (k_1) 
\end{gather}
and observe that it can, at most, produce a term of order $s$ (either by a subsequent contraction of $k_2^{\nu}$ in the first term or {\it via} the dot product in the second term). Let us consider the second term, approximating it by $-s \varepsilon_{\perp}^{\nu} (k_1)$, which therefore leads to
\begin{equation}
    -s^2 \bar{u} (k_2') \gamma^{\xi} u (k_2) \varepsilon_{\perp}^{\nu} (k_1) g_{\sigma \xi}^{\perp \perp} H_{\nu}^{\; \; \sigma} (-k_1-k,k-q) \; .
\end{equation}
Here, all contractions are between transverse components, but since $H_{\nu}^{\; \; \sigma}$ is a quadratic vertex in the momenta, it has a term proportional to $g_{\nu}^{\; \; \sigma}$ times a scalar product and leads~to 
\begin{equation}
    s^2 \bar{u} (k_2') \slashed{\varepsilon}_{\perp} (k_1) u (k_2) \left[ (k_1 + k) \cdot (k-q) \right] \;.
\end{equation}
The term $k_1 \cdot k$ contributes with 
\begin{equation}
    g^3 g_H f^{abc} (t^b t^c)_{ji} \bar{u} (k_2') \slashed{\varepsilon}_{\perp} (k_1) u (k_2) \; 2 s \; k_{1, \mu} \int \frac{d^D k}{(2 \pi)^D} \frac{k^{\mu}}{k^2 (k-q)^2 (k+k_1)^2 (k-k_2)^2} \; ,
    \label{Eq:Gribov:BoxDiagramHiggsDom}
\end{equation}
which, in the integration region $k^{\mu} \sim k_2^{\mu}$, takes a value of the order
\begin{equation}
    s^2 \int \frac{d^D k}{i(2 \pi)^D} \frac{1}{k^2 (k-q)^2 (k+k_1)^2 (k-k_2)^2} = s^2 D_0 (m_H^2, q^2, s) \sim s^2 \frac{1}{s} = s\; ,
\end{equation}
and hence is leading in the high-energy approximation. Here, we have used the fact that the integral $D_0$ behaves as $1/s$, as it can been seen in the Appendix~\ref{UsefulIntegrals}. There might still be a cancellation among terms that are dominant in energy, but a complete calculation using \texttt{FeynCalc}~\cite{Mertig:1991ca,Shtabovenko:2016cp} and \texttt{FeynHelpers}~\cite{Shtabovenko:2016whf}
shows that this is not the case and the full result for the non-Gribov part associated to diagrams $(g)$ and $(h)$ reads
\begin{equation}
      \delta_{gq \rightarrow H q}^{\text{n.G.} \; (g) + (h)} \equiv \frac{ \mathcal{A}_{gq \rightarrow Hq}^{(0)*} (\mathcal{A}_{gq \rightarrow Hq}^{\text{n.G.} \; (g)} + \mathcal{A}_{gq \rightarrow Hq}^{\text{n.G.} \; (h)})}{|\mathcal{A}_{gq \rightarrow Hq}^{(0)}|^2} = g^2 (-3 C_A) B_0 ( q^2 ) \; .
      \label{Eq:Gribov:NonGrib(g)(h)}
\end{equation}
This result will be useful in section~\ref{Sec:StrategyOfRapidityRegions}. \\

We now want to compare the previous result with the calculation of a non-Gribov contribution from the box diagram of the $\mathcal{A}_{gg \rightarrow gg}$ amplitude (right diagram in Fig.~\ref{Fig:Gribov:NonGribovBox}). Also in this case, we can take one of the terms, namely the one produced by approximating the two propagators as
\begin{equation*}
    g_{\rho \gamma} \rightarrow \frac{2 k_{2, \rho} k_{1, \gamma}}{s} \; , \hspace{1 cm} g_{ \sigma \xi} \rightarrow g^{\perp \perp}_{\sigma \xi} \; .
\end{equation*}
We have
\begin{gather*}
    \mathcal{A}_{gq \rightarrow gq}^{\text{n.G.}} = g^4 f^{adf} f^{dhg} (t^g t^f)_{ji} \varepsilon_{\perp, \mu} (k_1) \varepsilon_{\beta}^{*} (k_1') \frac{2}{s} \int \frac{d^D k}{i (2 \pi)^D} \frac{1}{k^2 (k-q)^2 (k+k_1)^2 (k-k_2)^2} \\
  \times  A^{\mu \rho \nu} (-k, k+k_1) k_{2, \rho} \; g_{\sigma \xi}^{\perp \perp} A_{\nu}^{\; \; \sigma \beta} (k-q,k_1+q) \bar{u} (k_2') \gamma^{\xi} (\slashed{k}_2-\slashed{k}) \slashed{k}_1 u (k_2) \; .
  \label{Eq:Gribov:BoxDiagramQCD}
\end{gather*}
We repeat the same steps as before and take
\begin{gather}
    \bar{u} (k_2') \gamma^{\xi} (\slashed{k}_2-\slashed{k}) \slashed{k}_1 u (k_2) \longrightarrow s \bar{u} (k_2') \gamma^{\xi} u (k_2) \; , \\
    \varepsilon_{\perp, \mu} (k_1) k_{2, \rho} A^{\mu \rho \nu} (-k, k+k_1)  \longrightarrow -s \varepsilon_{\perp}^{\nu} (k_1) \; .
\end{gather}
At this point, the contraction in the numerator is
\begin{equation}
    -s^2 \varepsilon_{\perp}^{\nu} (k_1) \left( \varepsilon_{\perp, \beta}^{*} (k_1') - \frac{\varepsilon_{\perp}^{*} (k_1') \cdot k_{1, \perp}'}{k_{1'} \cdot k_2} k_{2, \beta} \right) A_{\nu}^{\; \; \sigma \beta} (k-q,k_1+q) \bar{u} (k_2') \gamma_{\perp, \sigma} u (k_2) \; ,
\end{equation}
where we have specified the form of the polarization vector $ \varepsilon_{\beta}^{*} (k_1') $ in the gauge $ \varepsilon^{*} (k_1') \cdot k_2 = 0$. \\ The difference with what was obtained previously is clear. There is no possible contraction that can produce an additional factor $s$, since the momenta in the three-gluon vertex contract either transverse objects or a $k_2$ divided by a dot product of the order of $s$. Hence, the contribution is of the order
\begin{equation}
    s \int \frac{d^D k}{i(2 \pi)^D} \frac{1}{k^2 (k-q)^2 (k+k_1)^2 (k-k_2)^2} = s D_0 (m_H^2, q^2, s) \sim s^0 \; ,
\end{equation}
which shows that this kind of terms are suppressed in a pure QCD computation\footnote{Similar considerations hold for the $\mathcal{A}_{qq \rightarrow qq}$ or the $\mathcal{A}_{gg \rightarrow gg}$ amplitude.}. \\  

\subsection{The fate of the anomalous helicity structure}
\label{sec:anom-hel-cans} 
We now consider more closely the helicity structure of the Born amplitude, which is always preserved in the Gribov part of the one-loop amplitude, and reads
\begin{equation}
    \mathcal{H}_{{\rm{Born}}} \equiv \left( \varepsilon_{\perp} (k_1) \cdot q_{\perp} \right) \bar{u} (k_2-q) \frac{\slashed{k}_1}{s} u(k_2) \; .
\end{equation}
Using the Sudakov decomposition for the momentum transfer in the $t$-channel,
\begin{equation}
    q^{\mu} = \frac{q_{\perp}^2}{s} k_1^{\mu} + \frac{m_H^2 - q_{\perp}^2}{s} k_2^{\mu} + q_{\perp} \; ,
\end{equation}
we have
\begin{equation}
    \left( \varepsilon_{\perp} (k_1) \cdot q_{\perp} \right) \bar{u} (k_2-q) \frac{\slashed{q}-\slashed{q}_{\perp}}{q_{\perp}^2} u(k_2) = \left( \varepsilon_{\perp} (k_1) \cdot q_{\perp} \right) \bar{u} (k_2-q) \frac{\slashed{q}_{\perp}}{|q_{\perp}|^2} u(k_2) \; ,
\end{equation}
where in the last step we used $\bar{u} (k_2-q) \slashed{q} u(k_2) = \bar{u} (k_2-q) \left( \slashed{q} - \slashed{k}_2 + \slashed{k}_2 \right) u(k_2) = 0$.
Defining the following basis in transverse space:
\begin{equation}
    n_x^{\mu} = \frac{q_{\perp}^{\mu}}{|q_{\perp}|} \; , \; \hspace{1 cm}  n_{y}^{\mu} =\epsilon^{\mu\nu+-} \frac{q_{\perp \nu}}{|q_{\perp}|} \;,
\end{equation}
where the $+$ and $-$ components identify the directions along $k_1$ and $k_2$ respectively, then the ``transverse form'' of the helicity structure of the Born amplitude reads
\begin{equation}
    \mathcal{H}_{{\rm{Born}}} = \left( \varepsilon_{\perp} (k_1) \cdot n_x \right) \bar{u} (k_2-q) \slashed{n}_x u(k_2) \; . \label{eq:Born-HS}
\end{equation}
On the other hand, the above-discussed Lorentz structure produced by the non-Gribov contributions is
\begin{equation}
    \bar{u} (k_2-q) \slashed{\varepsilon}_{\perp} (k_1) u(k_2) =  -\bar{u} (k_2-q) \gamma_{\mu} u(k_2)\left( n_x^{\mu} n_x^{\nu} + n_{y}^{\mu} n_{y}^{\nu}  \right) \varepsilon_{\perp, \nu} = -\mathcal{H}_{{\rm{Born}}} - \mathcal{H}_{{\rm{anomalous}}} \; ,
    \label{FirstHelicity}
\end{equation}
where the {\it anomalous} helicity structure is
\begin{equation}
   \mathcal{H}_{{\rm{anomalous}}} \equiv \bar{u} (k_2-q) \slashed{n}_{y} u(k_2) \left( n_{y} \cdot \varepsilon_{\perp} \right) \; .
\end{equation}  
In the full $\mathcal{A}_{gq \rightarrow Hq}$ scattering amplitude, other types of Lorentz structures are also encountered, {\it e.g.}
\begin{gather}
    \bar{u} (k_2-q) \frac{\slashed{k}_1}{s} \slashed{q}_{\perp} \slashed{\varepsilon}_{\perp} (k_1) u(k_2) = \mathcal{H}_{{\rm{Born}}} + \mathcal{H}_{{\rm{anomalous}}} \; ,  
    \label{SecondThirdHelicity}
\end{gather}
but they are always reducible to the independent helicity structures $\mathcal{H}_{{\rm{Born}}}$ and $\mathcal{H}_{{\rm{anomalous}}}$.
When we take the interference between the one-loop and the Born amplitude, the contribution from the $\mathcal{H}_{{\rm{anomalous}}}$-part is proportional to 
\begin{equation}
  \sum_{{\rm{spin}}} \bar{u} (k_2-q) \slashed{n}_x u(k_2) \bar{u} (k_2) \slashed{n}_{y} u(k_2-q) = {\rm{Tr}} \left[ \slashed{k}_2 \slashed{n}_{y} ( \slashed{k}_2-\slashed{q} ) \slashed{n}_x \right] = 0 ,
\end{equation}
and thus vanishes. This implies that non-Gribov terms contribute to the interference between 1-loop and Born amplitude, but only through the Born helicity structure and explains why the Regge ansatz straightforwardly holds at the level of the interference~\cite{Celiberto:2022fgx}. \\

However, the Reggeization is a statement at the level of amplitudes, it is therefore reasonable to study the complete amplitude, preserving both Born and anomalous helicity structures. After producing the diagrams through \texttt{FeynArts}~\cite{Hahn:2000kx} and calculating their Regge limit by using \texttt{FeynCalc}~\cite{Mertig:1991ca,Shtabovenko:2016cp}, we used the relations~(\ref{FirstHelicity}), (\ref{SecondThirdHelicity}) to express the result in terms of the two independent helicity structures $\mathcal{H}_{{\rm{Born}}}$ and $\mathcal{H}_{{\rm{anomalous}}}$. We observed that the anomalous structure cancels out in the sum between diagrams ($g$), ($h$) and ($i$) in the Fig.~\ref{Fig:Gribov:GribovTrickHiggs}. This is in agreement with the result obtained in Ref.~\cite{Schmidt:1997wr} for helicity amplitudes. The outcome is therefore that, at one-loop level, although the amplitude receives non-Gribov contributions, it can still be cast into the Regge form.


\section{Computing the gluon-Reggeon-Higgs vertex through the strategy of rapidity regions}
\label{sec:RapidityRegions}
In Refs.~\cite{Fadin:1999qc,Fadin:2001dc} a powerful technique to extract NLO high-energy vertices has been developed. It makes possible to calculate complicated high-energy amplitudes at NLO in a compact and elegant way. The technique is based on two fundamental ingredients. The first is exactly the Gribov's trick, the second is the \textit{factorization} of effective high-energy vertices in different regions of rapidity. The latter property has been demonstrated for several effective vertices, which are however constructed from {\it renormalizable} Lagrangians.

We want to discuss how this technique can be used in presence of the gluon-Reggeon-Higgs vertex. In previous sections we saw that, in this case, the Gribov's trick is violated. Nonetheless, the contributions generated by this violation have been carefully isolated in Eqs.~(\ref{Eq:Gribov:NonGrib(i)}), (\ref{Eq:Gribov:NonGrib(g)(h)}) and sum up to
\begin{equation}
      \delta_{gq \rightarrow H q}^{\text{n.G.}}
      = g^2 (-2 C_A) B_0 ( q^2 ) = \frac{\bar{\alpha}_s}{4 \pi} \left( \frac{\vec{q}^{\; 2}}{\mu^2} \right)^{-\epsilon} \left[ -\frac{2 C_A}{\epsilon} - 4 C_A \right]  \; .
      \label{Eq:Gribov:NonGribTot}
\end{equation}
Treating this contribution separately, we can think of using the region technique on the Gribov terms, not only as an elegant and effective short-cut for the computation of amplitudes in the high-energy limit, but also as a tool to analyze factorization properties of the effective vertex at one-loop. Here, new features are expected in presence of an effective non-renormalizable dimension-5 operator. 

In the following, we briefly recall the method of regions in QCD, then we consider its application in presence of the gluon-Reggeon-Higgs vertex.

\label{Sec:StrategyOfRapidityRegions}
\subsection{Strategy of rapidity regions for QCD vertices}
\label{SubSec:StrategyOfRapidityRegionsQCD}
At one-loop order, Feynman diagrams for the process $A + B \rightarrow A' + B'$ can be divided into four classes. The first class includes corrections to the $t$-channel gluon propagator, the second and third are related to corrections to the interaction of the $t$-channel gluon with the particles $A,A'$ and $B,B'$ correspondingly, and the last one contains the diagrams with the two-gluon exchange in the $t$-channel. The first three classes fit into the diagrams that in Figs. \ref{Fig:Gribov:GribovTrickQuark}, \ref{Fig:Gribov:GribovTrickHiggs} we have labeled as \textit{single-gluon} $t$-channel diagrams. The contributions of the diagrams of the first three classes have the same dependence on $s$ as the Born amplitudes; moreover, the contributions of the first and second (first and third) classes depend on properties of the particles $B,B'$ ($A,A'$) in the same way as the Born amplitudes. It is evident therefore that the contribution of the diagrams of the second (third) class must be attributed to the vertex $\Gamma_{A'A}$ ($\Gamma_{B'B}$), whereas the contribution of the first class must be divided in equal parts between these vertices. As we have seen, these considerations also fit perfectly into the computation of the gluon-Reggeon-Higgs vertex\footnote{Remember that this class of diagrams have no problem related to the violation of the Gribov's trick.}. On the other side, the contributions of 
two-gluon exchange diagrams to the two vertices and to the trajectory are mixed and the problem of their separation arises.
For these diagrams, the fundamental idea is to decompose the loop momentum $k$ {\it \`a la} Sudakov, 
\begin{figure}
  \begin{center}
  \hspace{0.5 cm}\includegraphics[scale=0.55]{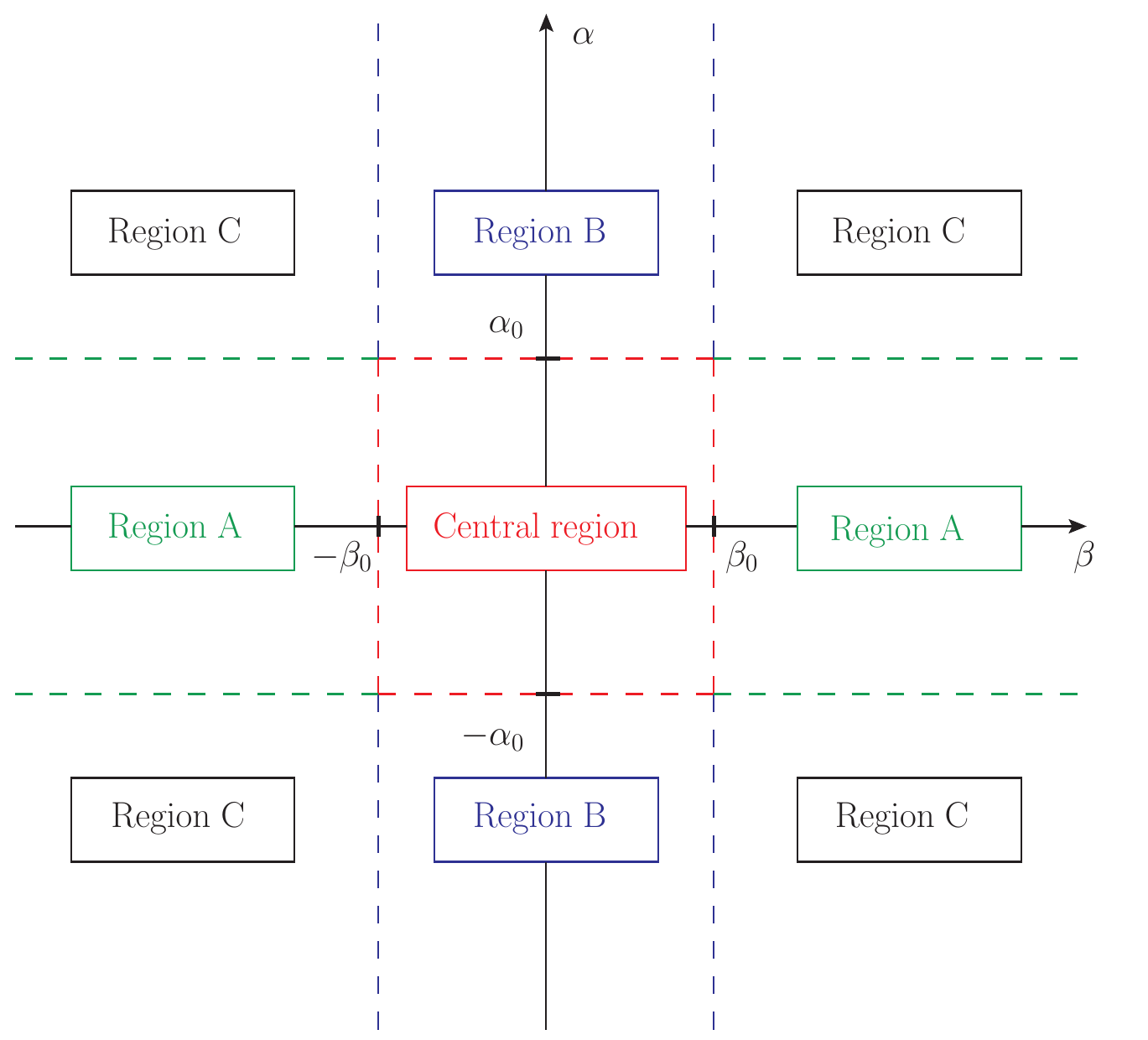}
  \end{center}
  \caption{Schematic representation of the four different rapidity regions.}
  \label{Fig:StrategyOfRegions}
\end{figure}
\begin{equation}
    k=\beta k_1 + \alpha k_2 + k_{\perp} \; ,
\label{Sudakov}
\end{equation}
and then to split the integration domain into four regions, introducing two rapidity parameters $\alpha_0$ and $\beta_0$ such that 
\begin{equation*}
    \alpha_0 \ll 1 \; , \hspace{0.5 cm} \beta_0 \ll 1 \; , \hspace{0.5 cm} s \alpha_0 \beta_0 \gg |t| .
\end{equation*}
The four regions are
\begin{equation*}
    \begin{cases}
    {\rm{\textbf{Central region}}} \hspace{1 cm} |\alpha| \lesssim \alpha_0 \; , \; |\beta| \lesssim \beta_0 \;,\\
    {\rm{\textbf{Region A}}} \hspace{1.97 cm} |\alpha| \lesssim \alpha_0 \; , \; |\beta| > \beta_0 \;,\\
    {\rm{\textbf{Region B}}} \hspace{2 cm} |\alpha| > \alpha_0 \; , \; |\beta| \lesssim \beta_0 \;,\\
    {{\rm{\textbf{Region C}}}} \hspace{2 cm} |\alpha| > \alpha_0 \; , \; |\beta| > \beta_0 \; ,
    \end{cases}
\end{equation*}
and are schematically depicted in Fig.~\ref{Fig:StrategyOfRegions}.
The factorization of vertices in different rapidity regions requires that in the region $|\alpha| \ll 1$ ($ |\beta| \ll 1$) we can factor out the vertex $\Gamma_{B'B}^{(0)}$ ($\Gamma_{A'A}^{(0)}$) from the diagrams. This assertion is proved for all vertices of QCD and for the vertex $\gamma^{(*)} R \rightarrow q \bar{q}$. \\

Let us discuss the general expected structure of the contribution in each of these four regions, postponing the details of the calculation to the particular case of the gluon-Reggeon-Higgs vertex.

\subsubsection*{Region C}
The first observation is that the contribution from the  region C is suppressed by a factor $|t|/ (\alpha_0 \beta_0 s) \ll 1$. 
This not only allows us to exclude this region from the calculation, but also to remove the $|\alpha| < \alpha_0$ ($|\beta| < \beta_0$) restriction in the region A (B), if necessary, since this would correspond to adding a power-suppressed quantity.

\subsubsection*{Central region}
In the central region the upper and the lower vertices factorize out and for the box diagram we obtain
    \begin{equation*}
        \mathcal{A}^{(8,-)}_{{\rm{box, Central}}} = - g^2 C_A s^2 \Gamma_{A'A}^{(0)} \; I^{\rm{central}} \; \Gamma_{B'B}^{(0)}\; ,
    \label{eq:GeneralOfStrat_Central}
    \end{equation*}
where
\begin{equation}
  \hspace{-0.15 cm}  I^{\rm{central}} = \frac{s}{2} \int_{-\alpha_0}^{\alpha_0}  \int_{-\beta_0}^{\beta_0} \int \frac{d^{D-2} k_{\perp}}{(2 \pi)^{D}i} \frac{d \alpha \; d \beta}{(\alpha \beta s + k_{\perp}^2+i0) (\alpha \beta s + (q-k)_{\perp}^2 + i0) (-\beta s + i 0) (\alpha s + i0)} \; .
  \label{eq:I_central}
\end{equation}
The calculation of this integral leads to\footnote{In Appendix~\ref{UsefulIntegrals}, we reproduce the calculation of this integral.}~\cite{Fadin:2001dc}
\begin{equation}
    \mathcal{A}^{(8,-)}_{{\rm{box, Central}}} = \Gamma_{A'A}^{(0)} \frac{2s}{t}  \Gamma_{B'B}^{(0)} \; \omega^{(1)}(t) \left[ \frac{1}{2} \ln \left( \frac{-s}{-t} \right) + \frac{\phi(\alpha_0)}{2} + \frac{\phi(\beta_0)}{2} \right] \; ,
\end{equation}
where
\begin{equation*}
    \phi(z) = \ln z + \frac{1}{2} \left( -\frac{1}{\epsilon} - \psi(1) + \psi(1+\epsilon) - 2 \psi(1-\epsilon) + 2 \psi(1-2\epsilon) \right) \; .
\end{equation*}
We need also to include the crossed diagram and this leads to the following result
\begin{equation}
        \mathcal{A}^{(8,-)}_{{\rm{Central}}} = \Gamma_{A'A}^{(0)} \frac{2s}{t}  \Gamma_{B'B}^{(0)} \; \omega^{(1)}(t) \left[ \frac{1}{2} \ln \left( \frac{-s}{t} \right) + \frac{1}{2} \ln \left( \frac{-s}{-t} \right) + \phi(\alpha_0) + \phi(\beta_0) \right] \; .
        \label{Eq:Regions:CentralAmp}
\end{equation}
This contributions is then split into a correction to the upper vertex,   
\begin{equation*}
        \Gamma_{A'A}^{({\rm{Central}})} = \Gamma_{A'A}^{(0)} \; \omega^{(1)}(t) \phi(\beta_0) \; ,
\end{equation*}
a correction to the lower vertex,
\begin{equation*}
        \Gamma_{B'B}^{({\rm{Central}})} = \Gamma_{B'B}^{(0)} \; \omega^{(1)}(t) \phi(\alpha_0) 
\end{equation*}
and a contribution containing energy logarithms which contribute to the one-loop Regge trajectory in Eq.~(\ref{Pole:Eq:ReggeForm}). We want to emphasize two important points: 
\begin{itemize}
    \item[\textbullet] The contribution calculated in this region is universal, {\it i.e.} it does not depend on the particular $\Gamma_{A'A}, \Gamma_{B'B}$ vertices.  
    \item[\textbullet] The separation into regions has induced rapidity divergences when $\alpha_0, \beta_0$ go to zero, which, as we shall see, will be cancelled by analogous divergences in the regions A and B.
\end{itemize}

\subsubsection*{Region A and Region B}
In the region A, the lower vertex $\Gamma_{B'B}$ factorizes and we find a correction that is assigned to the upper vertex,
   \begin{equation*}
        \Gamma_{A'A}^{{\rm{(A)}}} = \Gamma_{A'A}^{(0)} \delta_{{\rm{NLO}}}^{{\rm{(A)}}} =  \Gamma_{A'A}^{(0)} \left[ - \omega (t) \ln \beta_0 + \tilde{\delta}_{{\rm{NLO}}}^{{\rm{(A)}}} \right] \; ,
    \end{equation*}
where the term proportional to $\ln \beta_0$ is universal, while $\tilde{\delta}_{{\rm{NLO}}}^{({\rm{A}})}$ varies according to the specific vertex.
Similarly, region B provides a correction, which is naturally assigned to the lower vertex, of the form
    \begin{equation*}
        \Gamma_{B'B}^{{\rm{(B)}}} = \Gamma_{B'B}^{(0)} \; \delta_{{\rm{NLO}}}^{{\rm{(B)}}} = \Gamma_{B'B}^{(0)} \left[ - \omega (t) \ln \alpha_0 + \tilde{\delta}_{{\rm{NLO}}}^{{\rm{(B)}}} \right] \; .
    \end{equation*}
    
\subsection{Strategy of rapidity regions for the gluon-Reggeon-Higgs vertex}
\label{SubSec:StrategyOfRapidityRegionsHiggs}
\begin{figure}
      \centering
      \includegraphics[scale=0.50]{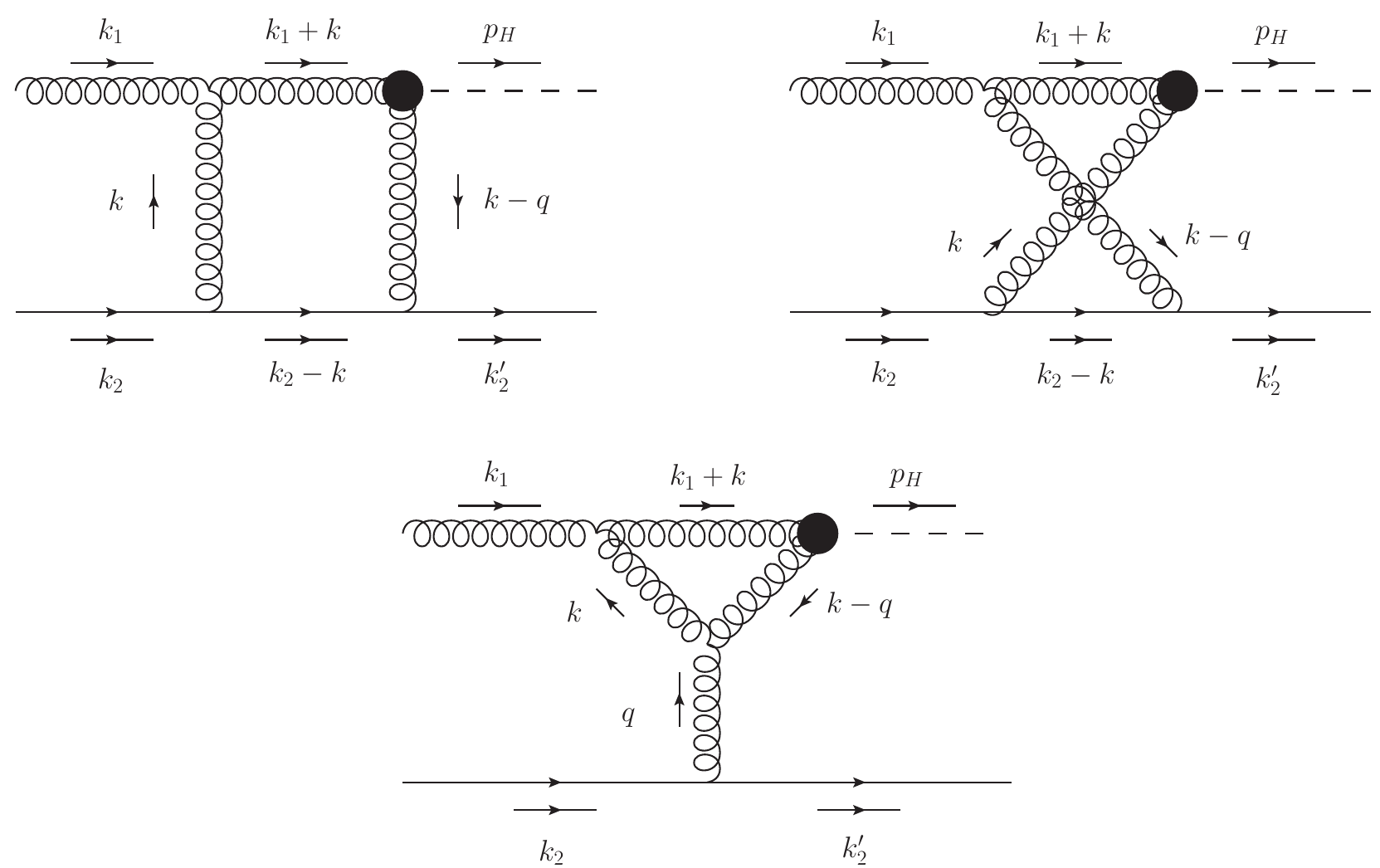}
      \caption{Box, crossed box and triangle diagrams with the corresponding momentum routings.}
      \label{BoxCrossTriangle}
\end{figure}
In this section, we intend to compute the Gribov part of diagrams with two gluons in the $t$-channel (in the top of Fig.~\ref{BoxCrossTriangle}) by the strategy of rapidity regions. These diagrams are complicated, but they greatly simplify when taken in combination with the triangular correction shown in the bottom of Fig.~\ref{BoxCrossTriangle}. \\

Let us start by considering the box and the crossed diagram in Fig.~\ref{BoxCrossTriangle}. The second one can be obtained from the first one by using the crossing symmetry
\begin{equation}
    \mathcal{A}_{{\rm{Cross}}} = - \mathcal{A}_{{\rm{Box}}} (s \rightarrow u \simeq -s) \; .
    \label{crossingsimmetry}
\end{equation}
Using the Gribov's trick for the two propagators connecting regions separated by large interval of rapidity, we find
\begin{gather}
    \mathcal{A}_{\rm{Box}} \simeq \frac{4}{s^2} g^3 g_H \epsilon_{\mu} (k_1) f^{abc} (t^b t^c)_{ji} \nonumber \\ \times \int \frac{d^D k}{(2 \pi)^{D}} \frac{ k_{2,\sigma} k_{2,\rho} A^{\mu \rho \nu}(-k,k+k_1) H_{\nu}^{\; \sigma} (-k-k_1, k-q)  \bar{u} (k_2-q) \slashed{k}_1 (\slashed{k}_2-\slashed{k})\slashed{k}_1 u(k_2)}{k^2 (k-q)^2 (k+k_1)^2 (k-k_2)^2} \; .
\end{gather}
From the Sudakov decomposition~(\ref{Sudakov}) for $k$, we get that
\begin{equation*}
    \bar{u} (k_2-q) \slashed{k}_1 (\slashed{k}_2-\slashed{k}) \slashed{k}_1 u(k_2) = s (1-\alpha) \bar{u} (k_2-q) \slashed{k}_1 u(k_2) \; .
\end{equation*}
Taking into account that the color factor is 
\begin{equation*}
    f^{a b c} (t^b t^c)_{ji} = \frac{1}{2} f^{a b c} (t^b t^c - t^c t^b)_{ji} = \frac{i}{2} C_A t_{ji}^a = \frac{i}{2} C_A \delta^{ac} t_{ji}^c \; ,
    \end{equation*}
we obtain
\begin{gather}
    \mathcal{A}_{{\rm{Box}}} \simeq - \frac{2}{s} g^3 g_H \epsilon_{\mu} (k_1) C_A \delta^{ac} t_{ji}^c \bar{u} (k_2-q) \slashed{k}_1 u(k_2) \nonumber \\ \times \int \frac{d^D k}{i (2 \pi)^{D}} \frac{(1-\alpha) k_{2,\sigma} k_{2,\rho} A^{\mu \rho \nu}(-k,k+k_1) H_{\nu}^{\; \sigma}(-k-k_1, k-q) }{k^2 (k-q)^2 (k+k_1)^2 (k-k_2)^2} \; .
\end{gather}

\subsubsection{Region A}
\begin{figure}
      \centering
      \begin{tabular}{cc}
      \parbox{0.5\textwidth}{\includegraphics[width=0.4\textwidth]{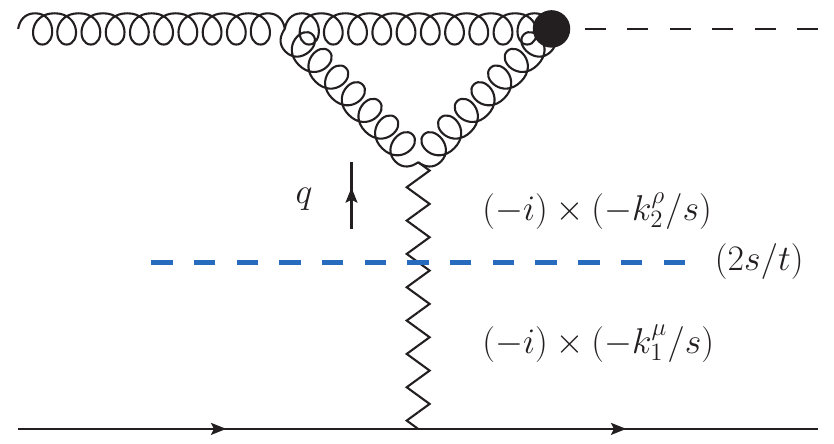}} &  \parbox{0.3\textwidth}  {\includegraphics[width=0.25 \textwidth]{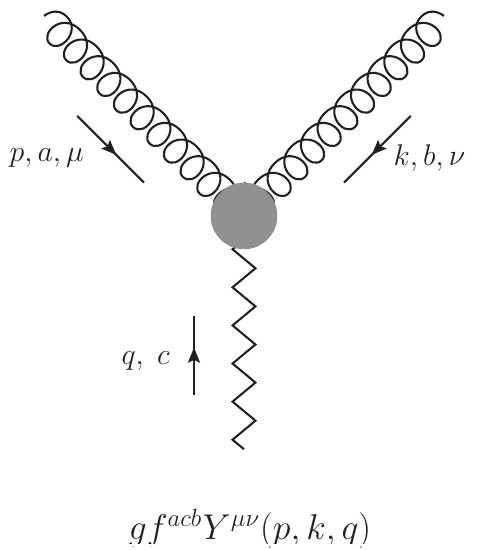}} \\
      \hspace{-1.7 cm} (a) & \hspace{-0.55 cm}(b)
      \end{tabular}
      \caption{ Diagrammatic representation of the Gribov prescriptions in the Regge limit for the triangle amplitude (a) and the full Reggeon-gluon-gluon vertex (b). }
      \label{HiggsNonSense}\label{ggReffective}
  \end{figure}


In the region A, due to crossing symmetry (\ref{crossingsimmetry}), the contribution from the crossed diagram is equal to the one from the box, so we have
\begin{equation}
    \mathcal{A}_{{\rm{A}}} = - 4 g^2 g_H \epsilon_{\mu} (k_1) C_A \delta^{ac} \Gamma_{qq'}^{c (0)} \int \frac{d^D k}{i (2 \pi)^{D}} \frac{k_{2,\sigma} k_{2,\rho} A^{\mu \rho \nu}(-k,k+k_1) H_{\nu}^{\; \sigma}(-k-k_1, k-q) }{k^2 (k-q)^2 (k+k_1)^2 (k-k_2)^2} \; .
\end{equation}
Moving to Sudakov variables, we express the integration measure as
\begin{equation}
    \int \frac{d^D k}{i(2 \pi)^D} = \frac{s}{2} \int \frac{d \alpha}{2 \pi i} \int d \beta  \int \frac{d^{D-2} k_{\perp}}{(2 \pi)^{D-1}} 
\end{equation}
and approximate the last propagator as
\begin{equation}
    (k-k_2)^2 \simeq - \beta s \; . 
\end{equation}
In this way, we easily find
\begin{gather}
    \mathcal{A}_{{\rm{A}}} = \frac{4}{s} g^2 g_H \delta^{ac} \epsilon_{\mu} (k_1) C_A \Gamma_{qq'}^{c(0)} \frac{s}{2} \int_{|\alpha| \;  \lesssim \; \alpha_0} \frac{d \alpha}{2 \pi i} \int_{|\beta| \;  > \; \beta_0} \frac{d\beta}{\beta} \int \frac{d^{D-2} k_{\perp}}{(2 \pi)^{D-1}} \nonumber \\ \times \frac{ k_{2,\sigma} k_{2,\rho} A^{\mu \rho \nu}(-k,k+k_1) H_{\nu}^{\; \sigma}(-k-k_1, k-q) }{k^2 (k-q)^2 (k+k_1)^2 } \; .
\end{gather}
Restoring formally the $1/t$ propagator factor, we can cast the contribution from the box and crossed diagram as 
\begin{gather}
    \mathcal{A}_{{\rm{A}}} = \Gamma_{qq'}^{c(0)} \left( \frac{2 s}{t} \right) g_H \delta^{ac} \epsilon_{\mu} (k_1) (- g^2 C_A) \; s \int_{|\alpha| \;  \lesssim \; \alpha_0} \frac{d \alpha}{2 \pi i} \int d \beta \int \frac{d^{D-2} k_{\perp}}{(2 \pi)^{D-1}} \nonumber \\ \times \frac{ A^{\mu}_{ \; \rho \gamma }(-k,k+k_1) \chi^{\rho}_{ \; \alpha} (-k,k-q,q^2,\beta_0,s) H^{\gamma \alpha}(-k-k_1, k-q) }{k^2 (k-q)^2 (k+k_1)^2 } \; , \label{eq:box-chi}
\end{gather}
where we single-out the factor $\left[ \frac{2 q^2}{s} \frac{k_{2, \alpha} \; k_2^{\rho}}{-\beta s} \right] \theta \left(  \left | \frac{-\beta s}{s} \right| - \beta_0 \right) $ from  the integrand, rewriting it in terms of the vertex
\begin{gather}
 \chi^{\rho}_{ \; \alpha} (p,k,q^2,\beta_0,s)  = \left[ \frac{2 q^2}{s} \frac{k_{2, \alpha} \; k_2^{\rho}}{k_2 (p-k)} \right] \theta \left(  \left | \frac{k_2 (p-k)}{s} \right| - \beta_0 \right) \; , \label{eq:chi-Rgg}
\end{gather}
which is analogous to the ``induced'' Reggeon-gluon-gluon coupling in Lipatov's EFT, discussed in section~\ref{subsec:IntLipatovEFT} (Eq.~(\ref{eq:Rgg-vertex-EFT})).
As anticipated above, adding together the region A of the box and crossed diagram with the triangle  greatly simplifies the computation.  To facilitate the addition of these  contributions we have introduced the vertex (\ref{eq:chi-Rgg}) in such a way that combining the Eq.~(\ref{eq:box-chi}) with the Gribov projection of the triangle diagram (Fig.~\ref{HiggsNonSense}(a)) boils down to the simple replacement of the vertex $\chi_\rho^\alpha$ in Eq.~(\ref{eq:box-chi}) by the complete effective Reggeon-gluon-gluon vertex (Fig.~\ref{ggReffective}(b)) with the tensorial structure
\begin{eqnarray}
&&\hspace{-5mm}Y^{\mu \nu}(p,k,q)= \frac{k_{2,\rho}}{s} A^{\mu \rho \nu} (p+k,-k) + \chi^{\mu\nu}(p,k,q^2,\beta_0,s) \label{Ymunu} \\ 
&&\hspace{-5mm}=\left[ g^{\mu \nu}  \frac{k_2 (k-p)}{s} + \frac{k_2^{\mu}}{s} (2 p + k)^{\nu} - \frac{k_2^{\nu}}{s} (2 k + p)^{\mu} + \frac{2 q^2}{s} \frac{k_2^{\mu} k_2^{\nu}}{k_2 (p-k)} \right] \theta \left(  \left | \frac{k_2(p-k)}{s} \right| - \beta_0 \right) \; . \nonumber
\end{eqnarray}
This vertex had been introduced e.g. in Fig.~3 of Ref.~\cite{Fadin:2001ap}. The theta function multiplies all four terms for convenience of algebraic simplifications in the numerator of the amplitude, but will affect only those terms in which the eikonal denominator $k_2(p-k)$ does not cancel, where it is needed to regularize the singularity in the point $\beta = 0$. 
  
  We also emphasize that while the region A is identified by the condition $|\alpha| < \alpha_0$, which restricts the $\alpha$-integration domain, one still can remove this restriction and integrate over $\alpha$ between $-\infty$ and $\infty$ as a consequence of the fact that the contribution from the region C is power-suppressed. 


\subsubsection{Triangular correction plus box contribution from the region A}




With the help of the effective vertex (\ref{Ymunu}) one can write the sum of triangle and region A of the box diagrams as
\begin{gather}
    \Gamma_{gH}^{ac({\rm{Tri}}+{\rm{A}})} = -i \epsilon_{\mu} (k_1) \int \frac{d^D k}{(2 \pi)^D} g f^{abh} A^{\mu \delta \nu} (-k, k_1+k) \left( \frac{-i g_{\delta \rho} \delta^{hg}}{k^2} \right) \\ \times (-g f^{g f c}) Y^{\rho \beta} (-k,k-q,q) \left( \frac{-i g_{\beta \alpha} \delta^{e f}}{(k-q)^2} \right) \left( \frac{-i g_{\nu \gamma} \delta^{bd}}{(k_1+k)^2} \right) i g_H \delta^{d e} H^{\gamma \alpha}(-k-k_1,k-q) \; , \nonumber
\end{gather}
which immediately gives
\begin{gather}
   \Gamma_{gH}^{ac({\rm{Tri}}+{\rm{A}})} = \frac{g_H \delta^{ac} \epsilon_{\mu} (k_1)}{2} (-g^2 C_A s) \int \frac{d^{D-2} k_{\perp}}{(2 \pi)^{D-1}} \int_{-\infty}^{\infty} d \beta \int_{-\infty}^{\infty} \frac{d \alpha}{2 \pi i} \nonumber \\ \times \frac{\left[A^{\mu}_{\; \rho \gamma} (-k, k_1+k) Y^{\rho }_{\; \; \alpha} (-k,k-q,q)  H^{\gamma \alpha}(-k-k_1,k-q) \right]}{(\alpha \beta s + k_{\perp}^2 + i 0) ( (\beta-\beta_q) (\alpha -\alpha_q) s + (q-k)_{\perp}^2 + i 0) (\alpha (1+\beta) s + k_{\perp}^2 + i 0)} \; ,
\label{TriInter}
\end{gather}
where we used the Sudakov decomposition for $q$,
\begin{equation}
    q = \beta_q k_1 + \alpha_q k_2 + k_{\perp} ,
\end{equation}
with
\begin{equation}
    \beta_q = \frac{- \vec{q}^{\; 2}}{s} \; , \hspace{1 cm}  \alpha_q = \frac{m_H^2+\vec{q}^{\; 2}}{s} \; .
\end{equation}
We want to use the residue theorem to perform the integration over the variable $\alpha$. We observe that there are three $\alpha$-poles in the integrand\footnote{The fourth term in $Y_{\; \; \alpha}^{\rho}$ does not bring further $\alpha$-poles.},
  \begin{equation}
      \alpha_1 = \frac{-k_{\perp}^2 - i0}{\beta s} \; , \hspace{1 cm} \alpha_2 = \alpha_q + \frac{-(q-k)_{\perp}^2 -i0}{(\beta-\beta_q) s} \; ,  \hspace{1 cm} \alpha_3 = \frac{-k_{\perp}^2 - i0}{(1+\beta) s} \; .
  \end{equation}
Let us note that the integration over $\alpha$ vanishes for $\beta >0$ or $\beta < -1$, having $\beta_q<0$. In the region $-1<\beta< \beta_q$, two poles are found in the region of the complex $\alpha$-plane where $\Im \alpha > 0$, while the other, $\alpha_3$, belongs to $\Im \alpha < 0$. Finally, in the region $\beta_q < \beta < 0$, one pole is found in the region of the complex $\alpha$-plane where $\Im \alpha > 0$, while the others, $\alpha_2$ and $\alpha_3$, belongs to $\Im \alpha < 0$. Closing the integration path in the region $\Im \alpha < 0$, we have a contribution from the pole $\alpha_3$ for $-1<\beta< 0$ and a contribution from the pole $\alpha_2$ for $ \beta_q <\beta< 0$. \\

At this point, a remark is in order. In the case of the term coming from the box and crossed diagrams in the region A, the contribution of the $\alpha_2$ pole is completely negligible. In fact, by definition, in the region A we have $\beta \gg \beta_q$ and $\alpha_2$ does not give any contribution. Nonetheless, the presence of the aforementioned pole is important for the triangular diagram that we are calculating in completely general kinematic conditions. We therefore proceed in this way: (i) we consider the contribution that comes from the $\alpha_3$ pole by combining all diagrams together\footnote{In this case, setting $\beta_q = 0$.} and (ii) we compute separately the contribution to the triangular diagram coming from the pole $\alpha_2$. 
  
\paragraph{Contribution from the $\alpha_3$-pole:}

We close the integration path in the region $\Im \alpha < 0$, and after trivial calculation (and making the substitution $\beta \rightarrow -\beta$), we find\footnote{The notation $\left[...\right]^{\beta \rightarrow -\beta}_{\alpha = \alpha_3}$ means that the contraction has to be calculated fixing $\alpha$ to $\alpha_3$ and performing the substitution $\beta \rightarrow - \beta$.}
\begin{gather}
    \Gamma_{gH}^{ac{({\rm{Tri}}-\alpha_3+{\rm{A}})}} =  \frac{g_H \delta^{ac} \epsilon_{\mu} (k_1)}{2} g^2 C_A \int \frac{d^{D-2} k_{\perp}}{(2 \pi)^{D-1}} \int_0^1 d \beta \\ \times \frac{(1-\beta) \left[A^{\mu}_{\; \rho \gamma} (-k, k_1+k) Y^{\rho }_{\; \; \alpha} (-k,k-q,q) H^{\gamma \alpha}(-k-k_1,k-q) \right]^{\beta \rightarrow -\beta}_{\alpha = \alpha_3}}{k_{\perp}^2 \left[ (k_{\perp}-(1-\beta) q_{\perp})^2 + \beta (1-\beta) m_H^2 \right]} \; .
\end{gather}
After very long but simple algebra, we end up with
\begin{gather*}
    \Gamma_{gH}^{ac{({\rm{Tri}}-\alpha_3+{\rm{A}})}} = \frac{g_H \varepsilon_{\perp, \mu} (k_1) \delta^{a c}}{2} \left( -\frac{g^2 C_A}{2} \right) \int_0^1 \frac{d \beta}{\beta} \theta (|\beta|-\beta_0) \int \frac{d^{D-2} k}{(2 \pi)^{D-1}} \frac{1}{k_{\perp}^2} \nonumber \\ \times \frac{1}{ \left[ (k_{\perp}-(1-\beta) q_{\perp})^2 + \beta (1-\beta) m_H^2 \right]} \Bigg \{ \hspace{-0.1 cm} k_{\perp}^{\mu} \hspace{-0.1 cm} \left[ (1-\beta) \bigg ( \hspace{-0.1 cm} \beta (3-\beta (2D-3)) ((1-\beta) q_{\perp}^2 - 2 k_{\perp} \cdot q_{\perp}) \right. \nonumber \\ \left. + \beta^2 m_H^2 (3 \beta + 2 (1- \beta) D - 5 ) + 2 q_{\perp}^2 \bigg) - 2 \beta^2 (D-1) k_{\perp}^2 \right] + q_{\perp}^{\mu} \bigg[ 2 (3 - 2 \beta) \beta k_{\perp}^2  \nonumber \\ \left.
     + (1-\beta) \bigg ( (1-\beta) (2-\beta) (-\beta m_H^2 - (1-\beta) q_{\perp}^2)-2 (3-\beta) \beta k_{\perp} \cdot q_{\perp}  \bigg) \right] \Bigg \} \; .
\end{gather*}
Adding and subtracting some quantities and using the fact the scaleless Feynman integral are zero in dimensional regularization\footnote{We work with same regulator for UV and IR-sector, $\epsilon_{{\rm{UV}}} = \epsilon_{{\rm{IR}}}$.}, we can greatly simplify the expression and obtain
\begin{gather}
   \Gamma_{gH}^{ac{({\rm{Tri}}-\alpha_3+{\rm{A}})}} = \frac{g_H \varepsilon_{\perp, \mu} (k_1) \delta^{a c}}{2} \left( -\frac{g^2 C_A}{2} \right) \int  \frac{d^{D-2} k_{\perp}}{(2 \pi)^{D-1}} \frac{1}{k_{\perp}^2} \int_0^1 \frac{d \beta}{\beta} \theta(|\beta|-\beta_0) \nonumber \\ \times \Bigg \{ k_{\perp}^{\mu} \, \frac{ 2 (D-4) \beta^2 (1-\beta) m_H^2 - \beta (3+\beta) k_{\perp}^2 + 2 (1-\beta) q_{\perp}^2 }{  (k_{\perp}-(1-\beta) q_{\perp})^2 + \beta (1-\beta) m_H^2 } \nonumber \\ - q_{\perp}^{\mu} \frac{ 2 (1-\beta) \left( \beta m_H^2 + (1-\beta) q_{\perp}^2 \right) - 3\beta (1-\beta) k_{\perp}^2 }{ (k_{\perp}-(1-\beta) q_{\perp})^2 + \beta (1-\beta) m_H^2 } \Bigg \} \; .
\end{gather}
We can split this into four contributions, three of which do not have any singularity for $\beta \sim 0$. The first is
\begin{gather}
 \Gamma_{gH,1}^{ac{({\rm{Tri}}-\alpha_3+{\rm{A}})}} = \Gamma_{gH}^{ac(0)} \frac{g^2 C_A}{2} \int \frac{d^{D-2} k_{\perp}}{(2 \pi)^{D-1}} \int_0^1 d \beta \frac{-3 (1-\beta)}{  (k_{\perp}-(1-\beta) q_{\perp})^2 + \beta (1-\beta) m_H^2 } \nonumber \\
 = \Gamma_{gH}^{ac(0)} \left[ \frac{- g^2 C_A \Gamma (1+\epsilon)}{(4 \pi)^{2-\epsilon}}  \right] \left( - \frac{3}{\epsilon} \frac{\Gamma (1-\epsilon) \Gamma (2-\epsilon)}{\Gamma (3-2\epsilon)} (-m_H^2)^{-\epsilon} \right) \; ,
 \label{1Tri}
\end{gather}
where we have used Eq.~(\ref{Eq:AppendixB:FeynmanA=1}).
The second contributions reads
\begin{equation}
\Gamma_{gH,2}^{ac{({\rm{Tri}}-\alpha_3+{\rm{A}})}} = \frac{g_H \varepsilon_{\perp, \mu} (k_1) \delta^{a c}}{2} \frac{g^2 C_A}{2} \int  \frac{d^{D-2} k_{\perp}}{(2 \pi)^{D-1}} \int_0^1 d \beta \frac{k_{\perp}^{\mu} (3+\beta) }{ (k_{\perp}-(1-\beta) q_{\perp})^2 + \beta (1-\beta) m_H^2} \; .
\end{equation}
By performing the translation $k_{\perp} \rightarrow k_{\perp} + (1-\beta) q_{\perp}$, and using again Eq.~(\ref{Eq:AppendixB:FeynmanA=1}), we find
\begin{equation}
    \Gamma_{gH,2}^{ac{({\rm{Tri}}-\alpha_3+{\rm{A}})}} = \Gamma_{gH}^{ac(0)} \left[ \frac{- g^2 C_A \Gamma (1+\epsilon)}{(4 \pi)^{2-\epsilon}}  \right] (-m_H^2)^{-\epsilon} \left( \frac{5}{3 \epsilon} + \frac{59}{18} \right) + {\cal O}(\epsilon) \; .
    \label{2Tri}
\end{equation}
The third term is
\begin{gather}
\Gamma_{gH,3}^{ac{({\rm{Tri}}-\alpha_3+{\rm{A}})}}= \frac{g_H \varepsilon_{\mu, \perp} (k_1) \delta^{a c}}{2} \left( -\frac{g^2 C_A}{2} \right) \int_0^1 \frac{d \beta}{\beta} \int \frac{d^{D-2} k_{\perp}}{(2 \pi)^{D-1}} \theta (|\beta|-\beta_0) \nonumber \\ \times \frac{k_{\perp}^{\mu} \left[  2 (1-\beta) q_{\perp}^2 \right] - q_{\perp}^{\mu} 2 (1-\beta) \left[ \beta m_H^2 + (1-\beta) q_{\perp}^2 \right]}{k_{\perp}^2 \left[ (k_{\perp}-(1-\beta) q_{\perp})^2 + \beta (1-\beta) m_H^2 \right]} 
\end{gather}
and it is the only term which is singular in $\beta=0$ and, thus, for which we cannot send $\beta_0$ to~$0$. We can calculate this third contribution by introducing the Feynman parametrization and, after that, performing the translation $k_{\perp} \rightarrow k_{\perp} + x (1-\beta) q_{\perp}$. In this way, we find
\begin{gather}
   \Gamma_{gH,3}^{ac{({\rm{Tri}}-\alpha_3+{\rm{A}})}} = \Gamma_{g H}^{ac(0)} \left( - g^2 C_A \right) \int_{\beta_0}^1 \frac{d \beta}{\beta} (1-\beta) \int_0^1 dx  \left[ (1-x)(1-\beta) \vec{q}^{\; 2} - \beta m_H^2 \right] \nonumber \\ \times \int \frac{d^{D-1} k_{\perp}}{(2 \pi)^{D-2}} \frac{1}{[ \vec{k}^{\; 2} + L ]^2} \; ,
\end{gather}
where 
\begin{equation}
    L = x (1-\beta) \left[ (1-x)(1-\beta) \vec{q}^{\; 2} - \beta m_H^2 \right] \; .
\end{equation}
Using Eqs.~(\ref{Eq:AppendixB:FeynmanA=1}) and the integral 
\begin{equation}
 \int_0^1 dx \; x^{-\epsilon-1} \left[ 1 -  \frac{(1-\beta) \vec{q}^{\; 2}}{(1-\beta) \vec{q}^{\; 2} - \beta m_H^2} x \right]^{-\epsilon}  = -\frac{1}{\epsilon} + \mathcal{O}(\epsilon) \; , 
\end{equation}
we obtain
\begin{equation}
    \Gamma_{gH,3}^{ac{({\rm{Tri}}-\alpha_3+{\rm{A}})}} = \Gamma^{ac(0)}_{gH} \frac{2 g^2 C_A \Gamma (1+\epsilon)}{(4 \pi)^{2-\epsilon}} \frac{1}{\epsilon} \int_{\beta_0}^1 \frac{d \beta}{\beta} (1-\beta)^{-\epsilon} \left[ (1-\beta) \vec{q}^{\; 2} - \beta m_H^2 \right]^{-\epsilon} + {\cal O} (\epsilon) \; . 
\end{equation}
Performing the $\epsilon$-expansion and setting $\beta_0=0$ where possible, within the required accuracy, we find
\begin{equation*}
  \Gamma_{gH,3}^{ac{({\rm{Tri}}-\alpha_3+{\rm{A}})}} = - \Gamma^{ac(0)}_{gH} \frac{2 g^2 C_A \Gamma (1+\epsilon) (\vec{q}^{\; 2})^{-\epsilon} }{(4 \pi)^{2-\epsilon} \epsilon} \left( \ln \beta_0 - \epsilon \left( \zeta(2) + {\rm{Li}}_2 \left( 1 + \frac{m_H^2}{\vec{q}^{\; 2}} \right) \right) \right) + {\cal O}(\epsilon) 
\end{equation*}
\begin{equation}
  = \Gamma^{ac(0)}_{gH} \left[ -\frac{ g^2 C_A \Gamma (1+\epsilon)}{(4 \pi)^{2-\epsilon}} (\vec{q}^{\; 2})^{-\epsilon} \right] \left(  - 2 \zeta(2) - 2 {\rm{Li}}_2 \left( 1 + \frac{m_H^2}{\vec{q}^{\; 2}} \right) \right) - \Gamma^{ac}_{gH} \omega^{(1)} (t) \ln \beta_0  + {\cal O}(\epsilon) .
  \label{3Tri}
\end{equation}
The fourth term is
\begin{gather}
   \Gamma_{gH,4}^{ac{({\rm{Tri}}-\alpha_3+{\rm{A}})}} = \frac{g_H \varepsilon_{\perp, \mu} (k_1) \delta^{a c}}{2} \left( -\frac{g^2 C_A}{2} \right) \int  \frac{d^{D-2} k_{\perp}}{(2 \pi)^{D-1}} \int_0^1 \frac{d \beta}{\beta} \nonumber \\ \times \frac{k_{\perp}^{\mu} \left[ 2 (D-4) \beta^2 (1-\beta) m_H^2 \right]}{k_{\perp}^2 \left[ (k_{\perp}-(1-\beta) q_{\perp})^2 + \beta (1-\beta) m_H^2 \right]} \theta(|\beta|-\beta_0) \; ,
\end{gather}
which, after trivial mathematical steps, becomes
\begin{gather}
   \Gamma_{gH,4}^{ac{({\rm{Tri}}-\alpha_3+{\rm{A}})}} = \Gamma^{ac(0)}_{gH} \left[ \frac{ 4 m_H^2 g^2 C_A \Gamma (1+\epsilon)}{(4 \pi)^{2-\epsilon}} \right] \epsilon \int_0^1 d\beta \; \beta (1-\beta)^{1-\epsilon} \nonumber \\ \times \int_0^1 dx \; x^{-\epsilon} \left[ (1-x)(1-\beta) \vec{q}^{\; 2} - \beta m_H^2 \right]^{-\epsilon-1} = {\cal O}(\epsilon) \; .
\end{gather}
Within constant terms in the $\epsilon$-expansion, we obtain the result by summing Eqs.~(\ref{1Tri}), (\ref{2Tri}) and (\ref{3Tri}). We can then define
\begin{gather}
  \delta_{{\rm{NLO}}}^{{({\rm{Tri}}-\alpha_3+{\rm{A}})}} \equiv \frac{\Gamma_{gH}^{ac{({\rm{Tri}}-\alpha_3+{\rm{A}})}}}{\Gamma^{ac(0)}_{gH}} = - \omega^{(1)} (t) \ln \beta_0 - \frac{g^2 C_A \Gamma (1+\epsilon)}{(4 \pi)^{2-\epsilon}} (\vec{q}^{\; 2})^{-\epsilon} \nonumber \\ \times \left[ - \frac{3}{\epsilon} \frac{\Gamma (1-\epsilon) \Gamma (2-\epsilon)}{\Gamma (3-2\epsilon)} \left( - \frac{m_H^2}{\vec{q}^{\; 2}} \right)^{-\epsilon} \hspace{-0.2 cm} + \left( \frac{5}{3 \epsilon} + \frac{59}{18} \right) \left( - \frac{m_H^2}{\vec{q}^{\; 2}} \right)^{-\epsilon} \hspace{-0.2 cm} - 2 \zeta (2) - 2 {\rm{Li}}_2 \left( 1 + \frac{m_H^2}{\vec{q}^{\; 2}} \right) \right] \; ,
  \label{Triangularcorrection}
\end{gather}
or, equivalently, 
\begin{gather}
 \delta_{{\rm{NLO}}}^{{({\rm{Tri}}-\alpha_3+{\rm{A}})}} = - \omega^{(1)} (t) \ln \beta_0 + \frac{\bar{\alpha}_s}{4 \pi} \left( \frac{\vec{q}^{\; 2}}{\mu^2} \right)^{-\epsilon} \nonumber \\ \times \left \{ - \frac{1}{6} \frac{C_A}{\epsilon} - \frac{5}{18} C_A + \frac{1}{6} C_A \ln \left( - \frac{m_H^2}{\vec{q}^{\; 2}} \right) + 2 C_A \left( \frac{\pi^2}{6} + {\rm{Li}}_2 \left( 1 + \frac{m_H^2}{\vec{q}^{\; 2}} \right)  \right) \right \} + {\cal O}(\epsilon) \; .
 \label{eq:DeltaTriAlpha1PlusA}
\end{gather}

\paragraph{Contribution from the $\alpha_2$-pole:}

Let us now turn our attention to the contribution from the pole $\alpha_2$-pole. Starting from Eq.~(\ref{TriInter}) and calculating the residue at the pole, we find
\begin{equation}
    \Gamma_{gH}^{ac({\rm{Tri}}-\alpha_2)} = \frac{g_H \varepsilon_{\perp, \mu} (k_1) \delta^{ac}}{2} \frac{g^2 C_A}{2} \int_0^1 d \beta \int \frac{d^{D-2} k_{\perp}}{(2 \pi)^{D-1}} \frac{k_{\perp}^{\mu} - 3 q_{\perp}^{\mu}}{(1-\beta) k_{\perp}^2 + \beta (k-q)_{\perp}^2 } \;,
\end{equation}
where we performed the re-scaling $\beta \rightarrow \beta_q \beta = (q_{\perp}^2/s) \beta$. Then, it is easy to see that this contribution reads
\begin{gather}
   \Gamma_{gH}^{ac({\rm{Tri}}-\alpha_2)} = \Gamma^{ac(0)}_{gH} \frac{g^2 C_A \Gamma (1+\epsilon) (\vec{q}^{\; 2})^{-\epsilon}}{(4 \pi)^{2-\epsilon}} \frac{1}{\epsilon} \frac{\Gamma^2 (1 - \epsilon)}{\Gamma (2-2 \epsilon)} \frac{5}{2}  = \Gamma^{ac(0)}_{gH} \frac{\bar{\alpha}_s}{4 \pi} \left( \frac{\vec{q}^{\; 2}}{ \mu^2 } \right)^{-\epsilon} \left[ \frac{5}{2} \frac{C_A}{\epsilon} + 5 C_A \right]
\end{gather}
and we can therefore define
\begin{equation}
    \delta_{{\rm{NLO}}}^{{({\rm{Tri}}-\alpha_2)}} = \frac{\bar{\alpha}_s}{4 \pi} \left( \frac{\vec{q}^{\; 2}}{ \mu^2 } \right)^{-\epsilon} \left[ \frac{5}{2} \frac{C_A}{\epsilon} + 5 C_A \right] \; .
    \label{eq:DeltaTriAlpha2}
\end{equation}

\subsubsection{Central region}
\label{Centralregion}

In the central region a minor surprise is avaiting us. We start considering only the box contribution, which reads
\begin{gather}
  \mathcal{A}_{{\rm{Box,Central}}} = \Gamma_{q q'}^{c(0)} \left( \frac{2s}{t} \right) g_H \epsilon_{\mu} (k_1) \delta^{ac} \left(-\frac{ g^2 C_A t}{s} \right) \nonumber \\ \times \frac{s}{2} \int_{-\alpha_0}^{\alpha_0} \frac{d \alpha}{2 \pi i} \int_{-\beta_0}^{\beta_0} d \beta \int \frac{d^{D-2} k}{(2 \pi)^{D-1}} \frac{ k_{2,\sigma} k_{2,\rho} A^{\mu \rho \nu}(-k,k+k_1) H_{\nu}^{\; \sigma}(-k-k_1, k-q)}{k^2 (k-q)^2 (k+k_1)^2 (k-k_2)^2}  \; .
\end{gather}
It is easy to see that in this region
\begin{equation}
    k_{2,\sigma} k_{2,\rho} A^{\mu \rho \nu}(-k,k+k_1) H_{\nu}^{\; \sigma}(-k-k_1, k-q) \simeq \frac{s^2}{2} ( q_{\perp}^{\mu}-k_{\perp}^{\mu})
\label{Eq:Regions:ContractionAtSmallBeta}
\end{equation}
and hence 
\begin{gather}
  \mathcal{A}_{{\rm{Box,Central}}} = \Gamma_{q q'}^{c(0)} \left( \frac{2s}{t} \right) g_H \epsilon_{\mu} (k_1) \delta^{ac} \left(- \frac{ g^2 C_A st}{2} \right) \nonumber \\ \times \frac{s}{2} \int_{-\alpha_0}^{\alpha_0} \frac{d \alpha}{2 \pi i} \int_{-\beta_0}^{\beta_0} d \beta \int \frac{d^{D-2} k}{(2 \pi)^{D-1}} \frac{q_{\perp}^{\mu}-k_{\perp}^{\mu}}{k^2 (k-q)^2 (k+k_1)^2 (k-k_2)^2} \; .
\end{gather}
The denominators can be written as
\begin{gather*}
    k^2 = \alpha \beta s + k_{\perp}^2 \; , \hspace{0.5 cm} (q-k)^2 = (\alpha_q-\alpha) (\beta_q-\beta) s + (q-k)_{\perp}^2 \; ,
\end{gather*}
\begin{equation}
    (k+k_1)^2 = \alpha (1+\beta) s + k_{\perp}^2 \; , \hspace{0.5 cm} (k-k_2)^2 = (\alpha-1) \beta s + k_{\perp}^2 \; .
\end{equation}
We are going to integrate in the $s|\alpha|$-complex plane (the integration in this case will be done from $-s\alpha_0$ and $s \alpha_0$, which tend to $-\infty$ and $+\infty$, respectively, when $s \rightarrow \infty$) and we have the following poles:
\begin{equation}
    s\alpha_1 = -\frac{k_{\perp}^2 + i0}{\beta} \; , \hspace{0.3 cm} s\alpha_2 = -\frac{(q-k)_{\perp}^2 + i0}{\beta} \; , \hspace{0.3 cm} s\alpha_3 = - \frac{k_{\perp}^2 + i0}{1+\beta} \; , \hspace{0.3 cm}  s\alpha_4 = -\frac{-s \beta + k_{\perp}^2 + i0}{\beta} \; .
\end{equation}
First, we observe that the integral will be non-zero only in the region ($-\beta_0 < \beta < 0$) and then that, when $s \rightarrow \infty$, $\beta_0 \rightarrow 0$ (and hence $\beta \sim 0$). This means that, when $s \rightarrow \infty$ while the pole $s \alpha_3$ is found at a ``fixed'' position in the $s|\alpha|$-complex plane, all other poles are found in the region in which $ \Re \{ s\alpha \} \rightarrow -\infty$, $\Im \{ s\alpha \} \rightarrow \infty $. Thanks to this, we can make a shift in the complex plane in a region in which $s|\alpha| \gg k_{\perp}^2$ always in the domain, in order to have
\begin{equation}
    (k+k_1)^2 = \alpha (1+\beta) s + k_{\perp}^2   \simeq \alpha s \; .
\end{equation}
An analogous shift in the $s|\beta|$-plane leads to the conclusion\footnote{Using the same argument we neglect $\alpha_q$ and $\beta_q$ with respect to $\alpha$ and $\beta$ respectively.}
\begin{equation}
    (k-k_2)^2 = (\alpha-1)\beta s + k_{\perp}^2 \simeq - \beta s \; .
\end{equation}
Then, we obtain
\begin{gather}
   \mathcal{A}_{{\rm{Box,Central}}} = \Gamma_{q q'}^{c(0)} \left( \frac{2s}{t} \right) g_H \epsilon_{\mu} (k_1) \delta^{ac} \left(- \frac{g^2 C_A st}{2} \right)   \\ \times \frac{s}{2} \int_{-\alpha_0}^{\alpha_0} \hspace{-0.3 cm} d \alpha \int_{-\beta_0}^{\beta_0} \hspace{-0.3 cm} d \beta \int \frac{d^{D-2} k_{\perp}}{(2 \pi)^{D}i} \frac{q_{\perp}^{\mu}-k_{\perp}^{\mu}}{(\alpha \beta s + k_{\perp}^2+i0) (\alpha \beta s + (q-k)_{\perp}^2 + i0) (-\beta s + i 0) (\alpha s + i0)} 
   \nonumber \; .
\label{eq_Central_Higgs_1}
\end{gather}
Here, we note the first important difference with respect to \cite{Fadin:2001dc}, that we reviewed in section~\ref{SubSec:StrategyOfRapidityRegionsQCD}: at this step of the computation the lower and upper Born effective vertices were factorized out of the integration, leaving the universal integral $I^{{\rm{central}}}$. In the present case, the numerator contains two terms. The first one is not problematic, because $q$ could be factored out and leave us with the aforementioned universal integral. However, the second one contains the loop variable and therefore cannot be taken out of the integral. This shows that the gluon-Reggeon-Higgs vertex does not factorize in the kinematic regions where the longitudinal component of the loop momentum $k$ along $k_1$, {\it i.e.} $\beta$, is small. \\

Nonetheless, in this region, we can use the symmetry of denominators in Eq.~(\ref{eq_Central_Higgs_1}) under the exchange $k_{\perp} \rightarrow q_{\perp}-k_{\perp}$, to replace the numerator by $\frac{1}{2} q_{\perp}$ and obtain
\begin{gather}
   \mathcal{A}_{{\rm{Box,Central}}} = \Gamma_{q q'}^{c(0)} \left( \frac{2s}{t} \right) \Gamma_{gH}^{ac(0)} \left(- \frac{g^2 C_A st}{2} \right)  \nonumber \\ \times \frac{s}{2} \int_{-\alpha_0}^{\alpha_0} \hspace{-0.3 cm} d \alpha \int_{-\beta_0}^{\beta_0} \hspace{-0.3 cm} d \beta \int \frac{d^{D-2} k}{(2 \pi)^{D}i} \frac{1}{(\alpha \beta s + k_{\perp}^2+i0) (\alpha \beta s + (q-k)_{\perp}^2 + i0) (-\beta s + i 0) (\alpha s + i0)} \; ,
\end{gather}
which is exactly in the form~(\ref{eq:GeneralOfStrat_Central}). Hence, the result is \cite{Fadin:2001dc}
\begin{gather}
   \mathcal{A}_{{\rm{Box,Central}}} = \Gamma_{q q'}^{c(0)} \left( \frac{2s}{t} \right) \Gamma_{gH}^{a c (0)}  \omega^{(1)}(t) \left[ \frac{1}{2} \ln \left( \frac{-s}{-t} \right) + \frac{\phi(\alpha_0)}{2} + \frac{\phi(\beta_0)}{2} \right] \; .
\end{gather}
Using the crossing symmetry in Eq.~(\ref{crossingsimmetry}), we find
\begin{gather}
    \mathcal{A}_{{\rm{Cross,Central}}} =  \Gamma_{q q'}^{c(0)} \left( \frac{2s}{t} \right) \Gamma_{gH}^{a c (0)}  \omega^{(1)}(t) \left[ \frac{1}{2} \ln \left( \frac{s}{-t} \right) + \frac{\phi(\alpha_0)}{2} + \frac{\phi(\beta_0)}{2} \right] 
\end{gather}
and, therefore,
\begin{equation}
    \mathcal{A}_{{\rm{Central}}} = \Gamma_{q q'}^{c(0)} \left( \frac{2s}{t} \right) \Gamma_{gH}^{a c (0)}  \omega^{(1)}(t) \left[ \frac{1}{2} \ln \left( \frac{s}{-t} \right) + \frac{1}{2} \ln \left( \frac{-s}{-t} \right) + \phi(\alpha_0) + \phi(\beta_0) \right] \; .
\end{equation}
Although the vertex did not factorize immediately at the integrand level, the result in the central region reproduces the standard result in Eq.~(\ref{Eq:Regions:CentralAmp}) and the vertex correction turns out to be\footnote{We recall that the term depending on $\phi(\alpha_0)$ should be assigned to the lower impact factor, while the one depending on the energy-logarithms to the Regge trajectory.}
\begin{equation}
    \Gamma_{g H}^{ac ({\rm{Central}})} = \Gamma_{g H}^{ac (0)}  \delta_{{\rm{NLO}}}^{({\rm{central}})} \;,
\end{equation}
with
\begin{equation}
     \delta_{{\rm{NLO}}}^{({\rm{central}})} = \omega^{(1)} (t) \phi (\beta_0) = \omega^{(1)} (t) \ln (\beta_0) + \frac{\bar{\alpha}_s}{4 \pi} \left( \frac{\vec{q}^{\; 2}}{\mu^2} \right)^{-\epsilon} \left \{ - \frac{C_A}{\epsilon^2} \right \} + {\cal O}(\epsilon^0) \; .
     \label{eq:DeltaCentral}
\end{equation}
Fortunately, it appears that our dimension-5 operator has no dramatic impact in this region, indeed (i) logarithms of energy are as we would expect, (ii) the correction to the vertex contains the correct logarithmic rapidity divergence in $\beta_0$ and the correct double pole\footnote{A double pole can only be generated from this region in this one-loop computation.} in Eq. (\ref{Eq:Gribov:DeltaNLOHiggs}).

\subsubsection{Region B and factorisation-violating eikonal terms}

In the region B, again, the contribution coming from the crossed diagram is equal to the one coming from the box, so we have
\begin{gather}
    \mathcal{A}_{{\rm{B}}} = - \frac{4}{s} g^3 g_H \epsilon_{\mu} (k_1) C_A \delta^{ac} t_{ji}^c \bar{u} (k_2-q) \slashed{k}_1 u(k_2) \nonumber \\ \times \int \frac{d^D k}{i (2 \pi)^{D}} \frac{(1-\alpha) k_{2,\sigma} k_{2,\rho} A^{\mu \rho \nu}(-k,k+k_1) H_{\nu}^{\; \sigma}(-k-k_1, k-q) }{k^2 (k-q)^2 (k+k_1)^2 (k-k_2)^2} \; .
\end{gather}
The third propagator can be approximated as
\begin{equation}
    (k+k_1)^2 = \alpha (1+\beta) s + k_{\perp}^2 \simeq \alpha s \; .
\end{equation}
In this region, the contraction in the numerator is exactly as in Eq.~(\ref{Eq:Regions:ContractionAtSmallBeta}) and hence
\begin{gather}
    \mathcal{A}_{{\rm{B}}} = \Gamma^{c(0)}_{q q'} \left( \frac{2s}{t} \right) \frac{\epsilon_{\mu} (k_1) \delta^{ac} g_H}{2} (-2 g^2 C_A t s) \int \frac{d^D k}{i (2 \pi)^{D}} \frac{(1-\alpha) ( q_{\perp}^{\mu}-k_{\perp}^{\mu}) }{k^2 (k-q)^2 \alpha s (k-k_2)^2} \nonumber \\
    = \Gamma^{c(0)}_{q q'} \left( \frac{2s}{t} \right)  \frac{\epsilon_{\mu} (k_1) \delta^{ac} g_H}{2} (- g^2 C_A t s ) \int_{| \alpha | > \alpha_0} \frac{d \alpha}{\alpha} (1-\alpha) \nonumber \\ \times \int \frac{d^{D-2} k_{\perp}}{(2 \pi)^{D-1}} (q_{\perp}^{\mu} - k_{\perp}^{\mu}) \int_{| \beta |< \beta_0} \frac{d \beta}{2 \pi i} \frac{1}{k^2 (k-q)^2 (k-k_2)^2} \; .
\end{gather}
Denominators can be expressed as
\begin{equation}
    \frac{1}{k^2} = \frac{1}{\alpha \beta s + k_{\perp}^2} \; , \hspace{0.5 cm} \frac{1}{(k-k_2)^2} = \frac{1}{\beta (\alpha-1) s + k_{\perp}^2} \; , \hspace{0.5 cm} \frac{1}{(k-q)^2} = \frac{1}{\alpha (\beta-\beta_q) s + (k-q)_{\perp}^2} \; . 
\end{equation}
This time, the longitudinal component with respect to $k_2$ of the exchanged momenta $q$, $\alpha_q$, is neglected with respect to $\alpha$, while $\beta_q = q_{\perp}^2/s$ is retained. We can again extend the integration in $\beta$ from $-\infty$ to $+ \infty$, by just adding a suppressed contribution from region C, and obtain 
\begin{equation*}
   \Gamma^{c(0)}_{q q'} \left( \frac{2s}{t} \right) \frac{\epsilon_{\mu} (k_1) \delta^{ac} g_H}{2} (- g^2 C_A t s)  \int_{|\alpha| > \alpha_0} \frac{d \alpha}{\alpha} (1-\alpha) \int \frac{d^{D-2} k_{\perp}}{(2 \pi)^{D-1}} (q_{\perp} - k_{\perp})^{\mu} 
\end{equation*}
\begin{equation}
   \times \int_{-\infty}^{\infty}  \frac{d \beta}{2 \pi i} \frac{1}{(\alpha \beta s + k_{\perp}^2) (\beta (\alpha-1) s + k_{\perp}^2) (\alpha (\beta-\beta_q) s + (k-q)_{\perp}^2)} \; .
\end{equation}
The integral is non-zero only for $\alpha_0 < \alpha < 1$ and, by using the residue theorem, we find
\begin{equation}
    \mathcal{A}_{{\rm{B}}} = \Gamma^{c(0)}_{q q'} \left( \frac{2s}{t} \right) \frac{\epsilon_{\mu} (k_1) \delta^{ac} g_H}{2} g^2 C_A t \int_{\alpha_0}^{1} \frac{d \alpha}{\alpha} (1-\alpha)^2 \int \frac{d^{D-2} k}{(2 \pi)^{D-1}} \frac{(q - k)_{\perp}^{\mu}}{k_{\perp}^2 (k_{\perp}-(1-\alpha) q_{\perp})^2} \; .
\end{equation}
Performing a tensor reduction, the correction from this region is
\begin{equation}
  \frac{\mathcal{A}_{B}}{\Gamma^{ac(0)}_{g H} \left( \frac{2s}{t} \right) \Gamma^{c(0)}_{q q'}} = \frac{g^2 C_A t}{2} \int_{\alpha_0}^{1} \frac{d \alpha}{\alpha} (1-\alpha)^2 \left( 1 + \alpha \right) \int \frac{d^{D-2} k_{\perp}}{(2 \pi)^{D-1}} \frac{1}{k_{\perp}^2 (k_{\perp} - (1-\alpha) q_{\perp})^2} \; .
\end{equation}
Now, after comparison with Ref.~\cite{Fadin:2001dc}, the first term in the bracket $\left( 1 + \alpha \right)$ gives the contribution which is assigned to the lower quark vertex, while the second gives
\begin{equation}
    \delta_{{\rm{NLO}}}^{({\rm{B}})} = \frac{\bar{\alpha}_s}{4 \pi} \left( \frac{\vec{q}^{\; 2}}{\mu^2} \right)^{-\epsilon} \left[ \frac{2 C_A}{\epsilon} + 4 C_A \right]  \; .
    \label{eq:DeltaB}
\end{equation}
Typically, the non-logarithmic contribution coming from the region B would be entirely attributed to the target (quark) impact-factor. However, we cannot modify the known one-loop quark-Reggeon-quark vertex and hence we are forced to assign the {\it factorisation-violating} contribution (\ref{eq:DeltaB}) to the gluon-Reggeon-Higgs vertex. In this region, the lack of factorization of the vertex has generated an additional term. In fact, it was the tensor reduction in the integral over $k_{\perp}$ that generated the factor $(1+\alpha)$, {\it i.e.} the anomalous term is related with the momentum dependence of the $Hgg$-vertex. \\

If we sum the results in Eqs.~(\ref{eq:DeltaTriAlpha1PlusA}), (\ref{eq:DeltaTriAlpha2}), (\ref{eq:DeltaCentral}), (\ref{eq:DeltaB}), and add the contributions coming from the remaining single-gluon $t$-channel diagrams, {\it i.e.} the diagrams $(b)$, $(c)$ and $(d)$ (this latter weighted with the factor 1/2) of Fig.~\ref{Fig:Gribov:GribovTrickHiggs} we obtain the ``Gribov'' (or ``eikonal'') part of the interference between the  one-loop and Born amplitudes. The elegant computation of the eikonal part  with the method of regions, which is presented in this section, is in complete agreement with the brute-force computation of the Regge limit of this amplitude with all $t$-channel propagators replaced by the standard Gribov's prescription (\ref{Gribov:Eq:GribovsTrick}), which we also have performed using \texttt{FeynCalc}. 

To obtain the correct result for the interference between one-loop and tree-level amplitude one has to add to the eikonal part discussed in the previous paragraph the full non-Gribov part isolated in Eq.~(\ref{Eq:Gribov:NonGribTot}). The result for the effective gluon-Reggeon-Higgs vertex, obtained from the complete amplitude, agrees with earlier results~\cite{Celiberto:2022fgx, Nefedov:2019mrg}. \\

{\bf The most important observation in this section} is that the  {\it factorisation-violating eikonal contribution}~(\ref{eq:DeltaB}) exactly cancels out the one generated by {\it non-Gribov contributions}~(\ref{Eq:Gribov:NonGribTot}). This means that at one loop not only the Regge form of the amplitude is preserved, but also that in a completely non-trivial way two anomalies generated by the non-renormalizable interaction cancel each other out. 

\section{The Regge limit of $gg\to gH$ amplitude}
\label{sec:EFT}
Below in section~\ref{sec:Checking}, we verify the non-ambiguity of the effective gluon-Reggeon-Higgs vertex. For this purpose, we extract the effective vertex starting from the $\mathcal{A}_{gg \rightarrow Hg}$ amplitude. The obtained result for the one-loop gluon-Reggeon-Higgs effective vertex~\cite{Celiberto:2022fgx} is in agreement with the computation performed within the Lipatov's EFT in Ref.~\cite{Nefedov:2019mrg}. This leads to the natural question: why does Lipatov's EFT leads straightforwardly to the right result? In the Lipatov's EFT, the Reggeized gluon is a scalar particle and $t$-channel Reggeon exchanges cannot lead to the above-mentioned anomalous helicity structure of the non-Gribov terms in the amplitude. The answer lies in the cancellation mechanism between the non-Gribov contributions and those that break the rapidity factorization. To show this, in section~\ref{sec:Checking} we isolate the non-Gribov terms and then, in section~\ref{sec:fact-viol-gg}, we calculate the contribution from the kinematical region B of the box and cross diagrams contributing to $\mathcal{A}_{gg \rightarrow Hg}$ and show that these two contributions cancel as they did in the case of $\mathcal{A}_{gq \rightarrow Hq}$.
Finally, in section~\ref{subsec:IntLipatovEFT}, in light of what has been described, we review the Lipatov's effective action framework and discuss the details of the computation done in \cite{Nefedov:2019mrg} within this framework.

\subsection{$gg \rightarrow gH$ amplitude at one loop}
\label{sec:Checking}
\begin{figure}
      \centering
      \includegraphics[scale=0.50]{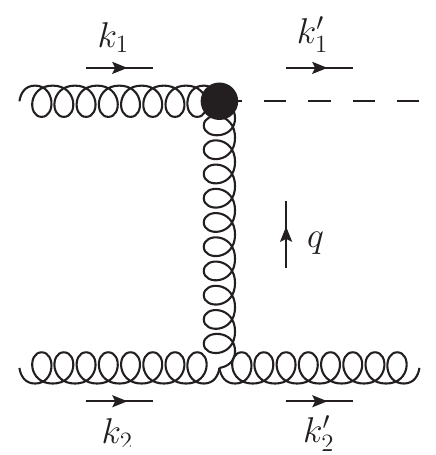}
      \caption{Dominant contribution to the gluon-gluon into Higgs-gluon scattering amplitude in the high-energy limit.}
      \label{Fig:Gribov:GluonHiggsGluonGluon}
\end{figure}
In this section, we make a one-loop check of the strategy illustrated in section~\ref{sec:GribovViolation} and study the non-Gribov contributions arising in the case of a gluon target. To do this, we consider the scattering of a gluon off a gluon to produce a Higgs plus a gluon, $\mathcal{A}_{gg \rightarrow Hg}$. The relevant Born diagram is shown in Fig.~\ref{Fig:Gribov:GluonHiggsGluonGluon} and leads to the following amplitude  
\begin{gather}
    \mathcal{A}_{gg \rightarrow Hg}^{(0)} = g_H \delta^{ac} \varepsilon_{\perp, \mu} (k_1) H^{\mu \rho} \left( -k_1,-q \right) \frac{1}{t} \left( g_{\rho \eta}^{\perp \perp} + \frac{2 k_{1, \rho} k_{2, \eta}}{s} + \frac{2 k_{1, \eta} k_{2, \rho}}{s} \right) g T^c_{bd} \nonumber \\
  \times  A^{\eta \gamma \delta} (-k_2, k_2-q) \varepsilon_{\perp, \gamma} (k_2) \varepsilon_{ \delta} (k_2') \; ,
  \label{gg_to_Hg}
\end{gather}
where we employ the gauge choices
\begin{gather}
     \varepsilon (k_1) \cdot k_2 = 0 \implies \varepsilon (k_1) = \varepsilon_{\perp} (k_1) \; , \\
     \varepsilon (k_2) \cdot k_1 = 0 \implies \varepsilon (k_2) = \varepsilon_{\perp} (k_2) \; , \\
      \varepsilon^{*} (k_2') \cdot k_1 = 0 \implies \varepsilon^{*} (k_2') = - \frac{\varepsilon_{\perp}^{*} (k_2') \cdot k_2' }{k_2' \cdot k_1 } k_1 + \varepsilon_{\perp}^{*} (k_2') \; .
     \label{GaugeChoice2}
\end{gather}
The second term in the round bracket of~(\ref{gg_to_Hg}) gives a vanishing contribution, while the first gives a contribution suppressed like $1/s$ with respect to the expected leading behaviour. Hence, we consider only the third term, the Gribov one, and get
\begin{gather}
    \mathcal{A}_{gg \rightarrow Hg}^{(0)} = \frac{g_H \delta^{ac} \varepsilon_{\perp} (k_1) \cdot q_{\perp}}{2} \left( \frac{2s}{t} \right) g T^c_{bd} (-\varepsilon_{\perp} (k_2) \cdot \varepsilon_{\perp} (k_2')) = \Gamma_{gH}^{ac(0)} \left( \frac{2s}{t} \right) \Gamma_{gg}^{cbd(0)} \; ,
\end{gather}
where
\begin{equation}
     \Gamma_{gg}^{cbd(0)} = g T^c_{bd} (-\varepsilon_{\perp} (k_2) \cdot \varepsilon_{\perp} (k_2')) = g T^c_{bd} \delta_{\lambda_{2'} \lambda_2} \;,
\end{equation}
and which has the expected Regge form. \\

\begin{figure}
  \begin{center}
  \includegraphics[scale=0.45]{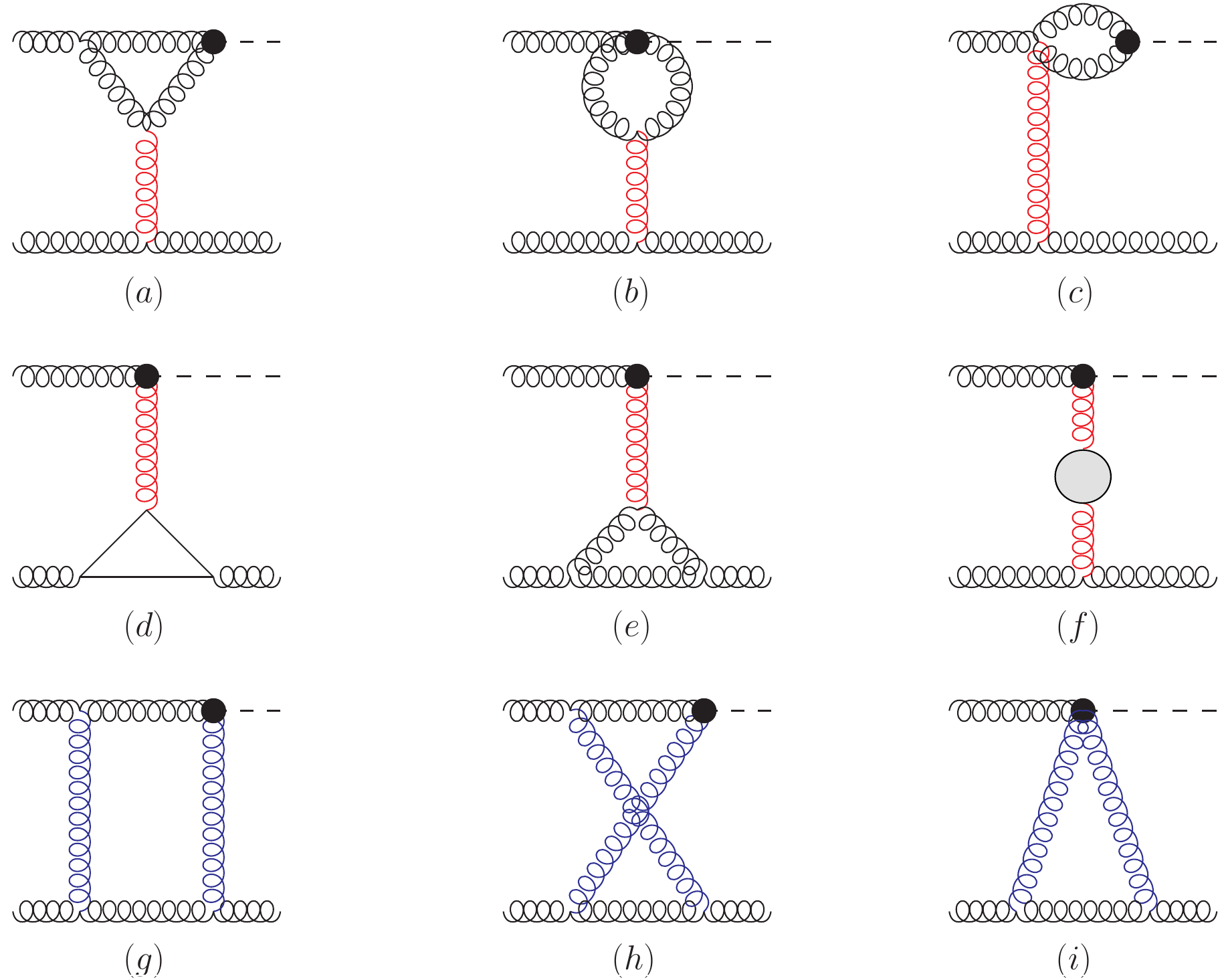}
  \end{center}
  \caption{Dominant contribution to the gluon-gluon into Higgs-gluon scattering amplitude in the high-energy limit at the next-to-leading order. The blob in diagram $(f)$ contains the summed contribution of quarks, gluons and ghosts, while the triangle in the diagram $(d)$ should be understood as the sum of quark, anti-quark, ghost and anti-ghost contributions. All propagators in red are the one that can be approximated {\it \`a la} Gribov, while the propagators in blue must be approximated as in Eq.~(\ref{Gribov:Eq:GribovsTrickMod}).}
  \label{Fig:Gribov:GribovTrickHiggsgggH}
\end{figure}
We now turn our attention to the one-loop computation of the $\mathcal{A}_{gg \rightarrow Hg}$. Although there are results available in the literature, we re-compute this amplitude from scratch to establish if the modified Gribov prescription suggested in Eq.~(\ref{Gribov:Eq:GribovsTrickMod}) works in the case of the $gg \rightarrow gH$ amplitude. The idea is to generate all contributing diagrams using \texttt{FeynArts}~\cite{Kublbeck:1990xc,Hahn:2000kx} and compute them in a fully general kinematics. We then take the high-energy limit to compare it with what expected from a computation based on high-energy techniques. At one-loop, in Feynman gauge, there are 60 diagrams, 46 of which are zero or suppressed in the high-energy limit. The $s$-leading diagrams are shown in Fig.~\ref{Fig:Gribov:GribovTrickHiggsgggH}. For the first three diagrams we find
\begin{gather}
     \frac{\mathcal{A}_{gg \rightarrow Hg}^{(1), (a)} \; \mathcal{A}_{gg \rightarrow Hg}^{(0)*}}{|\mathcal{A}_{gg \rightarrow Hg}^{(0)}|^2} = \frac{g^2 C_A }{4 (D-2) (1-D)
   \left(m_H^2+\vec{q}^{\; 2}\right)^2} \nonumber \\
   \times \Bigg \{ 2 (D-1) \left[ \left(-5 (D-2) m_H^4-4(3 D-8) m_H^2
   \vec{q}^{\; 2}+(16-7 D) (\vec{q}_2^{\; 2})^2 \right) B_0(q^{2}) \right. \nonumber \\
   \left. +2 (D-2) m_H^2 \left(2
   m_H^4 +3 m_H^2 \vec{q}^{\; 2} + (\vec{q}^{\; 2})^2\right)
   C_0\left(q^{ 2},m_H^2\right) \right] + \left( -D(D-2) (\vec{q}^{\; 2})^2 \right. \nonumber \\
   + \left. (D (D (4 D-35)+92)-60)
   m_H^4-2 (D (-2 (D-9) D-41)+24) m_H^2 \vec{q}^{\; 2}\right)
   B_0(m_H^2)\Bigg\} \; , \\
   \frac{\mathcal{A}_{gg \rightarrow Hg}^{(1), (b)} \; \mathcal{A}_{gg \rightarrow Hg}^{(0)*}}{|\mathcal{A}_{gg \rightarrow Hg}^{(0)}|^2} = - \frac{3}{2} g^2 C_A B_0(q^{ 2}) \; ,
   \end{gather}
   \begin{equation}
   \frac{\mathcal{A}_{gg \rightarrow Hg}^{(1), (c)} \; \mathcal{A}_{gg \rightarrow Hg}^{(0)*} }{|\mathcal{A}_{gg \rightarrow Hg}^{(0)}|^2} = g^2 C_A \frac{(1- \epsilon)}{2(3-2 \epsilon)}  B_0(m_H^2) \; .
   \end{equation}
These results coincide with those found in~\cite{Celiberto:2022fgx} for the corresponding diagrams in which the lower gluon line is replaced by the quark line. This confirms that these diagrams are dominated by the Gribov's approximation on the propagator shown in red in diagrams $(a)$, $(b)$, $(c)$. In fact, when the Gribov's trick is applied to those propagators, the lower Born vertex factorizes completely and the final result becomes insensitive to the specific target. The sum of the diagrams $(d)$ and $(e)$ in the high-energy approximation reduces to
\begin{gather}
    \frac{\mathcal{A}_{gg \rightarrow Hg}^{(1), (d)+(e)} \; \mathcal{A}_{gg \rightarrow Hg}^{(0)*} }{|\mathcal{A}_{gg \rightarrow Hg}^{(0)}|^2} \nonumber \\ = \omega^{(1)} (t) \left[ - \frac{3}{8 \epsilon} + \frac{1}{2(1-\epsilon)} - \frac{5}{4(1-2 \epsilon)} - \frac{1}{3-2 \epsilon} + \frac{n_f}{2 C_A} \frac{2 (1 - \epsilon)^3 + \epsilon^2}{(1-\epsilon)^2(1-2 \epsilon) (3-2 \epsilon)} \right] \; ,
\end{gather}
which coincides with the helicity-conserving part\footnote{The helicity-non-conserving part vanishes when we take the interference with the Born amplitude.} of the irreducible three-gluon vertex contribution in Eq.~(56) of Ref.~\cite{Fadin:2001dc}. This confirms that also in these diagrams the Gribov contribution is the only dominant one in the high-energy limit. Similarly, diagram $(f)$ gives the usual $t$-channel gluon self energy
\begin{equation}
     \frac{\mathcal{A}_{gg \rightarrow Hg}^{(1), (f)} \; \mathcal{A}_{gg \rightarrow Hg}^{(0)*} }{|\mathcal{A}_{gg \rightarrow Hg}^{(0)}|^2} = \omega^{(1)} (t) \frac{(5-3 \epsilon) C_A - (1-\epsilon) n_f}{2 (1-2 \epsilon) (3-2 \epsilon) C_A} \; .
\end{equation}
We therefore conclude that all single $t$-channel gluon exchange diagrams are dominated by the usual Gribov's approximation, in complete analogy to what is found for the $\mathcal{A}_{gq \rightarrow Hq}$ amplitude. \\

We now turn our attention to the two-gluon $t$-channel exchange diagrams. The three diagrams that do not vanish are identical to those of the amplitude $\mathcal{A}_{gq \rightarrow Hq}$ if the lower gluon line is replaced by the quark one. Again, the diagram $(i)$, which would be zero in a pure Gribov approximation, does not vanish and gives 
\begin{equation}
   \frac{\mathcal{A}_{gg \rightarrow Hg}^{(1), (i)} \; \mathcal{A}_{gg \rightarrow Hg}^{(0)*} }{|\mathcal{A}_{gg \rightarrow Hg}^{(0)}|^2} = g^2 C_A \left( \frac{ 8 \epsilon^3 - 18 \epsilon^2 + 7 \epsilon +2 }{8 \epsilon (\epsilon-1)^2} \right) B_0 (q^{2}) \;.
\end{equation}
In this result, the transverse terms in Eq.~(\ref{Gribov:Eq:GribovsTrickMod}) must be kept for both propagators. Diagrams $(g)$ and $(h)$, as expected, contain terms beyond the Gribov approximation; their sum of these latter reads 
\begin{equation}
   \frac{\mathcal{A}_{gg \rightarrow Hg}^{(1), (g)+(h)} \; \mathcal{A}_{gg \rightarrow Hg}^{(0)*} }{|\mathcal{A}_{gg \rightarrow Hg}^{(0)}|^2} = -g^2 C_A \left( \frac{24 \epsilon^2 -50 \epsilon +25}{8 (\epsilon-1)^2} \right) B_0 (q^{2}) \;.
\end{equation}
In this case, we observe that no $t$-channel propagator can be approximated {\it \`a la} Gribov. \\ 

It remains to be verified whether the correction found allows us to uniquely define the virtual contribution to the Higgs impact factor at one-loop. The sum of Regge limits for all interferences of Born and one-loop diagrams without approximations reads
\begin{gather}
  \mathcal{A}_{gg \rightarrow Hg}^{(1)} \; \mathcal{A}_{gg \rightarrow Hg}^{(0)*} =|\mathcal{A}_{gg \rightarrow Hg}^{(0)}|^2 C_A \frac{\Gamma^2 (1-\epsilon)}{\Gamma (1-2 \epsilon)} \frac{\bar{\alpha}_s}{4 \pi} \left( \frac{\vec{q}^{\; 2}}{\mu^2} \right)^{-\epsilon} \nonumber \\ \times \left \{ - \frac{3}{\epsilon^2}  + \frac{1}{\epsilon} \left( \ln \left( \frac{-s}{-t} \right) + \ln \left( \frac{s}{-t} \right)  \right) + 2 {\rm{Li}}_2 \left( 1 + \frac{m_H^2}{\vec{q}^{\; 2}} \right) + 4 \zeta (2) \right \} \; , \label{eq:gg-Hg_QCD-result}
\end{gather}
where
\begin{equation}
    \bar{\alpha}_s = \frac{g^2 \Gamma (1+\epsilon) \mu^{-2 \epsilon}}{(4 \pi)^{1-\epsilon}} \; . 
\end{equation}
This result agrees with previous results that can be found in literature~\cite{Schmidt:1997wr,Nefedov:2019mrg}. We can immediately cast the correction into the following form:
\begin{equation}
\begin{split}
     \frac{\mathcal{A}_{gg \rightarrow Hg}^{(1)} \; \mathcal{A}_{gg \rightarrow Hg}^{(0)*}}{|\mathcal{A}_{gg \rightarrow Hg}^{(0)}|^2} = & \; \frac{\bar{\alpha}_s}{4 \pi} \frac{\Gamma^2 (1-\epsilon)}{\Gamma (1-2 \epsilon)} \left( \frac{\vec{q}^{\; 2}}{\mu^2} \right)^{-\epsilon} \left \{ - \frac{3}{\epsilon^2} C_A + 2 C_A \left( {\rm{Li}}_2 \left( 1 + \frac{m_H^2}{\vec{q}^{\; 2}} \right) + 2 \zeta (2) \right) \right \} \\ 
     & + \frac{\omega^{(1)} (t)}{2} \left( \ln \left( \frac{s}{-t} \right) + \ln \left( \frac{-s}{-t} \right) \right) \; .
\end{split}
\end{equation}
To extract the correction to be associated to the gluon-Reggeon-Higgs vertex, we must subtract the contribution that goes into the Regge trajectory, as well as the contribution associated to the one-loop corrections to the gluon-Reggeon-gluon effective vertex, $\Gamma_{gg}^{c(1)}$. The latter contribution can be constructed using the one-loop result in Ref.~\cite{Fadin:2001dc} and gives
\begin{gather}
\frac{\Gamma_{gg}^{c(1)}}{\Gamma_{gg}^{c(0)}} \equiv  \delta_{G}^{(+)} = \frac{\bar{\alpha}_s}{4 \pi} \frac{\Gamma^2 (1-\epsilon)}{\Gamma (1-2 \epsilon)} \left( \frac{\vec{q}^{\; 2}}{\mu^2} \right)^{-\epsilon} \nonumber \\ \times \left \{ - \frac{2}{\epsilon^2} C_A - \left( \frac{11 C_A - 2 n_f}{6 \epsilon} \right) + \frac{5}{9} n_f - \frac{67}{18} C_A + 3 \zeta(2) C_A \right \} \; .
\end{gather}
Hence, we find that the one-loop correction to the Higgs vertex is
\begin{gather}
   \frac{\Gamma_{gH}^{ac(1)}}{\Gamma_{gH}^{ac(0)}} = \left( \frac{ \mathcal{A}_{gg \rightarrow Hg}^{(1)} \; \mathcal{A}_{gg \rightarrow Hg}^{(0)*} }{|\mathcal{A}_{gg \rightarrow Hg}^{(0)}|^2} - \frac{\omega^{(1)} (t)}{2} \left( \ln \left( \frac{s}{-t} \right) + \ln \left( \frac{-s}{-t} \right) \right) - \delta_{G}^{(+)} \right) \nonumber \\
   = \frac{\bar{\alpha}_s}{4 \pi} \left( \frac{\vec{q}^{\; 2}}{\mu^2} \right)^{-\epsilon} \left \{ - \frac{C_A}{\epsilon^2} + \frac{11 C_A - 2 n_f}{6 \epsilon} - \frac{5 n_f}{9} + C_A \left( 2 {\rm{Li}}_2 \left( 1 + \frac{m_H^2}{\vec{q}^{\; 2}} \right) +  2 \zeta (2) + \frac{67}{18} \right)  \right \} \; ,
   \label{Eq:Gribov:DeltaNLOHiggs}
\end{gather}
in agreement with \cite{Celiberto:2022fgx}. This calculation tells us that, using the idea proposed in Ref.~\cite{Celiberto:2022fgx}, we obtain a consistent result between the two channels $gq \rightarrow Hq$ and $gg \rightarrow gH$. It also confirms the full agreement between the calculation in~\cite{Celiberto:2022fgx} and the one in the Lipatov's EFT framework~\cite{Nefedov:2019mrg}. \\

Subtracting from the Eq.~(\ref{eq:gg-Hg_QCD-result}) the result of the same computation but with the Gribov prescription for all $t$-channel propagators, we obtain the non-Gribov contribution to the interference: 
\begin{equation}
    \delta_{gg \rightarrow Hg}^{\text{n.G.}} = \frac{\bar{\alpha}_s}{4 \pi} \left( \frac{\vec{q}^{\; 2}}{\mu^2} \right)^{-\epsilon} \frac{C_A}{4} \left[ \frac{1}{\epsilon^2} - \frac{5}{\epsilon} - 9 - \zeta(2) \right] + \mathcal{O} (\epsilon) \; .
    \label{Non-Gribov_gggH}
\end{equation}

At this point, it is natural to expect that this amplitude also has the factorisation-breaking contribution in the kinematical region B of the Gribov part. This calculation is presented in the next subsection.

\subsection{The region B of $\mathcal{A}_{gg \rightarrow Hg}$ and factorisation-violating eikonal terms}
\label{sec:fact-viol-gg}

In this section, we calculate the contribution of the kinematical region B\footnote{Recall the definition of this region at the beginning of section \ref{SubSec:StrategyOfRapidityRegionsQCD}.} of diagrams $(g)$ and $(h)$. The sum of the two contributions gives
\begin{gather}
    \mathcal{M}_B = \frac{4 g_H g^3}{s^2} C_A f^{abc} \varepsilon_{\perp, \mu} (k_1) \varepsilon_{\perp, \nu} (k_2) \varepsilon_{\gamma }^{*} (k_2') \int \frac{d^D k}{(2 \pi)^D} k_{2, \rho} k_{2, \zeta} k_{1, \lambda} k_{1, \delta} \nonumber \\ 
    \times \frac{  A^{\mu \rho}_{\; \; \; \sigma} \hspace{-0.10 cm} \left( -k, k_1+k \right) \hspace{-0.10 cm} H^{\sigma \zeta} \hspace{-0.10 cm} \left( -k_1-k, k-q \right) \hspace{-0.10 cm} A^{\lambda \nu}_{\; \; \; \beta} \hspace{-0.10 cm} \left( -k_2, k_2 - k \right) \hspace{-0.10 cm} A^{\delta \beta \gamma} \left( k-k_2, k_2-q \right) }{k^2 (k-q)^2 (k+k_1)^2 (k-k_2)^2} \; . 
\end{gather}
Performing the contractions and introducing the Sudakov variables, we get
\begin{gather}
    \Gamma_{gg'}^{(0)dbc} \left( \frac{2s}{t} \right) \frac{g_H \delta^{ad} \varepsilon_{\perp, \mu} (k_1)}{2} \left( -q^2 g^2 C_A \right) \int_{-\infty}^{\infty} \frac{d \alpha}{\alpha} \left( 1 - \frac{\alpha}{2} \right)^2 \int \frac{d^{D-2} k_{\perp}}{(2 \pi)^{D-1}} (q-k)^{\mu}_{\perp} \nonumber \\ \times \int_{-\infty}^{\infty} \frac{d (\beta s)}{2 \pi i} \frac{1}{(\alpha \beta s + k_{\perp}^2) (\beta (\alpha-1) s + k_{\perp}^2) (\alpha (\beta-\beta_q) s + (k-q)_{\perp}^2)} \; . 
\end{gather}
Performing the integration over $\beta$ and the tensor reduction on the $k_{\perp}$-integral, we obtain
\begin{equation}
    \Gamma_{gg'}^{(0)dbc} \left( \frac{2s}{t} \right) \Gamma_{gH}^{(0)ad} \frac{g^2 C_A q_{\perp}^2}{2} \int \frac{d^{D-2} k_{\perp}}{(2 \pi)^{D-1}} \frac{1}{k_{\perp}^2 (k_{\perp} - q_{\perp})^2} \int_{\alpha_0}^1 \frac{d \alpha}{\alpha} (1+\alpha) (1-\alpha)^{D-5} \left( 1 - \frac{\alpha}{2} \right)^2 \; .
\end{equation}
Now, after comparison with Ref.~\cite{Fadin:2001dc}, the first term in the bracket $\left( 1 + \alpha \right)$ gives the contribution which is assigned to the lower gluon vertex, while the second must be again assigned to the Higgs vertex and reads
\begin{gather}
    \delta_{gg\to Hg}^{(\rm{B})} = \omega^{(1)} (t) \int_{\alpha_0}^1 d \alpha \; (1-\alpha)^{-1-2 \epsilon} \left( 1 - \frac{\alpha}{2} \right)^2 
   \nonumber \\ = \omega^{(1)} (t) \left[ - \frac{1}{2 \epsilon} - \frac{\Gamma (-2 \epsilon)}{\Gamma (2-2 \epsilon)} + \frac{\Gamma (-2 \epsilon)}{2 \Gamma (3-2 \epsilon)} \right] \nonumber \\
    \simeq \frac{\bar{\alpha}_s}{4 \pi} \left( \frac{\vec{q}^{\; 2}}{\mu^2} \right)^{-\epsilon} \frac{C_A}{4} \left[ - \frac{1}{\epsilon^2} + \frac{5}{\epsilon} + 9 + \zeta(2) \right] + \mathcal{O} (\epsilon) \; .
\end{gather}
Recalling Eq.~(\ref{Non-Gribov_gggH}), we realize that
\begin{equation}
  \delta_{gg \rightarrow Hg}^{\text{n.G.}} + \delta_{gg\to Hg}^{(\rm{B})} = 0. 
\end{equation}
As it happened for the quark target, the two anomalies cancel each other out.

\subsection{Comparison with the Lipatov's EFT framework}
\label{subsec:IntLipatovEFT}
 
To facilitate understanding of this section, we provide a brief introduction to the gauge-invariant effective field theory for multi-Regge processes in QCD, also known as Lipatov's EFT framework. We closely follow Ref.~\cite{Nefedov:2019mrg} and adapt our notation to it.
In the EFT literature the following Sudakov decomposition for an arbitrary four-vector is used:
\begin{equation}
k^\mu=\frac{1}{2}\left(k_+n_-^\mu + k_-n_+^\mu \right) + k_T^\mu,  
\label{Eq:Sud-dec}
\end{equation} 
 where $n^\mu_-=(n^-)^\mu=2 k_2^\mu/\sqrt{s}$, $n^\mu_+=(n^+)^\mu=2 k_1^\mu/\sqrt{s}$ (with the momenta $k_{1,2}$ of Fig.~\ref{Fig:Gribov:GluonHiggs}), $s=(k_1+k_2)^2$, so that $n_{\pm}^2=0$, $n_+n_-=2$. We use the convention, where the position of $\pm$ indices has no meaning: $k_{\pm}=k^{\pm}=n_{\pm}k$ and $n_\pm k_T=0$. The square of a four-vector is given by $k^2=k_+k_- - {\bf k}_T^2$. \\ 

The fundamental idea of the High-Energy EFT~\cite{Lipatov:1995pn} is to slice the whole rapidity range into intervals, corresponding to clusters of particles, highly separated in rapidity. At leading power in energy, only Reggeized gluon exchanges in $t$-channels contribute to the amplitudes we are interested in, so we drop the Reggeized quark contributions to the Lagrangian.  For each interval of rapidity, a separate copy of the QCD Lagrangian is defined. The complete Lagrangian of the EFT is
\begin{equation}
  L_{\rm eff}= 4{\rm tr}\left[ R_+ \partial_T^2 R_- \right] + \sum\limits_i \left[ L_{\rm QCD}(A_\mu^{[y_i, y_{i+1}]}, \psi_q^{[y_i, y_{i+1}]}) + L_{\rm Rg}(A_\mu^{[y_i, y_{i+1}]}, R_+, R_-)   \right],   \label{Eq:LEFT}
 \end{equation}
where the index $[y_i, y_{i+1}]$ of the field means, that the real part of the rapidity of its momentum modes is restricted to lie within the interval $y_i\leq {\rm Re}(y) \leq y_{i+1}$ and $R_{\pm}$ are the Reggeized gluon fields which are scalar fields in the adjoint representation of the color group $SU(N_c)$ in this EFT. The kinetic part of the Lagrangian (\ref{Eq:LEFT}) leads to bare propagators, connecting the $R_+$-field with $R_-$: $-i/(2{q}_T^2)$, where $q_T$ is the transverse part of the momentum of the Reggeon. Due to MRK-kinematics, fields $R_{\pm}$ are subject to the following constraints:
 \begin{eqnarray}
 \partial_+ R_- = \partial_- R_+ = 0, \label{Eq:kin-constr-R}
  \end{eqnarray}
where $\partial_\pm=n_{\pm}^\mu\partial_\mu=2\partial/\partial x_{\mp}$. 

The fields $R_{\pm}$ are gauge invariant and this requirement mostly fixes the form of their ``tree-level'' interactions with the QCD gluons up to various $i\varepsilon$ prescriptions for the eikonal denominators. The latter prescriptions become important at loop level and can be determined from the requirement of factorisation and negative signature of the one-Reggeon exchange contributions~\cite{Hentschinski:2011xg}. The following Hermitian form of the effective action~\cite{Lipatov:1996ts,Bondarenko:2018pvv} satisfies requirements of definite signature and factorisation, introduced in the Ref.~\cite{Hentschinski:2011xg}:
\begin{equation}
  L_{Rg}(x)=\frac{i}{g_s}{\rm tr}\left[R_+(x) \partial_\rho^2 \partial_- \left(W_x[A_-]-W_x^\dagger[A_-]\right) + R_-(x) \partial_\rho^2\partial_+ \left(W_x[A_+]-W_x^\dagger[A_+]\right) \right], \label{Eq:L-Rg}
\end{equation}
where $W_x[A_{\pm}]$ is (past-pointing) half-infinite Wilson line, stretching in the $(+)$ or $(-)$ light-cone direction from the point $x$:
\begin{eqnarray}
  W_x[A_{\pm}]&=& P\exp\left[\frac{-ig_s}{2} \int\limits_{-\infty}^{x_{\mp}} dx'_{\mp} A_{\pm}\left(x_{\pm}, x'_{\mp}, {\bf x}_{T}\right)  \right] \nonumber \\
  &=& 1-ig_s\left(\partial_\pm^{-1}A_{\pm} \right) + (-ig_s)^2\left(\partial_\pm^{-1}A_{\pm}\partial_\pm^{-1}A_{\pm}\right)+\ldots , \label{Eq:WL-def}
  \end{eqnarray}
where we have defined operators $\partial^{-1}_{\pm}$ to act as $\partial^{-1}_{\pm}f(x)=\int\limits_{-\infty}^{x^{\mp}} dx'_{\mp}/2\ f(x_{\pm},x'_{\mp},{\bf x}_T)$ so that on the level of Feynman rules, they correspond to eikonal denominators with definite $i\varepsilon$-prescription:$-i/(k_{\pm}+i\varepsilon)$.  \\

The Lagrangian (\ref{Eq:L-Rg}) generates an infinite sequence of induced vertices of interaction of Reggeized gluon with $n$ QCD gluons. The simplest of them are the $R_+g$-transition vertex and $R_+gg$ interaction vertex, corresponding respectively to $O(g_s^0)$ and $O(g_s^1)$ terms in (\ref{Eq:L-Rg}):
\begin{eqnarray}
 \Delta_{+\mu_1}^{ab_1}(q,k_1)&=&(-iq^2)n^-_{\mu_1}\delta_{ab_1}, \label{eq:Rg-vertex-EFT} \\
 \Delta_{+\mu_1 \mu_2}^{ab_1 b_2}(q,k_1, k_2) &=&  g_s q^2 (n^-_{\mu_1} n^-_{\mu_2}) \frac{f^{ab_1b_2}}{[k_1^-]}, \label{eq:Rgg-vertex-EFT}
\end{eqnarray} 
where $q$ is the (incoming) momentum of the Reggeon with the color index $a$, $k_{1,2}$ are the (incoming) momenta of gluons coupled to the vertex, with their color (Lorentz) indices denoted as $b_{1,2}$($\mu_{1,2}$) and $k_1^-+k_2^-=0$ due to the MRK constraint (\ref{Eq:kin-constr-R}) and $1/[k_-] = [1/(k_-+i\varepsilon) + 1/(k_--i\varepsilon)]/2$ is the PV pole prescription. The vertex (\ref{eq:Rg-vertex-EFT}) is nothing but the Gribov's prescription, while the vertex (\ref{eq:Rgg-vertex-EFT}) coincides with the vertex (\ref{eq:chi-Rgg}) which we have introduced in the region analysis. \\

As seen from the QCD computations by the method of regions, when a rapidity separation is introduced, the various regions are affected by a new type of singularities, usually called \textit{rapidity divergences}. In the original formulation~\cite{Lipatov:1995pn} of the EFT such divergences are regulated by explicit cutoffs on the real part of rapidity that are imposed in each region: $y_i \leq {\rm Re} \{ y \} \leq y_{i+1}$. The dependence on the regulators, $y_i$'s, cancels between the contributions of neighboring clusters in each order of perturbation theory. 
The regularisation by the ``hard'' rapidity cutoff is inconvenient for the practical computations of Feynman integrals, therefore  in Ref.~\cite{Nefedov:2019mrg}, following the techniques proposed in~\cite{Hentschinski:2011tz,Chachamis:2012cc}, the approach based on \textit{tilted Wilson lines} was adopted, which boils down to the shift of the direction vectors of Wilson lines in the Lagrangian (\ref{Eq:L-Rg}):
\begin{equation}
    n_{\pm}^\mu \to \tilde{n}_{\pm}^\mu = n_{\pm}^\mu + r n_{\mp}^\mu,\ \frac{1}{[k_{\pm}]} \to \frac{1}{[\tilde{k}_{\pm}]}=\frac{1}{[k_{\pm}+ r k_{\mp}]},\ 0 < r \ll 1, \label{eq:r-reg}
\end{equation}
to the finite rapidity $\pm (\ln r)/2$, note that now $\tilde{n}_{\pm}^2=4r$ and $\tilde{n}_+ \tilde{n}_-=2(1+r)$. To keep the effective action gauge invariant at finite $r$, the MRK kinematic constraints (\ref{Eq:kin-constr-R}) also have to be modified to~\cite{Nefedov:2019mrg}
\begin{equation}
    (\partial_+ +r\partial_-) R_- = (\partial_- +r\partial_+) R_+ = 0. \label{eq:MRK-r}
\end{equation}


In Ref.~\cite{Nefedov:2019mrg} the Higgs$\to gR$ scattering vertex at one loop had been computed in the Lipatov's EFT formalism and it was found that, when properly combined with the one-loop correction to the Reggeized gluon propagator and $g\to Rg$ vertex, the obtained result reproduces the Regge limit of one-loop QCD amplitude (\ref{eq:gg-Hg_QCD-result}). From the discussion in the previous sections, it is clear that the correctness of the calculation made through the Lipatov's EFT is non-trivial. Indeed, the Reggeized gluon is a scalar particle in Lipatov's EFT and $t$-channel Reggeon exchanges can not lead to the anomalous helicity structures of the non-Gribov terms in the amplitude. In this case, the Gribov's trick (eikonal approximation) is already incorporated into the construction of the Reggeized gluon fields, leading to the vertex of Eq.~(\ref{eq:Rg-vertex-EFT}). On the other hand, effective interaction vertices with Reggeons in the EFT are obtained from expansions of Wilson lines that are insensitive to the specific ``projectile vertex''. This means that the factorisation breaking eikonal contribution of the region B is also excluded from the EFT calculation and this therefore explains the agreement with the QCD calculation. 

The explanation above relies on the fact that the loop corrections to both impact factors in the EFT are gauge-invariant and completely independent. In the computation of the  Higgs$\to gR$ vertex in Ref.~\cite{Nefedov:2019mrg} it turned out to be important to keep the $O(r)$ terms in the numerator of the amplitude, arising from the regularisation (\ref{eq:r-reg}) and modified kinematic constraint (\ref{eq:MRK-r}) to preserve the gauge-invariance of the result even at $O(r^0)$ due to the following mechanism. At one of the stages of tensor reduction of the diagrams (4), (6) and (7) in the Fig. 3 of Ref.~\cite{Nefedov:2019mrg}, the following integral appears:
\begin{equation}
 \int \frac{d^D l\ l^\mu}{l^2 (q-l)^2 [\tilde{l}_-]} = \frac{\tilde{n}_-^\mu}{4r} \int \frac{d^D l}{l^2 (q-l)^2}, \label{eq:sing-TR}
\end{equation}
with the momentum of the Reggeon $q$ satisfying the constraint $\tilde{q}_-=0$ following from Eq.~(\ref{eq:MRK-r}). The term proportional to $q^\mu$ in the r.h.s. of Eq.~(\ref{eq:sing-TR}) is absent because $\int d^D l/\{ l^2 (q-l)^2 [\tilde{l}_-] \} =0$. The singularity of the Eq.~(\ref{eq:sing-TR}) at $r\to 0$ cancels with some $O(r)$ terms in the numerator of the amplitude to produce contributions $\propto n_-\cdot \varepsilon(k_1)$, which are important for the gauge invariance of the $g\to R+$Higgs vertex.  

\section{Conclusions and outlook}
\label{sec:conclusions}

The resummation of large energy contributions to Higgs boson production channels, in a full next-to-leading logarithmic approximation, has received attention in recent years~\cite{Nefedov:2019mrg, Hentschinski:2020tbi, Celiberto:2020tmb, Celiberto:2022fgx, Andersen:2022zte,Rinaudo:2022nar}. A very useful tool, widely used in the literature, allowing to considerably simplify the computation, is the infinite-top-mass approximation. Nevertheless, in the infinite-top-mass limit, the gluon-gluon-Higgs coupling is described in terms of a local non-renormalizable operator of dimension five and therefore the \textit{Regge ansatz} for partonic scattering amplitudes should be questioned. \\

In this work, we carefully investigated the high-energy behavior of the one-loop $\mathcal{A}_{gq \rightarrow Hq}$ and $\mathcal{A}_{gg \rightarrow Hg}$ amplitudes in the aforementioned limit and found several non-trivial differences with respect to standard full QCD amplitudes. First of all, the presence of leading-$s$ non-eikonal contributions, {\it i.e.} terms which are dominant in the $s$-expansion of amplitudes, but do not come from the usual Gribov's effective polarization for the $t$-channel gluons. In the $\mathcal{A}_{gq \rightarrow Hq}$ amplitude such contributions couple helicities of the incoming gluon and quarks. We demonstrated that the aformentioned terms can be organized as the sum of a contribution that preserves the helicity structure of the Born amplitude and one that violates it. This latter contribution would invalidate the Regge form of the amplitude, but it disappears due to a cancellation across different diagrams, thus preserving the one-loop Regge form. However the non-Gribov terms still contribute even to the part of the amplitude proportional to the Born-level helicity structure. We also carefully analyzed the Gribov's part of the $\mathcal{A}_{gq \rightarrow Hq}$ amplitude using the \textit{strategy of rapidity regions}~\cite{Fadin:2001dc}. We  showed that a naive implementation of this technique fails and identified the culprit for this failure in the lack of rapidity regions partitioning of the amplitude. This latter is due to the non-factorizability of the gluon-Reggeon-Higgs effective vertex in the kinematic region where the longitudinal component of the loop momentum $k$ along $k_1$ is small (the region B in section~\ref{SubSec:StrategyOfRapidityRegionsQCD}) which leads to the appearance of the {\it factorisation-violating eikonal terms}. Interestingly, the two anomalous contributions, {\it i.e.} the {\it non-Gribov terms with the Born helicity structure} and {\it factorisation-violating eikonal terms}, cancel each other out in the full amplitude. Furthermore, we showed the universality of the one-loop correction and demonstrated the agreement between the result obtained in the standard BFKL approach and the one obtained using the Lipatov effective action. Even in the $\mathcal{A}_{gg \rightarrow Hg}$ amplitude, the {\it non-Gribov terms with the Born helicity structure} and {\it factorisation-violating eikonal terms} appear which again cancel each other out. In the EFT calculation both of these contributions are automatically excluded from the beginning due to the properties of factorisation and gauge invariance of impact-factors which are enforced by the Lipatov's EFT formalism. \\

There are several possible continuations for this work. The most natural is to calculate the effective vertex for the Higgs production at the physical value of $m_t$. In this case, given the absence of non-renormalizable interactions, one would not expect any anomalous behavior. On the other side, the Higgs impact factor in the complete theory still presents a peculiarity: it has a loop already at the leading order. A quark or gluon (but also a scalar particle) that interacts with a Reggeon can be thought of as a projectile that interacts with an infinite number of gluons as it propagates. In the high-energy limit, all these subsequent interactions are eikonal and therefore these gluon propagators in the $t$-channel can be always approximated {\it \`a la} Gribov. In the impact factor for the production of the Higgs, a quark/gluon line is replaced by an object containing a loop. In fact, the projectile is represented by a gluon fluctuating in a heavy $Q$-$\bar{Q}$ pair which, after the infinite series of interactions with the gluons in the $t$-channel, recouples to produce the Higgs. The eikonality of the subsequent interactions does not seem obvious \textit{a priori}. A successful determination of this impact factor would be of great practical use. Although very complex, its leading term in the $m_t$-expansion could provide an important confirmation of the calculation discussed here and in Ref.~\cite{Celiberto:2022fgx}~\footnote{From the phenomenological point of view, in Ref.~\cite{Celiberto:2020tmb}, it can be clearly seen that, for LHC center-of-mass energy, in the regime of validity of the BFKL approach, the infinite-top-mass is a reliable approximation.}. 

Another interesting development would be to test the Regge form of the $\mathcal{A}_{gq \rightarrow Hq}$ and $\mathcal{A}_{gg \rightarrow Hg}$ amplitudes at higher loops to understand whether the mechanism that preserves the Regge form is valid also beyond the one-loop order of perturbation theory.

\acknowledgments

We are indebted to Victor S. Fadin for helpful hints and comments. We also want to thank Renaud Boussarie, Francesco G. Celiberto, Saad Nabebaccus, Mohammed M. A. Mohammed, D. Yu. Ivanov, Lech Szymanowski and Samuel Wallon for discussions and Andreas van Hameren, Cyrille Marquet, Jean-Philippe Lansberg, Emmet Byrne and Giulio Falcioni for stimulating questions. 
The work of A.P. was supported by the INFN/QFT@COLLIDERS Project, Italy. The work of M.F. is supported by the Agence Nationale de la Recherche under the contract
ANR-17-CE31-0019. The work of M.N. has been supported by the Marie Skłodowska-Curie action “RadCor4HEF” under
grant agreement No. 101065263.
All the pictures in this work have been done using JaxoDraw~\cite{Binosi:2008ig}. \\

\appendix

\section{Higgs effective field theory framework}
\label{AppendixA}

\begin{figure}
  \begin{center}
  \includegraphics[scale=0.65]{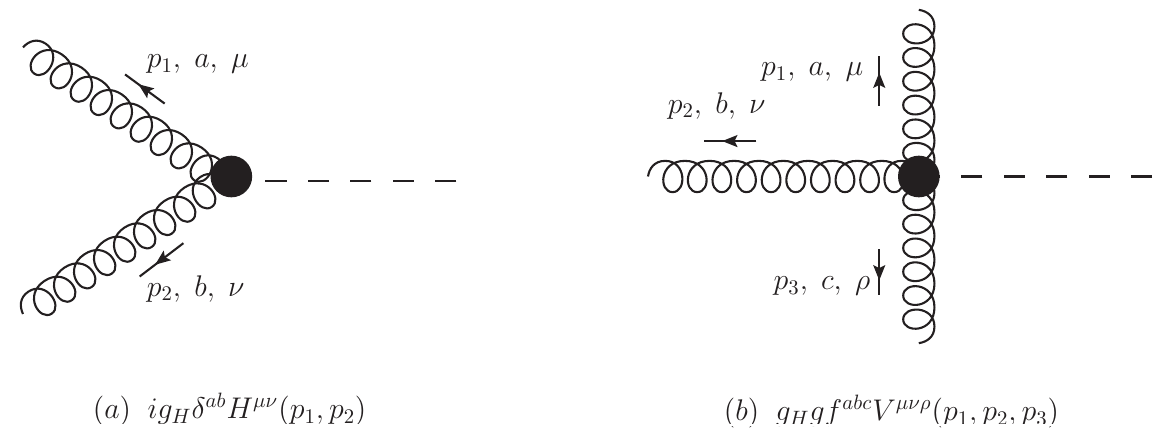}
  \end{center}
  \caption{Feynman rule for the $(a)$ $ggH$ vertex and $(b)$ $gggH$ vertex.}
  \label{HiggsDiagrams}
\end{figure}

In the approximation of the infinite mass of the top quark, the Higgs field couples to QCD {\it via} the effective Lagrangian,
\begin{equation}
\mathcal{L}_{ggH} = - \frac{g_H}{4} F_{\mu \nu}^{a} F^{\mu \nu,a} H \; ,
\label{EffLagrangia}
\end{equation}
where $H$ is the Higgs field, $F_{\mu \nu}^a = \partial_{\mu} A_{\nu}^a - \partial_{\nu} A_{\mu}^a  + g f^{abc} A_{\mu}^b A_{\nu}^c$ is the field strength tensor.
The Feynman rules associated to the Lagrangian~(\ref{EffLagrangia}) and used in this work are shown in Fig.~\ref{HiggsDiagrams}. The tensor structures appearing in Fig.~\ref{HiggsDiagrams} are
\begin{equation}
    H^{\mu \nu} (p_1,p_2) = g^{\mu \nu} (p_1 \cdot p_2) - p_1^{\nu} p_2^{\mu} \;,
\label{ggH}
\end{equation}
\begin{equation}
    V^{\mu \nu \rho} (p_1, p_2, p_3) = (p_1 - p_2)^{\rho} g^{\mu \nu} + (p_2 - p_3)^{\mu} g^{\nu \rho} + (p_3 - p_1)^{\nu} g^{\rho \mu} \;.
\label{gggH}
\end{equation}

\section{Useful integrals}
\label{UsefulIntegrals}
In this appendix we give a list of useful results.
\subsection*{One-loop scalar integrals in $D$ and $D-2$ dimensions}
We need the scalar integrals 
\begin{equation}
    B_0(-\vec{q}^{\; 2}) = \int \frac{d^D k}{i(2 \pi)^D} \frac{1}{k^2 (k+q)^2} = - \frac{1}{(4 \pi)^{2-\epsilon}} \frac{\Gamma
   (1+\epsilon) \Gamma^2 (-\epsilon)}{2 (1-2 \epsilon ) \Gamma (-2 \epsilon )} (\vec{q}^{\; 2})^{-\epsilon } \; ,
\end{equation}
\begin{equation*}
   C_0( q^{\; 2}, m_H^2) = \int \frac{d^D k}{i(2 \pi)^D} \frac{1}{k^2 (k+q)^2 (k+p_H)^2} 
\end{equation*}
\begin{equation}
    = \frac{1}{(4 \pi)^{2-\epsilon}} \frac{ \Gamma (1+\epsilon) \Gamma^2(-\epsilon)}{2 \Gamma (-2 \epsilon )} \frac{1}{\epsilon} \frac{\left((\vec{q}^{\;2})^{-\epsilon }-\left(-m_H^2\right)^{-\epsilon }\right)}{m_H^2 + \vec{q}^{\; 2}} \; , 
\end{equation}
\begin{equation*}
   D_0(m_H^2,q^{2},s) = \int \frac{d^D k}{i(2 \pi)^D} \frac{1}{k^2 (k-q)^2 (k+k_1)^2 (k-k_2)^2} 
\end{equation*}
\begin{equation}   
= \frac{\Gamma (1+\epsilon) \Gamma^2(-\epsilon)}{(4 \pi)^{2-\epsilon} \vec{q}^{\; 2} s} 
\frac{1}{\epsilon \Gamma (-2 \epsilon )} \left[(\vec{q}^{\; 2})^{-\epsilon }+(-s)^{-\epsilon}-\left(-m_H^2\right)^{-\epsilon}
\right.\end{equation}
\begin{equation*}
\left. -\epsilon^2 \left(\text{Li}_2\left(1+\frac{m_H^2}{\vec{q}^{\; 2}}\right)+\frac{1}{2} \ln
   ^2\left(-\frac{\vec{q}^{\; 2}}{s}\right)+\frac{\pi ^2}{3}\right) \right]
   + {\cal O}(\epsilon) \; .
\end{equation*}
In the last integral, the kinematic constraints $(k_1+q)^2=p_H^2=m_H^2$ and $(q-k_2)^2=(k_2')^2=0$ have been used. \\

We also extensively use the Feynman integral  
\begin{equation}
    \int \frac{d^{D-2} \vec{k}}{(2 \pi)^{D-2}} \frac{1}{[ \vec{k}^{\; 2} + L ]^a} = \frac{1}{(4 \pi)^{\frac{D-2}{2}}}\frac{\Gamma \left( a - \frac{D-2}{2} \right)} {\Gamma (a)} L^{\frac{D-2}{2}-a} \; .  
    \label{Eq:AppendixB:FeynmanA=1}
\end{equation}
\subsection*{Universal integral of the central region}
We calculate here the integral
\begin{gather}
  \hspace{-0.15 cm}  I^{\rm{central}} = \frac{s}{2} \int_{-\alpha_0}^{\alpha_0}  \int_{-\beta_0}^{\beta_0} \int \frac{d^{D-2} k}{(2 \pi)^{D}i} \frac{d \alpha \; d \beta}{(\alpha \beta s + k_{\perp}^2+i0) (\alpha \beta s + (q-k)_{\perp}^2 + i0) (-\beta s + i 0) (\alpha s + i0)} \;. \nonumber 
\end{gather}
We first perform the longitudinal integrations; we thus define the auxiliary integral
\begin{gather}
    I \equiv \int_{-\alpha_0}^{\alpha_0} d \alpha \int_{-\beta_0}^{\beta_0} d \beta \frac{1}{\left[ \vec{k}^{2} - (\vec{q}-\vec{k})^2 \right]} \frac{1}{(-\beta s + i 0) (\alpha s + i0)}  \nonumber \\ \times \left[ \frac{1}{(\alpha \beta s + k_{\perp}^2+i0)} - \frac{1}{(\alpha \beta s + (q-k)_{\perp}^2 + i0)} \right] \equiv I_1 + I_1 (k_{\perp} \rightarrow (q-k)_{\perp}) \; .
\end{gather}
$I_1$ can be further decomposed as
\begin{gather}
    I_1 = \frac{1}{\left[ \vec{k}^{\; 2} - (\vec{q}-\vec{k})^2 \right]} \int_{-\alpha_0}^{\alpha_0} d \alpha \int_{-\beta_0}^{\beta_0} d \beta  \frac{1}{(\alpha \beta s + k_{\perp}^2+i0)} \frac{1}{(-\beta s + i 0) (\alpha s + i0)} \nonumber \\ \equiv - \frac{1}{\vec{k}^{\; 2}} \frac{1}{\left[ \vec{k}^{\; 2} - (\vec{q}-\vec{k})^2 \right]} \left( J(\vec{k}^{\; 2}) - J') \right) \; ,
\end{gather}
where the integral $J'$ is\footnote{We calculate it in the sense of the principal value.}
\begin{equation}
    J' = \int_{-\alpha_0}^{\alpha_0} d \alpha \frac{1}{\alpha s + i0} \int_{-\beta_0}^{\beta_0} d \beta \frac{1}{\beta s - i0} = \frac{\pi^2}{s^2} \; ,
\end{equation}
while 
\begin{equation}
    J(\vec{k}^{\; 2}) = \int_{-\alpha_0}^{\alpha_0} d \alpha \int_{-\beta_0}^{\beta_0} d \beta \frac{\alpha}{(\alpha s + i0) (\alpha \beta s + k_{\perp}^2+i0)} \; .
\end{equation}
To compute this integral, we split the $\alpha$-domain in three regions: 1) $[-\alpha_0, - \frac{\vec{k}^{\, 2}}{s \beta_0}]$, 2) $[-\frac{\vec{k}^{\, 2}}{s \beta_0}, \frac{\vec{k}^{\, 2}}{s \beta_0}]$ and 3)$[\frac{\vec{k}^{\, 2}}{s \beta_0},\alpha_0]$. In this way, we have\footnote{$P$ indicates the principal value.}
\begin{gather*}
    J^{(1)}(\vec{k}^{\, 2}) = \int_{-\alpha_0}^{- \frac{\vec{k}^{\, 2}}{s \beta_0}} d \alpha \int_{-\beta_0}^{\beta_0} d \beta \frac{\alpha}{(\alpha s + i0) (\alpha \beta s + k_{\perp}^2+i0)} \\ = \int_{-\alpha_0}^{- \frac{\vec{k}^{\, 2}}{s \beta_0}} d \alpha \frac{\alpha}{(\alpha s + i0)} \int_{-\beta_0}^{\beta_0} d \beta \left[ P \left( \frac{1}{\alpha \beta s + k_{\perp}^2} \right) - i \pi \delta(\alpha \beta s + k_{\perp}^2) \right] 
\end{gather*}
\begin{equation}
    =\frac{1}{s^2} \int_{-\alpha_0}^{- \frac{\vec{k}^2}{s \beta_0}} d \alpha \frac{1}{\alpha} \ln \left( \frac{-\alpha \beta_0 s -k_{\perp}^{2}}{-\alpha \beta_0 s + k_{\perp}^{2}} \right) + \frac{i \pi}{s^2} \ln \left( \frac{-k_{\perp}^{2}}{\alpha_0 \beta_0 s} \right) 
\end{equation}
and similarly 
\begin{gather}
    J^{(3)}(\vec{k}^{\, 2}) = \int_{\frac{\vec{k}^{\, 2}}{s \beta_0}}^{{\alpha_0}} d \alpha \int_{-\beta_0}^{\beta_0} d \beta \frac{\alpha}{(\alpha s + i0) (\alpha \beta s + k_{\perp}^2+i0)} \nonumber \\ = - \frac{1}{s^2} \int_{-\alpha_0}^{- \frac{\vec{k}^2}{s \beta_0}} d \alpha \frac{1}{\alpha} \ln \left( \frac{-\alpha \beta_0 s +k_{\perp}^{2}}{-\alpha \beta_0 s - k_{\perp}^{2}} \right) + \frac{i \pi}{s^2} \ln \left( \frac{-k_{\perp}^{2}}{\alpha_0 \beta_0 s} \right) \; .
\end{gather}
The central region gives
\begin{equation}
    J^{(2)}(\vec{k}^{\, 2}) = \int_{-\frac{\vec{k}^{\, 2}}{s \beta_0}}^{\frac{\vec{k}^{\, 2}}{s \beta_0}} d \alpha \int_{-\beta_0}^{\beta_0} d \beta \frac{\alpha}{(\alpha s + i0) (\alpha \beta s + k_{\perp}^2+i0)} = - \frac{\pi^2}{2 s^2} \; .
\end{equation}
Summing the three contributions, the result is
\begin{gather}
    J(\vec{k}^{\, 2}) = - \frac{\pi^2}{2 s^2} + \frac{2 \pi i}{s^2} \ln \left( \frac{-k_{\perp}^{2}}{\alpha_0 \beta_0 s} \right) + \frac{1}{s^2} \int_{-\alpha_0}^{- \frac{\vec{k}^{\, 2}}{s \beta_0}} d \alpha \frac{1}{\alpha} \ln \left( \frac{(-\alpha \beta_0 s -k_{\perp}^{2})}{(-\alpha \beta_0 s + k_{\perp}^{2})} \frac{(-\alpha \beta_0 s -k_{\perp}^{2})}{(-\alpha \beta_0 s + k_{\perp}^{2})} \right) \nonumber \\ \simeq - \frac{\pi^2}{s^2} + \frac{2 \pi i}{s^2} \ln \left( \frac{-k_{\perp}^{2}}{\alpha_0 \beta_0 s} \right) = - \frac{\pi^2}{s^2} - \frac{2 \pi i}{s^2} \ln \left( \frac{\alpha_0 \beta_0 s}{-k_{\perp}^{2}} \right) \; .
\end{gather}
Using
\begin{equation}
    \ln \left( \frac{s}{|t|} \right) - i \pi = \ln \left( \frac{s}{t} \right) \; ,
\end{equation}
we can reconstruct
\begin{gather}
    I = I_1 + I_1 (k_{\perp} \rightarrow (q-k)_{\perp}) =  \frac{1}{\vec{k}^{\, 2} - (\vec{q}-\vec{k})^{2}} \nonumber \\ \times \left \{ \frac{1}{\vec{k}^{\, 2}} \frac{2 \pi i}{s^2} \ln \left( \frac{-\alpha_0 \beta_0 s}{-k_{\perp}^{\, 2}} \right) - \frac{1}{(\vec{q}-\vec{k})^{2}} \frac{2 \pi i}{s^2} \ln \left( \frac{-\alpha_0 \beta_0 s}{-(q-k)_{\perp}^{2}} \right) \right \} \; .
\end{gather}
Finally, we come back to the original integral in Eq.~(\ref{eq:I_central}), which reads
\begin{gather}
  I^{{\rm{central}}} = \frac{1}{2s} \int \frac{d^{D-2} k}{(2 \pi)^{D-1}}  \frac{1}{\vec{k}^{\, 2} - (\vec{q}-\vec{k})^{2}} \left \{ \frac{1}{\vec{k}^{\, 2}} \ln \left( \frac{-\alpha_0 \beta_0 s}{-k_{\perp}^{2}} \right) - \frac{1}{(\vec{q}-\vec{k})^{2}}  \ln \left( \frac{-\alpha_0 \beta_0 s}{-(q-k)_{\perp}^{2}} \right) \right \} ,
\end{gather}
and perform the transverse momentum integration, to get
\begin{gather}
I^{{\rm{central}}} =  \frac{ \Gamma(1+\epsilon)(\vec{q}^{\; 2})^{-\epsilon}}{(4 \pi)^{2-\epsilon}} \frac{\Gamma^2(-\epsilon)}{\Gamma(-2\epsilon)} \frac{1}{st} \left[ \ln \left( \frac{-s}{-t} \right) + \phi \left( \alpha_0 \right) +  \phi \left( \beta_0 \right) \right] \; .
\end{gather}

\bibliographystyle{apsrev}
\bibliography{references}

\begin{thebibliography}{71}
\expandafter\ifx\csname natexlab\endcsname\relax\def\natexlab#1{#1}\fi
\expandafter\ifx\csname bibnamefont\endcsname\relax
  \def\bibnamefont#1{#1}\fi
\expandafter\ifx\csname bibfnamefont\endcsname\relax
  \def\bibfnamefont#1{#1}\fi
\expandafter\ifx\csname citenamefont\endcsname\relax
  \def\citenamefont#1{#1}\fi
\expandafter\ifx\csname url\endcsname\relax
  \def\url#1{\texttt{#1}}\fi
\expandafter\ifx\csname urlprefix\endcsname\relax\def\urlprefix{URL }\fi
\providecommand{\bibinfo}[2]{#2}
\providecommand{\eprint}[2][]{\url{#2}}

\bibitem[{\citenamefont{Fadin et~al.}(1975)\citenamefont{Fadin, Kuraev, and Lipatov}}]{Fadin:1975cb}
\bibinfo{author}{\bibfnamefont{V.~S.} \bibnamefont{Fadin}}, \bibinfo{author}{\bibfnamefont{E.}~\bibnamefont{Kuraev}}, \bibnamefont{and} \bibinfo{author}{\bibfnamefont{L.}~\bibnamefont{Lipatov}}, \bibinfo{journal}{Phys. Lett. B} \textbf{\bibinfo{volume}{60}}, \bibinfo{pages}{50} (\bibinfo{year}{1975}).

\bibitem[{\citenamefont{Kuraev et~al.}(1976)\citenamefont{Kuraev, Lipatov, and Fadin}}]{Kuraev:1976ge}
\bibinfo{author}{\bibfnamefont{E.~A.} \bibnamefont{Kuraev}}, \bibinfo{author}{\bibfnamefont{L.~N.} \bibnamefont{Lipatov}}, \bibnamefont{and} \bibinfo{author}{\bibfnamefont{V.~S.} \bibnamefont{Fadin}}, \bibinfo{journal}{Sov. Phys. JETP} \textbf{\bibinfo{volume}{44}}, \bibinfo{pages}{443} (\bibinfo{year}{1976}).

\bibitem[{\citenamefont{Kuraev et~al.}(1977)\citenamefont{Kuraev, Lipatov, and Fadin}}]{Kuraev:1977fs}
\bibinfo{author}{\bibfnamefont{E.}~\bibnamefont{Kuraev}}, \bibinfo{author}{\bibfnamefont{L.}~\bibnamefont{Lipatov}}, \bibnamefont{and} \bibinfo{author}{\bibfnamefont{V.~S.} \bibnamefont{Fadin}}, \bibinfo{journal}{Sov.\ Phys.\ JETP} \textbf{\bibinfo{volume}{45}}, \bibinfo{pages}{199} (\bibinfo{year}{1977}).

\bibitem[{\citenamefont{Balitsky and Lipatov}(1978)}]{Balitsky:1978ic}
\bibinfo{author}{\bibfnamefont{I.}~\bibnamefont{Balitsky}} \bibnamefont{and} \bibinfo{author}{\bibfnamefont{L.}~\bibnamefont{Lipatov}}, \bibinfo{journal}{Sov.\ J.\ Nucl.\ Phys.} \textbf{\bibinfo{volume}{28}}, \bibinfo{pages}{822} (\bibinfo{year}{1978}).

\bibitem[{\citenamefont{Grisaru et~al.}(1973{\natexlab{a}})\citenamefont{Grisaru, Schnitzer, and Tsao}}]{Grisaru:1973vw}
\bibinfo{author}{\bibfnamefont{M.~T.} \bibnamefont{Grisaru}}, \bibinfo{author}{\bibfnamefont{H.~J.} \bibnamefont{Schnitzer}}, \bibnamefont{and} \bibinfo{author}{\bibfnamefont{H.-S.} \bibnamefont{Tsao}}, \bibinfo{journal}{Phys. Rev. Lett.} \textbf{\bibinfo{volume}{30}}, \bibinfo{pages}{811} (\bibinfo{year}{1973}{\natexlab{a}}).

\bibitem[{\citenamefont{Grisaru et~al.}(1973{\natexlab{b}})\citenamefont{Grisaru, Schnitzer, and Tsao}}]{Grisaru:1973wbb}
\bibinfo{author}{\bibfnamefont{M.~T.} \bibnamefont{Grisaru}}, \bibinfo{author}{\bibfnamefont{H.~J.} \bibnamefont{Schnitzer}}, \bibnamefont{and} \bibinfo{author}{\bibfnamefont{H.-S.} \bibnamefont{Tsao}}, \bibinfo{journal}{Phys. Rev. D} \textbf{\bibinfo{volume}{8}}, \bibinfo{pages}{4498} (\bibinfo{year}{1973}{\natexlab{b}}).

\bibitem[{\citenamefont{Lipatov}(1976)}]{Lipatov:1976zz}
\bibinfo{author}{\bibfnamefont{L.~N.} \bibnamefont{Lipatov}}, \bibinfo{journal}{Sov. J. Nucl. Phys.} \textbf{\bibinfo{volume}{23}}, \bibinfo{pages}{338} (\bibinfo{year}{1976}).

\bibitem[{\citenamefont{Balitsky et~al.}(1979)\citenamefont{Balitsky, Lipatov, and Fadin}}]{Balitsky:1979ap}
\bibinfo{author}{\bibfnamefont{I.~I.} \bibnamefont{Balitsky}}, \bibinfo{author}{\bibfnamefont{L.~N.} \bibnamefont{Lipatov}}, \bibnamefont{and} \bibinfo{author}{\bibfnamefont{V.~S.} \bibnamefont{Fadin}} (\bibinfo{year}{1979}).

\bibitem[{\citenamefont{Fadin et~al.}(2015)\citenamefont{Fadin, Kozlov, and Reznichenko}}]{Fadin:2015ym}
\bibinfo{author}{\bibfnamefont{V.}~\bibnamefont{Fadin}}, \bibinfo{author}{\bibfnamefont{M.}~\bibnamefont{Kozlov}}, \bibnamefont{and} \bibinfo{author}{\bibfnamefont{A.}~\bibnamefont{Reznichenko}}, \bibinfo{journal}{Physical Review D} \textbf{\bibinfo{volume}{92}} (\bibinfo{year}{2015}).

\bibitem[{\citenamefont{Fadin et~al.}(2006)\citenamefont{Fadin, Fiore, Kozlov, and Reznichenko}}]{Fadin:2006pr}
\bibinfo{author}{\bibfnamefont{V.}~\bibnamefont{Fadin}}, \bibinfo{author}{\bibfnamefont{R.}~\bibnamefont{Fiore}}, \bibinfo{author}{\bibfnamefont{M.}~\bibnamefont{Kozlov}}, \bibnamefont{and} \bibinfo{author}{\bibfnamefont{A.}~\bibnamefont{Reznichenko}}, \bibinfo{journal}{Physics Letters B} \textbf{\bibinfo{volume}{639}}, \bibinfo{pages}{74–81} (\bibinfo{year}{2006}), ISSN \bibinfo{issn}{0370-2693}, \urlprefix\url{http://dx.doi.org/10.1016/j.physletb.2006.03.031}.

\bibitem[{\citenamefont{Fadin and Lipatov}(1998)}]{Fadin:1998py}
\bibinfo{author}{\bibfnamefont{V.~S.} \bibnamefont{Fadin}} \bibnamefont{and} \bibinfo{author}{\bibfnamefont{L.~N.} \bibnamefont{Lipatov}}, \bibinfo{journal}{Phys. Lett. B} \textbf{\bibinfo{volume}{429}}, \bibinfo{pages}{127} (\bibinfo{year}{1998}), \eprint{hep-ph/9802290}.

\bibitem[{\citenamefont{Ciafaloni and Camici}(1998)}]{Ciafaloni:1998gs}
\bibinfo{author}{\bibfnamefont{M.}~\bibnamefont{Ciafaloni}} \bibnamefont{and} \bibinfo{author}{\bibfnamefont{G.}~\bibnamefont{Camici}}, \bibinfo{journal}{Phys. Lett. B} \textbf{\bibinfo{volume}{430}}, \bibinfo{pages}{349} (\bibinfo{year}{1998}), \eprint{hep-ph/9803389}.

\bibitem[{\citenamefont{Fadin et~al.}(1999)\citenamefont{Fadin, Fiore, and Papa}}]{Fadin:1998jv}
\bibinfo{author}{\bibfnamefont{V.~S.} \bibnamefont{Fadin}}, \bibinfo{author}{\bibfnamefont{R.}~\bibnamefont{Fiore}}, \bibnamefont{and} \bibinfo{author}{\bibfnamefont{A.}~\bibnamefont{Papa}}, \bibinfo{journal}{Phys. Rev. D} \textbf{\bibinfo{volume}{60}}, \bibinfo{pages}{074025} (\bibinfo{year}{1999}), \eprint{hep-ph/9812456}.

\bibitem[{\citenamefont{Fadin and Gorbachev}(2000{\natexlab{a}})}]{Fadin:2000kx}
\bibinfo{author}{\bibfnamefont{V.~S.} \bibnamefont{Fadin}} \bibnamefont{and} \bibinfo{author}{\bibfnamefont{D.~A.} \bibnamefont{Gorbachev}}, \bibinfo{journal}{JETP Lett.} \textbf{\bibinfo{volume}{71}}, \bibinfo{pages}{222} (\bibinfo{year}{2000}{\natexlab{a}}).

\bibitem[{\citenamefont{Fadin and Gorbachev}(2000{\natexlab{b}})}]{Fadin:2000hu}
\bibinfo{author}{\bibfnamefont{V.~S.} \bibnamefont{Fadin}} \bibnamefont{and} \bibinfo{author}{\bibfnamefont{D.~A.} \bibnamefont{Gorbachev}}, \bibinfo{journal}{Phys. Atom. Nucl.} \textbf{\bibinfo{volume}{63}}, \bibinfo{pages}{2157} (\bibinfo{year}{2000}{\natexlab{b}}).

\bibitem[{\citenamefont{Fadin and Fiore}(2005{\natexlab{a}})}]{Fadin:2004zq}
\bibinfo{author}{\bibfnamefont{V.~S.} \bibnamefont{Fadin}} \bibnamefont{and} \bibinfo{author}{\bibfnamefont{R.}~\bibnamefont{Fiore}}, \bibinfo{journal}{Phys. Lett. B} \textbf{\bibinfo{volume}{610}}, \bibinfo{pages}{61} (\bibinfo{year}{2005}{\natexlab{a}}), \bibinfo{note}{[Erratum: Phys.Lett.B 621, 320 (2005)]}, \eprint{hep-ph/0412386}.

\bibitem[{\citenamefont{Fadin and Fiore}(2005{\natexlab{b}})}]{Fadin:2005zj}
\bibinfo{author}{\bibfnamefont{V.~S.} \bibnamefont{Fadin}} \bibnamefont{and} \bibinfo{author}{\bibfnamefont{R.}~\bibnamefont{Fiore}}, \bibinfo{journal}{Phys. Rev. D} \textbf{\bibinfo{volume}{72}}, \bibinfo{pages}{014018} (\bibinfo{year}{2005}{\natexlab{b}}), \eprint{hep-ph/0502045}.

\bibitem[{\citenamefont{Byrne et~al.}(2022)\citenamefont{Byrne, Del~Duca, Dixon, Gardi, and Smillie}}]{Byrne:2022wzk}
\bibinfo{author}{\bibfnamefont{E.~P.} \bibnamefont{Byrne}}, \bibinfo{author}{\bibfnamefont{V.}~\bibnamefont{Del~Duca}}, \bibinfo{author}{\bibfnamefont{L.~J.} \bibnamefont{Dixon}}, \bibinfo{author}{\bibfnamefont{E.}~\bibnamefont{Gardi}}, \bibnamefont{and} \bibinfo{author}{\bibfnamefont{J.~M.} \bibnamefont{Smillie}}, \bibinfo{journal}{JHEP} \textbf{\bibinfo{volume}{08}}, \bibinfo{pages}{271} (\bibinfo{year}{2022}), \eprint{2204.12459}.

\bibitem[{\citenamefont{Del~Duca et~al.}(2022)\citenamefont{Del~Duca, Marzucca, and Verbeek}}]{DelDuca:2021vjq}
\bibinfo{author}{\bibfnamefont{V.}~\bibnamefont{Del~Duca}}, \bibinfo{author}{\bibfnamefont{R.}~\bibnamefont{Marzucca}}, \bibnamefont{and} \bibinfo{author}{\bibfnamefont{B.}~\bibnamefont{Verbeek}}, \bibinfo{journal}{JHEP} \textbf{\bibinfo{volume}{01}}, \bibinfo{pages}{149} (\bibinfo{year}{2022}), \eprint{2111.14265}.

\bibitem[{\citenamefont{Caola et~al.}(2022)\citenamefont{Caola, Chakraborty, Gambuti, von Manteuffel, and Tancredi}}]{Caola:2021izf}
\bibinfo{author}{\bibfnamefont{F.}~\bibnamefont{Caola}}, \bibinfo{author}{\bibfnamefont{A.}~\bibnamefont{Chakraborty}}, \bibinfo{author}{\bibfnamefont{G.}~\bibnamefont{Gambuti}}, \bibinfo{author}{\bibfnamefont{A.}~\bibnamefont{von Manteuffel}}, \bibnamefont{and} \bibinfo{author}{\bibfnamefont{L.}~\bibnamefont{Tancredi}}, \bibinfo{journal}{Phys. Rev. Lett.} \textbf{\bibinfo{volume}{128}}, \bibinfo{pages}{212001} (\bibinfo{year}{2022}), \eprint{2112.11097}.

\bibitem[{\citenamefont{Falcioni et~al.}(2022)\citenamefont{Falcioni, Gardi, Maher, Milloy, and Vernazza}}]{Falcioni:2021dgr}
\bibinfo{author}{\bibfnamefont{G.}~\bibnamefont{Falcioni}}, \bibinfo{author}{\bibfnamefont{E.}~\bibnamefont{Gardi}}, \bibinfo{author}{\bibfnamefont{N.}~\bibnamefont{Maher}}, \bibinfo{author}{\bibfnamefont{C.}~\bibnamefont{Milloy}}, \bibnamefont{and} \bibinfo{author}{\bibfnamefont{L.}~\bibnamefont{Vernazza}}, \bibinfo{journal}{Phys. Rev. Lett.} \textbf{\bibinfo{volume}{128}}, \bibinfo{pages}{132001} (\bibinfo{year}{2022}), \eprint{2112.11098}.

\bibitem[{\citenamefont{Fadin et~al.}(2023)\citenamefont{Fadin, Fucilla, and Papa}}]{Fadin:2023roz}
\bibinfo{author}{\bibfnamefont{V.~S.} \bibnamefont{Fadin}}, \bibinfo{author}{\bibfnamefont{M.}~\bibnamefont{Fucilla}}, \bibnamefont{and} \bibinfo{author}{\bibfnamefont{A.}~\bibnamefont{Papa}}, \bibinfo{journal}{JHEP} \textbf{\bibinfo{volume}{04}}, \bibinfo{pages}{137} (\bibinfo{year}{2023}), \eprint{2302.09868}.

\bibitem[{\citenamefont{Fadin et~al.}(2000{\natexlab{a}})\citenamefont{Fadin, Fiore, Kotsky, and Papa}}]{Fadin:1999de}
\bibinfo{author}{\bibfnamefont{V.~S.} \bibnamefont{Fadin}}, \bibinfo{author}{\bibfnamefont{R.}~\bibnamefont{Fiore}}, \bibinfo{author}{\bibfnamefont{M.~I.} \bibnamefont{Kotsky}}, \bibnamefont{and} \bibinfo{author}{\bibfnamefont{A.}~\bibnamefont{Papa}}, \bibinfo{journal}{Phys. Rev. D} \textbf{\bibinfo{volume}{61}}, \bibinfo{pages}{094005} (\bibinfo{year}{2000}{\natexlab{a}}), \eprint{hep-ph/9908264}.

\bibitem[{\citenamefont{Fadin et~al.}(2000{\natexlab{b}})\citenamefont{Fadin, Fiore, Kotsky, and Papa}}]{Fadin:1999df}
\bibinfo{author}{\bibfnamefont{V.~S.} \bibnamefont{Fadin}}, \bibinfo{author}{\bibfnamefont{R.}~\bibnamefont{Fiore}}, \bibinfo{author}{\bibfnamefont{M.~I.} \bibnamefont{Kotsky}}, \bibnamefont{and} \bibinfo{author}{\bibfnamefont{A.}~\bibnamefont{Papa}}, \bibinfo{journal}{Phys. Rev. D} \textbf{\bibinfo{volume}{61}}, \bibinfo{pages}{094006} (\bibinfo{year}{2000}{\natexlab{b}}), \eprint{hep-ph/9908265}.

\bibitem[{\citenamefont{Ciafaloni}(1998)}]{Ciafaloni:1998kx}
\bibinfo{author}{\bibfnamefont{M.}~\bibnamefont{Ciafaloni}}, \bibinfo{journal}{Phys. Lett. B} \textbf{\bibinfo{volume}{429}}, \bibinfo{pages}{363} (\bibinfo{year}{1998}), \eprint{hep-ph/9801322}.

\bibitem[{\citenamefont{Ciafaloni and Colferai}(1999)}]{Ciafaloni:1998hu}
\bibinfo{author}{\bibfnamefont{M.}~\bibnamefont{Ciafaloni}} \bibnamefont{and} \bibinfo{author}{\bibfnamefont{D.}~\bibnamefont{Colferai}}, \bibinfo{journal}{Nucl. Phys. B} \textbf{\bibinfo{volume}{538}}, \bibinfo{pages}{187} (\bibinfo{year}{1999}), \eprint{hep-ph/9806350}.

\bibitem[{\citenamefont{Ciafaloni and Rodrigo}(2000)}]{Ciafaloni:2000sq}
\bibinfo{author}{\bibfnamefont{M.}~\bibnamefont{Ciafaloni}} \bibnamefont{and} \bibinfo{author}{\bibfnamefont{G.}~\bibnamefont{Rodrigo}}, \bibinfo{journal}{JHEP} \textbf{\bibinfo{volume}{05}}, \bibinfo{pages}{042} (\bibinfo{year}{2000}), \eprint{hep-ph/0004033}.

\bibitem[{\citenamefont{Bartels et~al.}(2002{\natexlab{a}})\citenamefont{Bartels, Colferai, and Vacca}}]{Bartels:2001ge}
\bibinfo{author}{\bibfnamefont{J.}~\bibnamefont{Bartels}}, \bibinfo{author}{\bibfnamefont{D.}~\bibnamefont{Colferai}}, \bibnamefont{and} \bibinfo{author}{\bibfnamefont{G.~P.} \bibnamefont{Vacca}}, \bibinfo{journal}{Eur. Phys. J. C} \textbf{\bibinfo{volume}{24}}, \bibinfo{pages}{83} (\bibinfo{year}{2002}{\natexlab{a}}), \eprint{hep-ph/0112283}.

\bibitem[{\citenamefont{Bartels et~al.}(2003)\citenamefont{Bartels, Colferai, and Vacca}}]{Bartels:2002yj}
\bibinfo{author}{\bibfnamefont{J.}~\bibnamefont{Bartels}}, \bibinfo{author}{\bibfnamefont{D.}~\bibnamefont{Colferai}}, \bibnamefont{and} \bibinfo{author}{\bibfnamefont{G.~P.} \bibnamefont{Vacca}}, \bibinfo{journal}{Eur. Phys. J. C} \textbf{\bibinfo{volume}{29}}, \bibinfo{pages}{235} (\bibinfo{year}{2003}), \eprint{hep-ph/0206290}.

\bibitem[{\citenamefont{Caporale et~al.}(2012)\citenamefont{Caporale, Ivanov, Murdaca, Papa, and Perri}}]{Caporale:2011cc}
\bibinfo{author}{\bibfnamefont{F.}~\bibnamefont{Caporale}}, \bibinfo{author}{\bibfnamefont{D.~{\relax Yu}.} \bibnamefont{Ivanov}}, \bibinfo{author}{\bibfnamefont{B.}~\bibnamefont{Murdaca}}, \bibinfo{author}{\bibfnamefont{A.}~\bibnamefont{Papa}}, \bibnamefont{and} \bibinfo{author}{\bibfnamefont{A.}~\bibnamefont{Perri}}, \bibinfo{journal}{JHEP} \textbf{\bibinfo{volume}{02}}, \bibinfo{pages}{101} (\bibinfo{year}{2012}), \eprint{1112.3752}.

\bibitem[{\citenamefont{Ivanov and Papa}(2012{\natexlab{a}})}]{Ivanov:2012ms}
\bibinfo{author}{\bibfnamefont{D.~{\relax Yu}.} \bibnamefont{Ivanov}} \bibnamefont{and} \bibinfo{author}{\bibfnamefont{A.}~\bibnamefont{Papa}}, \bibinfo{journal}{JHEP} \textbf{\bibinfo{volume}{05}}, \bibinfo{pages}{086} (\bibinfo{year}{2012}{\natexlab{a}}), \eprint{1202.1082}.

\bibitem[{\citenamefont{Colferai and Niccoli}(2015)}]{Colferai:2015zfa}
\bibinfo{author}{\bibfnamefont{D.}~\bibnamefont{Colferai}} \bibnamefont{and} \bibinfo{author}{\bibfnamefont{A.}~\bibnamefont{Niccoli}}, \bibinfo{journal}{JHEP} \textbf{\bibinfo{volume}{04}}, \bibinfo{pages}{071} (\bibinfo{year}{2015}), \eprint{1501.07442}.

\bibitem[{\citenamefont{Ivanov and Papa}(2012{\natexlab{b}})}]{Ivanov:2012iv}
\bibinfo{author}{\bibfnamefont{D.~{\relax Yu}.} \bibnamefont{Ivanov}} \bibnamefont{and} \bibinfo{author}{\bibfnamefont{A.}~\bibnamefont{Papa}}, \bibinfo{journal}{JHEP} \textbf{\bibinfo{volume}{07}}, \bibinfo{pages}{045} (\bibinfo{year}{2012}{\natexlab{b}}), \eprint{1205.6068}.

\bibitem[{\citenamefont{Bartels et~al.}(2001)\citenamefont{Bartels, Gieseke, and Qiao}}]{Bartels:2000gt}
\bibinfo{author}{\bibfnamefont{J.}~\bibnamefont{Bartels}}, \bibinfo{author}{\bibfnamefont{S.}~\bibnamefont{Gieseke}}, \bibnamefont{and} \bibinfo{author}{\bibfnamefont{C.~F.} \bibnamefont{Qiao}}, \bibinfo{journal}{Phys. Rev. D} \textbf{\bibinfo{volume}{63}}, \bibinfo{pages}{056014} (\bibinfo{year}{2001}), \bibinfo{note}{[Erratum: Phys.Rev.D 65, 079902 (2002)]}, \eprint{hep-ph/0009102}.

\bibitem[{\citenamefont{Bartels et~al.}(2002{\natexlab{b}})\citenamefont{Bartels, Gieseke, and Kyrieleis}}]{Bartels:2001mv}
\bibinfo{author}{\bibfnamefont{J.}~\bibnamefont{Bartels}}, \bibinfo{author}{\bibfnamefont{S.}~\bibnamefont{Gieseke}}, \bibnamefont{and} \bibinfo{author}{\bibfnamefont{A.}~\bibnamefont{Kyrieleis}}, \bibinfo{journal}{Phys. Rev. D} \textbf{\bibinfo{volume}{65}}, \bibinfo{pages}{014006} (\bibinfo{year}{2002}{\natexlab{b}}), \eprint{hep-ph/0107152}.

\bibitem[{\citenamefont{Bartels et~al.}(2002{\natexlab{c}})\citenamefont{Bartels, Colferai, Gieseke, and Kyrieleis}}]{Bartels:2002uz}
\bibinfo{author}{\bibfnamefont{J.}~\bibnamefont{Bartels}}, \bibinfo{author}{\bibfnamefont{D.}~\bibnamefont{Colferai}}, \bibinfo{author}{\bibfnamefont{S.}~\bibnamefont{Gieseke}}, \bibnamefont{and} \bibinfo{author}{\bibfnamefont{A.}~\bibnamefont{Kyrieleis}}, \bibinfo{journal}{Phys. Rev. D} \textbf{\bibinfo{volume}{66}}, \bibinfo{pages}{094017} (\bibinfo{year}{2002}{\natexlab{c}}), \eprint{hep-ph/0208130}.

\bibitem[{\citenamefont{Bartels}(2003)}]{Bartels:2003zi}
\bibinfo{author}{\bibfnamefont{J.}~\bibnamefont{Bartels}}, \bibinfo{journal}{Nucl. Phys. B Proc. Suppl.} \textbf{\bibinfo{volume}{116}}, \bibinfo{pages}{126} (\bibinfo{year}{2003}).

\bibitem[{\citenamefont{Bartels and Kyrieleis}(2004)}]{Bartels:2004bi}
\bibinfo{author}{\bibfnamefont{J.}~\bibnamefont{Bartels}} \bibnamefont{and} \bibinfo{author}{\bibfnamefont{A.}~\bibnamefont{Kyrieleis}}, \bibinfo{journal}{Phys. Rev. D} \textbf{\bibinfo{volume}{70}}, \bibinfo{pages}{114003} (\bibinfo{year}{2004}), \eprint{hep-ph/0407051}.

\bibitem[{\citenamefont{Fadin et~al.}(2002)\citenamefont{Fadin, Ivanov, and Kotsky}}]{Fadin:2001ap}
\bibinfo{author}{\bibfnamefont{V.~S.} \bibnamefont{Fadin}}, \bibinfo{author}{\bibfnamefont{D.~{\relax Yu}.} \bibnamefont{Ivanov}}, \bibnamefont{and} \bibinfo{author}{\bibfnamefont{M.~I.} \bibnamefont{Kotsky}}, \bibinfo{journal}{Phys. Atom. Nucl.} \textbf{\bibinfo{volume}{65}}, \bibinfo{pages}{1513} (\bibinfo{year}{2002}), \eprint{hep-ph/0106099}.

\bibitem[{\citenamefont{Balitsky and Chirilli}(2013)}]{Balitsky:2012bs}
\bibinfo{author}{\bibfnamefont{I.}~\bibnamefont{Balitsky}} \bibnamefont{and} \bibinfo{author}{\bibfnamefont{G.~A.} \bibnamefont{Chirilli}}, \bibinfo{journal}{Phys. Rev. D} \textbf{\bibinfo{volume}{87}}, \bibinfo{pages}{014013} (\bibinfo{year}{2013}), \eprint{1207.3844}.

\bibitem[{\citenamefont{Hentschinski et~al.}(2021)\citenamefont{Hentschinski, Kutak, and van Hameren}}]{Hentschinski:2020tbi}
\bibinfo{author}{\bibfnamefont{M.}~\bibnamefont{Hentschinski}}, \bibinfo{author}{\bibfnamefont{K.}~\bibnamefont{Kutak}}, \bibnamefont{and} \bibinfo{author}{\bibfnamefont{A.}~\bibnamefont{van Hameren}}, \bibinfo{journal}{Eur. Phys. J. C} \textbf{\bibinfo{volume}{81}}, \bibinfo{pages}{112} (\bibinfo{year}{2021}), \bibinfo{note}{[Erratum: Eur. Phys. J. C 81, 262 (2021)]}, \eprint{2011.03193}.

\bibitem[{\citenamefont{Celiberto et~al.}(2022{\natexlab{a}})\citenamefont{Celiberto, Fucilla, Ivanov, Mohammed, and Papa}}]{Celiberto:2022fgx}
\bibinfo{author}{\bibfnamefont{F.~G.} \bibnamefont{Celiberto}}, \bibinfo{author}{\bibfnamefont{M.}~\bibnamefont{Fucilla}}, \bibinfo{author}{\bibfnamefont{D.~{\relax Yu}.} \bibnamefont{Ivanov}}, \bibinfo{author}{\bibfnamefont{M.~M.~A.} \bibnamefont{Mohammed}}, \bibnamefont{and} \bibinfo{author}{\bibfnamefont{A.}~\bibnamefont{Papa}}, \bibinfo{journal}{JHEP} \textbf{\bibinfo{volume}{08}}, \bibinfo{pages}{092} (\bibinfo{year}{2022}{\natexlab{a}}), \eprint{2205.02681}.

\bibitem[{\citenamefont{Nefedov}(2019)}]{Nefedov:2019mrg}
\bibinfo{author}{\bibfnamefont{M.~A.} \bibnamefont{Nefedov}}, \bibinfo{journal}{Nucl. Phys. B} \textbf{\bibinfo{volume}{946}}, \bibinfo{pages}{114715} (\bibinfo{year}{2019}), \eprint{1902.11030}.

\bibitem[{\citenamefont{Celiberto}(2021)}]{Celiberto:2020wpk}
\bibinfo{author}{\bibfnamefont{F.~G.} \bibnamefont{Celiberto}}, \bibinfo{journal}{Eur. Phys. J. C} \textbf{\bibinfo{volume}{81}}, \bibinfo{pages}{691} (\bibinfo{year}{2021}), \eprint{2008.07378}.

\bibitem[{\citenamefont{Celiberto}(2022)}]{Celiberto:2022keu}
\bibinfo{author}{\bibfnamefont{F.~G.} \bibnamefont{Celiberto}}, \bibinfo{journal}{Phys. Lett. B} \textbf{\bibinfo{volume}{835}}, \bibinfo{pages}{137554} (\bibinfo{year}{2022}), \eprint{2206.09413}.

\bibitem[{\citenamefont{Bolognino et~al.}(2018)\citenamefont{Bolognino, Celiberto, Ivanov, and Papa}}]{Bolognino:2018rhb}
\bibinfo{author}{\bibfnamefont{A.~D.} \bibnamefont{Bolognino}}, \bibinfo{author}{\bibfnamefont{F.~G.} \bibnamefont{Celiberto}}, \bibinfo{author}{\bibfnamefont{D.~{\relax Yu}.} \bibnamefont{Ivanov}}, \bibnamefont{and} \bibinfo{author}{\bibfnamefont{A.}~\bibnamefont{Papa}}, \bibinfo{journal}{Eur. Phys. J.} \textbf{\bibinfo{volume}{C78}}, \bibinfo{pages}{1023} (\bibinfo{year}{2018}), \eprint{1808.02395}.

\bibitem[{\citenamefont{Celiberto et~al.}(2021)\citenamefont{Celiberto, Ivanov, Mohammed, and Papa}}]{Celiberto:2020tmb}
\bibinfo{author}{\bibfnamefont{F.~G.} \bibnamefont{Celiberto}}, \bibinfo{author}{\bibfnamefont{D.~{\relax Yu}.} \bibnamefont{Ivanov}}, \bibinfo{author}{\bibfnamefont{M.~M.~A.} \bibnamefont{Mohammed}}, \bibnamefont{and} \bibinfo{author}{\bibfnamefont{A.}~\bibnamefont{Papa}}, \bibinfo{journal}{Eur. Phys. J. C} \textbf{\bibinfo{volume}{81}}, \bibinfo{pages}{293} (\bibinfo{year}{2021}), \eprint{2008.00501}.

\bibitem[{\citenamefont{Celiberto et~al.}(2022{\natexlab{b}})\citenamefont{Celiberto, Fucilla, Mohammed, and Papa}}]{Celiberto:2022zdg}
\bibinfo{author}{\bibfnamefont{F.~G.} \bibnamefont{Celiberto}}, \bibinfo{author}{\bibfnamefont{M.}~\bibnamefont{Fucilla}}, \bibinfo{author}{\bibfnamefont{M.~M.~A.} \bibnamefont{Mohammed}}, \bibnamefont{and} \bibinfo{author}{\bibfnamefont{A.}~\bibnamefont{Papa}}, \bibinfo{journal}{Phys. Rev. D} \textbf{\bibinfo{volume}{105}}, \bibinfo{pages}{114056} (\bibinfo{year}{2022}{\natexlab{b}}), \eprint{2205.13429}.

\bibitem[{\citenamefont{Celiberto and Papa}(2023)}]{Celiberto:2023rtu}
\bibinfo{author}{\bibfnamefont{F.~G.} \bibnamefont{Celiberto}} \bibnamefont{and} \bibinfo{author}{\bibfnamefont{A.}~\bibnamefont{Papa}} (\bibinfo{year}{2023}), \eprint{2305.00962}.

\bibitem[{\citenamefont{Celiberto et~al.}(2023{\natexlab{a}})\citenamefont{Celiberto, Delle~Rose, Fucilla, Gatto, and Papa}}]{Celiberto:2023uuk}
\bibinfo{author}{\bibfnamefont{F.~G.} \bibnamefont{Celiberto}}, \bibinfo{author}{\bibfnamefont{L.}~\bibnamefont{Delle~Rose}}, \bibinfo{author}{\bibfnamefont{M.}~\bibnamefont{Fucilla}}, \bibinfo{author}{\bibfnamefont{G.}~\bibnamefont{Gatto}}, \bibnamefont{and} \bibinfo{author}{\bibfnamefont{A.}~\bibnamefont{Papa}}, in \emph{\bibinfo{booktitle}{{57th Rencontres de Moriond on QCD and High Energy Interactions}}} (\bibinfo{year}{2023}{\natexlab{a}}), \eprint{2305.05052}.

\bibitem[{\citenamefont{Celiberto et~al.}(2023{\natexlab{b}})\citenamefont{Celiberto, Delle~Rose, Fucilla, Gatto, and Papa}}]{Celiberto:2023eba}
\bibinfo{author}{\bibfnamefont{F.~G.} \bibnamefont{Celiberto}}, \bibinfo{author}{\bibfnamefont{L.}~\bibnamefont{Delle~Rose}}, \bibinfo{author}{\bibfnamefont{M.}~\bibnamefont{Fucilla}}, \bibinfo{author}{\bibfnamefont{G.}~\bibnamefont{Gatto}}, \bibnamefont{and} \bibinfo{author}{\bibfnamefont{A.}~\bibnamefont{Papa}}, in \emph{\bibinfo{booktitle}{{16th International Symposium on Radiative Corrections: Applications of Quantum Field Theory to Phenomenology}}} (\bibinfo{year}{2023}{\natexlab{b}}), \eprint{2309.11573}.

\bibitem[{\citenamefont{Catani et~al.}(1990)\citenamefont{Catani, Ciafaloni, and Hautmann}}]{Catani:1990xk}
\bibinfo{author}{\bibfnamefont{S.}~\bibnamefont{Catani}}, \bibinfo{author}{\bibfnamefont{M.}~\bibnamefont{Ciafaloni}}, \bibnamefont{and} \bibinfo{author}{\bibfnamefont{F.}~\bibnamefont{Hautmann}}, \bibinfo{journal}{Phys. Lett. B} \textbf{\bibinfo{volume}{242}}, \bibinfo{pages}{97} (\bibinfo{year}{1990}).

\bibitem[{\citenamefont{Catani et~al.}(1991)\citenamefont{Catani, Ciafaloni, and Hautmann}}]{Catani:1990eg}
\bibinfo{author}{\bibfnamefont{S.}~\bibnamefont{Catani}}, \bibinfo{author}{\bibfnamefont{M.}~\bibnamefont{Ciafaloni}}, \bibnamefont{and} \bibinfo{author}{\bibfnamefont{F.}~\bibnamefont{Hautmann}}, \bibinfo{journal}{Nucl. Phys. B} \textbf{\bibinfo{volume}{366}}, \bibinfo{pages}{135} (\bibinfo{year}{1991}).

\bibitem[{\citenamefont{Hautmann}(2002)}]{Hautmann:2002tu}
\bibinfo{author}{\bibfnamefont{F.}~\bibnamefont{Hautmann}}, \bibinfo{journal}{Phys. Lett. B} \textbf{\bibinfo{volume}{535}}, \bibinfo{pages}{159} (\bibinfo{year}{2002}), \eprint{hep-ph/0203140}.

\bibitem[{\citenamefont{Fadin and Martin}(1999)}]{Fadin:1999qc}
\bibinfo{author}{\bibfnamefont{V.~S.} \bibnamefont{Fadin}} \bibnamefont{and} \bibinfo{author}{\bibfnamefont{A.~D.} \bibnamefont{Martin}}, \bibinfo{journal}{Phys. Rev. D} \textbf{\bibinfo{volume}{60}}, \bibinfo{pages}{114008} (\bibinfo{year}{1999}), \eprint{hep-ph/9904505}.

\bibitem[{\citenamefont{Fadin and Fiore}(2001)}]{Fadin:2001dc}
\bibinfo{author}{\bibfnamefont{V.~S.} \bibnamefont{Fadin}} \bibnamefont{and} \bibinfo{author}{\bibfnamefont{R.}~\bibnamefont{Fiore}}, \bibinfo{journal}{Phys. Rev. D} \textbf{\bibinfo{volume}{64}}, \bibinfo{pages}{114012} (\bibinfo{year}{2001}), \eprint{hep-ph/0107010}.

\bibitem[{\citenamefont{Mertig et~al.}(1991)\citenamefont{Mertig, Böhm, and Denner}}]{Mertig:1991ca}
\bibinfo{author}{\bibfnamefont{R.}~\bibnamefont{Mertig}}, \bibinfo{author}{\bibfnamefont{M.}~\bibnamefont{Böhm}}, \bibnamefont{and} \bibinfo{author}{\bibfnamefont{A.}~\bibnamefont{Denner}}, \bibinfo{journal}{Comput. Physics Commun.} \textbf{\bibinfo{volume}{64}}, \bibinfo{pages}{345} (\bibinfo{year}{1991}).

\bibitem[{\citenamefont{Shtabovenko et~al.}(2016)\citenamefont{Shtabovenko, Mertig, and Orellana}}]{Shtabovenko:2016cp}
\bibinfo{author}{\bibfnamefont{V.}~\bibnamefont{Shtabovenko}}, \bibinfo{author}{\bibfnamefont{R.}~\bibnamefont{Mertig}}, \bibnamefont{and} \bibinfo{author}{\bibfnamefont{F.}~\bibnamefont{Orellana}}, \bibinfo{journal}{Comput. Physics Commun.} \textbf{\bibinfo{volume}{207}}, \bibinfo{pages}{432} (\bibinfo{year}{2016}).

\bibitem[{\citenamefont{Shtabovenko}(2017)}]{Shtabovenko:2016whf}
\bibinfo{author}{\bibfnamefont{V.}~\bibnamefont{Shtabovenko}}, \bibinfo{journal}{Comput. Phys. Commun.} \textbf{\bibinfo{volume}{218}}, \bibinfo{pages}{48} (\bibinfo{year}{2017}), \eprint{1611.06793}.

\bibitem[{\citenamefont{Hahn}(2001)}]{Hahn:2000kx}
\bibinfo{author}{\bibfnamefont{T.}~\bibnamefont{Hahn}}, \bibinfo{journal}{Comput. Phys. Commun.} \textbf{\bibinfo{volume}{140}}, \bibinfo{pages}{418} (\bibinfo{year}{2001}), \eprint{hep-ph/0012260}.

\bibitem[{\citenamefont{Schmidt}(1997)}]{Schmidt:1997wr}
\bibinfo{author}{\bibfnamefont{C.~R.} \bibnamefont{Schmidt}}, \bibinfo{journal}{Phys. Lett. B} \textbf{\bibinfo{volume}{413}}, \bibinfo{pages}{391} (\bibinfo{year}{1997}), \eprint{hep-ph/9707448}.

\bibitem[{\citenamefont{Kublbeck et~al.}(1990)\citenamefont{Kublbeck, Bohm, and Denner}}]{Kublbeck:1990xc}
\bibinfo{author}{\bibfnamefont{J.}~\bibnamefont{Kublbeck}}, \bibinfo{author}{\bibfnamefont{M.}~\bibnamefont{Bohm}}, \bibnamefont{and} \bibinfo{author}{\bibfnamefont{A.}~\bibnamefont{Denner}}, \bibinfo{journal}{Comput. Phys. Commun.} \textbf{\bibinfo{volume}{60}}, \bibinfo{pages}{165} (\bibinfo{year}{1990}).

\bibitem[{\citenamefont{Lipatov}(1995)}]{Lipatov:1995pn}
\bibinfo{author}{\bibfnamefont{L.}~\bibnamefont{Lipatov}}, \bibinfo{journal}{Nucl. Phys. B} \textbf{\bibinfo{volume}{452}}, \bibinfo{pages}{369} (\bibinfo{year}{1995}), \eprint{hep-ph/9502308}.

\bibitem[{\citenamefont{Hentschinski}(2012)}]{Hentschinski:2011xg}
\bibinfo{author}{\bibfnamefont{M.}~\bibnamefont{Hentschinski}}, \bibinfo{journal}{Nucl. Phys. B} \textbf{\bibinfo{volume}{859}}, \bibinfo{pages}{129} (\bibinfo{year}{2012}), \eprint{1112.4509}.

\bibitem[{\citenamefont{Lipatov}(1997)}]{Lipatov:1996ts}
\bibinfo{author}{\bibfnamefont{L.~N.} \bibnamefont{Lipatov}}, \bibinfo{journal}{Phys. Rept.} \textbf{\bibinfo{volume}{286}}, \bibinfo{pages}{131} (\bibinfo{year}{1997}), \eprint{hep-ph/9610276}.

\bibitem[{\citenamefont{Bondarenko and Zubkov}(2018)}]{Bondarenko:2018pvv}
\bibinfo{author}{\bibfnamefont{S.}~\bibnamefont{Bondarenko}} \bibnamefont{and} \bibinfo{author}{\bibfnamefont{M.~A.} \bibnamefont{Zubkov}}, \bibinfo{journal}{Eur. Phys. J. C} \textbf{\bibinfo{volume}{78}}, \bibinfo{pages}{617} (\bibinfo{year}{2018}), \eprint{1801.08066}.

\bibitem[{\citenamefont{Hentschinski and Sabio~Vera}(2012)}]{Hentschinski:2011tz}
\bibinfo{author}{\bibfnamefont{M.}~\bibnamefont{Hentschinski}} \bibnamefont{and} \bibinfo{author}{\bibfnamefont{A.}~\bibnamefont{Sabio~Vera}}, \bibinfo{journal}{Phys. Rev. D} \textbf{\bibinfo{volume}{85}}, \bibinfo{pages}{056006} (\bibinfo{year}{2012}), \eprint{1110.6741}.

\bibitem[{\citenamefont{Chachamis et~al.}(2013)\citenamefont{Chachamis, Hentschinski, Madrigal~Mart\'\i{}nez, and Sabio~Vera}}]{Chachamis:2012cc}
\bibinfo{author}{\bibfnamefont{G.}~\bibnamefont{Chachamis}}, \bibinfo{author}{\bibfnamefont{M.}~\bibnamefont{Hentschinski}}, \bibinfo{author}{\bibfnamefont{J.~D.} \bibnamefont{Madrigal~Mart\'\i{}nez}}, \bibnamefont{and} \bibinfo{author}{\bibfnamefont{A.}~\bibnamefont{Sabio~Vera}}, \bibinfo{journal}{Phys. Rev. D} \textbf{\bibinfo{volume}{87}}, \bibinfo{pages}{076009} (\bibinfo{year}{2013}), \eprint{1212.4992}.

\bibitem[{\citenamefont{Andersen et~al.}(2023)\citenamefont{Andersen, Hassan, Maier, Paltrinieri, Papaefstathiou, and Smillie}}]{Andersen:2022zte}
\bibinfo{author}{\bibfnamefont{J.~R.} \bibnamefont{Andersen}}, \bibinfo{author}{\bibfnamefont{H.}~\bibnamefont{Hassan}}, \bibinfo{author}{\bibfnamefont{A.}~\bibnamefont{Maier}}, \bibinfo{author}{\bibfnamefont{J.}~\bibnamefont{Paltrinieri}}, \bibinfo{author}{\bibfnamefont{A.}~\bibnamefont{Papaefstathiou}}, \bibnamefont{and} \bibinfo{author}{\bibfnamefont{J.~M.} \bibnamefont{Smillie}}, \bibinfo{journal}{JHEP} \textbf{\bibinfo{volume}{03}}, \bibinfo{pages}{001} (\bibinfo{year}{2023}), \eprint{2210.10671}.

\bibitem[{\citenamefont{Rinaudo}(2023)}]{Rinaudo:2022nar}
\bibinfo{author}{\bibfnamefont{A.}~\bibnamefont{Rinaudo}}, \bibinfo{journal}{Acta Phys. Polon. Supp.} \textbf{\bibinfo{volume}{16}}, \bibinfo{pages}{47} (\bibinfo{year}{2023}), \eprint{2212.02959}.

\bibitem[{\citenamefont{Binosi et~al.}(2009)\citenamefont{Binosi, Collins, Kaufhold, and Theussl}}]{Binosi:2008ig}
\bibinfo{author}{\bibfnamefont{D.}~\bibnamefont{Binosi}}, \bibinfo{author}{\bibfnamefont{J.}~\bibnamefont{Collins}}, \bibinfo{author}{\bibfnamefont{C.}~\bibnamefont{Kaufhold}}, \bibnamefont{and} \bibinfo{author}{\bibfnamefont{L.}~\bibnamefont{Theussl}}, \bibinfo{journal}{Comput. Phys. Commun.} \textbf{\bibinfo{volume}{180}}, \bibinfo{pages}{1709} (\bibinfo{year}{2009}), \eprint{0811.4113}.

\end{thebibliography}

\end{document}